\documentstyle[12pt]{article} \textheight 24cm \topmargin -0.5cm
\def\ba{\begin{eqnarray}}
\def\ea{\end{eqnarray}}

\def\O{\hbox{\bf O}}

\def\lb{\label}
\def\be{\begin{equation}}
\def\ee{\end{equation}}

\def\G{\Gamma}

\begin{document}
\title{Manifolds with reduced holonomy in superstring theories}
\author{O.P.Santillan
\thanks{Bogoliubov Laboratory of Theoretical Physics, JINR, Dubna, Russia,
email: firenzecita@hotmail.com and osvaldo@theor.jinr.ru}}
\date{}
\maketitle

\tableofcontents

\section{Introduction}

   The Weinberg-Salam theory is a quantum field theory that
explain successfully two of the four known interactions of the
nature, the electromagnetic and the weak interactions. The strong
interaction is widely believed to be described by quantum
chromodynamics, which is a theory that fits successfully with deep
inelastic lepton nucleon scattering data. The remaining interaction,
namely gravity, is believed to be described at classical level, by
Einstein's General Theory of Relativity. The Weinberg-Salam model
plus quantum chromodynamics and Einstein gravity are incompatible
for small distances or, which is the same, for high energy scales of
the order of the Planck mass $M_{pl}\approx 10^{19}$ GeV. The
physics at this scale of energy is not known yet.

      Superstring theory attempts to explain the physical phenomena at this scale.
The present work deals with applications of manifolds with reduced
holonomy to supersymmetric and superstring theories. The problem of
classifying the possible holonomy groups for Riemannian and pseudo
Riemannian spaces was possed by Cartan in \cite{Cartan} and
partially solved by Berger in \cite{Berger}, who presented a list of
all the possible holonomy groups. Two special cases of this list,
relevant for are purposes, are the quaternion Kahler and hyperkahler
spaces. By definition these spaces are $4n$-dimensional with
holonomies included in the Lie groups $Sp(n)\times Sp(1)$ and
$Sp(n)$, respectively. Some of their properties has been
investigated in \cite{Wolf}, \cite{Ishihara}, and \cite{Proeyen} but
they are not completely classified at the present.

     It is convenient to explain what we mean when we speak
about an holonomy group. Consider a field $\psi$ defined over a
given point $p$ on a $n$-dimensional manifold $M$, and let us
parallel transport it around a given closed curve $\gamma$. In
general the result of this transport will be a new field
$\psi_{\gamma}$ at $p$ that is different than the original one
$\psi$. Let us define the transformation matrix $H_{\gamma}$ by
\be\lb{holonom} \psi_{\gamma}=H_{\gamma}\psi. \ee If we take into
account all the possible closed $C^1$-piecewise curves passing
through the point $p$ it can be shown that the corresponding
matrices $H(p)$ form a group. This is the full holonomy group of the
manifold at the point $p$ and is a subgroup of the orthogonal group
$O(n)$. If instead we consider only curves $\gamma$ that are
contractible to a point, we will obtain the so called restricted
holonomy group $H^0(p)$ which is a subgroup of $SO(n)$.

    If we change from the point $p$ to a point $q$ and fix some $C^1$-piecewise
curve $\beta$ connecting both point we will have the transformation
$$
H(q)= A(\beta)H(p)A(\beta)^{-1}
$$
and the same for $H^0(p)$. The matrix $A$ belongs to the holonomy
group $H(p)$ and it follows that $H(p)$ and $H(q)$ are isomorphic.
Therefore we can talk about holonomy groups without reference to
some specific point $p$. An space is orientable if and only if the
holonomy group is in $SO(n)$. If the fundamental group $\pi_1(M)=0$,
then $H=H^0$. We recall that there exist non simply connected
manifolds for which $\pi_1(M)=0$, an example is the flat tori
\cite{Besse}.

   From the definition of the curvature tensor
$R_{abcd}$ \cite{Landau} it is known the variation of $\psi$ under
an infinitesimal closed curve $\gamma$ is given by \be\lb{vario}
\delta \psi=G_{ab}\delta A^{ab}\psi. \ee Here $A^{ab}$ is the area
spanned by the curve $\gamma$ and $G_{ab}=R_{abcd}\gamma^{cd}$ being
$\gamma^{cd}$ some generators of $SO(n)$ in the representation of
the field (for applications to string theory we deal usually with
spinor fields $\psi$, but this is not important in our discussion).
Then it is clear that the "infinitesimal" holonomy group $H^i$ is
determined completely by the curvature tensor. Under infinitesimal
parallel transport $\psi$ satisfies \be\lb{infino}
\partial_i \psi=-\omega_{iab} \gamma^{ab}\psi,
\ee being $\omega^a_b$ the spin connection defined by
$$
de^a+\omega^a_b\wedge e^b=0.
$$
From (\ref{infino}) we infer that the transformation around a (not
necessarily infinitesimal) curve $\gamma$ is given by the operator
\be\lb{defhol} U(\gamma)=P \exp(-\int_{\gamma}\omega_{iab}
\gamma^{ab}dx^i) \ee where $P$ denote path-ordering of the
exponential \cite{Ambrose}. By considering all this curves we obtain
the full holonomy group, which could be larger than the
infinitesimal one.

    If the curve $\gamma$ is homotopic to the identity then the
matrices $U(\gamma)$ are generically in $SO(n)$ with the tangent
indices of $\omega^a_b$ taking values in the Lie algebra $so(n)$.
But there exists infinitely many examples in which the holonomy is
reduced to a subgroup of $SO(n)$. A characteristic feature of this
reduction is the presence of globally defined tensors $\alpha$ that
are invariant under parallel transport. This property is equivalent
to the annulation of their covariant derivative, i.e, $D\alpha=0$
\cite{Besse}. A globally parallel transported field is an analogous
concept of an invariant subspace for the Lie group $SO(n)$. The
presence of an invariant subspace implies the reduction of $SO(n)$
to another Lie group with lower dimension. A common feature of
reduced holonomy is in general the presence of covariantly constant
$p$-forms that play an analog role than those invariant subspaces.
For instance in $D=4$ we have the isomorphism $SO(4)\simeq
SU(2)_L\times SU(2)_R$, and $SU(2)$ holonomy spaces are
characterized by a covariantly constant right-handed (or
left-handed) spinor $\epsilon_{R}$ defined over the whole manifold.
The presence of $\epsilon_{R}$ implies that the curvature is
self-dual and that there exist a rotation of the frame taking the
anti-self-dual part of $\omega^a_b$ to zero. Examples with self-dual
curvature are the Eguchi-Hanson and Taub-Nut spaces. Also for $G_2$
holonomy manifolds one can choose an orthogonal frame $e^i$ in which
the octonionic 3-form
$$
\Phi=e^1 \wedge e^2 \wedge e^7 + e^1 \wedge e^3 \wedge e^6 + e^1
\wedge e^4 \wedge e^5 + e^2 \wedge e^3 \wedge e^5 + e^4 \wedge e^2
\wedge e^6
$$
$$
+\; e^3 \wedge e^4 \wedge e^7 + e^5 \wedge e^6 \wedge e^7
$$
and its dual $\ast\Phi$ are closed \cite{Gibb2}. This forms are
indeed invariant under the action of $G_2$ and their closure is
equivalent to the presence of a spinor $\eta$ such that $D_i\eta=0$.
We also recall that $G_2$ is indeed a subgroup of $SO(7)$ with a one
dimensional invariant subspace (as the space generated by $\eta$).

    During the last twenty years it became clear that quaternion Kahler
and hyperkahler geometry is an effective tool in field theory with
and without the presence of supersymmetries. For example the moduli
space of magnetic monopoles and the moduli space of Yang-Mills
instantons in flat space are hyperkahler \cite{Atiyo} , \cite{AHDM}.
Hyperkahler geometry also describe the couplings of supersymmetric
sigma models with N=4 rigid supersymmetries
\cite{Hitchon},\cite{Galicki} and \cite{Alvarez}. If a Wess-Zumino
type term is included, then the resulting geometry will be
hyperkahler torsion (HKT) \cite{Witten}. When the supersymmetry is
local the hypermultiplets appears coupled to gravity and the
resulting target space is a quaternionic Kahler manifold
\cite{Bagger}. Quaternionic manifolds also characterize the
hypermultiplet geometry of classical and perturbative moduli spaces
of type II strings compactified on a Calabi-Yau manifold
\cite{Rocco}, and even taking into account non perturbative
corrections due to D-instantons \cite{Ketov}. Therefore there
appears an interesting problem to construct quaternion Kahler spaces
with singularities \cite{Pedersen}. Other applications of this
geometry can be found in \cite{Fre}-\cite{deWit} and references
therein.

   One of the latest achievements for constructing this type of manifolds
is the hyperkahler quotient \cite{Hitchon}, \cite{Kronheimer}, a
simple example of this quotient is realized in the ADHM construction
\cite{AHDM}. This is a construction of hyperkahler manifolds of a
given dimension taking the quotient of a higher dimensional
hyperkahler one by certain group generating tri-holomorphic
isometries, and it have a physical origin related to dualities in
supersymmetric sigma models \cite{Hitchon}. A sort of inverse method
is due to Swann \cite{Swann} who shows how a quaternionic Kahler
metric in D=$4n$ can be extended to a quaternionic Kahler and
hyperkahler examples in D=$4(n+1)$. The Swann construction was
applied recently to construct hyperkahler cones in \cite{Anguelova},
relevant in theories with N=2 rigid supersymmetries, and to
construct certain scalar manifolds in M-theory on a Calabi-Yau
threefold in the vicinity of a flop transition \cite{mohap}.

    Hypergeometry (that is quaternion Kahler and hyperkahler
geometry) is not the only case of the Berger list with applications
in modern theoretical physics. Spaces of $SU(3)$, $G_2$ and
$Spin(7)$ holonomy play a crucial role as internal spaces of
heterotic string and M-theory preserving certain amount of
supersymmetries \cite{Papadopolus}, \cite{Duff}. The main feature of
such manifolds is the presence of parallel spinors fields $\eta_i$.
The number of supersymmetries preserved by the usual Kaluza-Klein
reduction is related to the number of such spinors over the internal
space. In general, if a given Riemannian metric with dimension $n$
admits at least one covariantly constant spinor $\eta$ satisfying
$D_i\eta=0$ the holonomy group will be $SU(\frac{n}{2})$,
$Sp(\frac{n}{4})$, $G_2$ or $Spin(7)$. The last two cases
corresponds to seven and eight dimensions. For $G_2$ (and $Spin(7)$)
holonomy manifolds there is exactly one $\eta$.

      In fact after the completion of the Berger work it
remained as an open question the existence of metrics with
exceptional holonomies $G_2$ and $Spin(7)$. This problem was solved
by Bryant and Salamon thirty years after the appearance of the
Berger list \cite{Bryant}. They showed that such spaces indeed exist
by providing a family of examples \cite{Salamon}. The construction
of this family was based on an interesting link between quaternion
kahler and special holonomy spaces. This link allows to find a large
class of spaces with special holonomy as extensions of four
dimensional quaternion Kahler manifolds. This is called the
Bryant-Salamon extension, and has certain analogies with the Swann
one.

    Spaces with special holonomy has vanishing Ricci tensor, i.e,
they are Ricci-flat. Their curvature tensor $R_{abcd}$ satisfies a
generalization of self-duality condition to $D=7$ namely
$$
R_{ab}=\pm\frac{c_{abcd}}{2}R_{cd}.
$$
The octonion constants $c_{abcd}$ in $D=7$ play an analogous role of
the Kronecker symbols $\varepsilon_{abcd}$ in $D=4$. Self-duality
implies Ricci-flatness and $G_2$ restricted holonomy \cite{Acharya}
(analogous considerations hold in eight dimensions for $Spin(7)$).
Although self-duality gives a non linear system of equations they
have been solved in cases with suitable symmetries \cite{Gibb2}.

      After the Bryant work explicit compact and non-compact
$G_2$ holonomy metrics were constructed in \cite{Lolo} and
\cite{Salamon}, and complete ones in \cite{Gomis} and
\cite{Branbauer}. Recently, new examples have been found in
\cite{Joyce3}, \cite{Konishi}, \cite{Gibbi}, \cite{Gibbano},
\cite{Chong}, \cite{Lu} and \cite{Gibbe}. There are in the
literature examples of "weak $G_2$ holonomy" \cite{Duff}, which are
again backgrounds of the M-theory that give rise to N=1
supersymmetry in $D=4$. In this case, there is an spinor field
$\eta$ which is not covariantly constant but satisfies
$D_i\eta\sim\lambda\gamma_i \eta$. The Ricci flatness condition is
replaced by $R_{ij}\sim \lambda g_{ij}$. In the limit
$\lambda\rightarrow 0$ one obtain $G_2$ as restricted holonomy
group. Hitchin has shown that under certain conditions is possible
to construct this kind of manifolds starting with an $Spin(7)$
holonomy one \cite{Hitcho}. Physically weak $G_2$ spaces are
supersymmetric backgrounds in presence of fluxes.

   Although there is a big progress in the study of special holonomy spaces, there
are also some open problems, in particular related to Kaluza-Klein
theories. A realistic compactification of M-theory should give rise
to chiral matter and this can not be obtained by applying
Kaluza-Klein procedure over $G_2$ smooth manifolds \cite{Witten4}.
In the smooth case the harmonic Kaluza-Klein decomposition of the
eleven-dimensional supergravity is the N=1 four dimensional
supergravity coupled Abelian vector multiplets. But the chiral
matter fields can emerge only if the manifold develops a
singularity, as pointed out by Witten and Acharya in \cite{Acharya2}
and \cite{Witten2}. It turns out that to obtain a realistic model
one should investigate the dynamics of the M-theory over orbifolds.
Another request is that the internal space should be compact. The
existence of compact spaces with special holonomy and orbifold
singularities was proved rigorously by Joyce in \cite{Joyce3}. But
no explicit metric over such spaces is known yet.

     Fortunately, some examples with weak $G_2$ holonomy that
are compact and admitting certain kind of singularities are known
\cite{Bilal2}. Moreover Witten has shown that the physics near a
singularity is local in essence and independent on the global
properties of the internal space \cite{Witten}. For this reason
there is still much interest in construct special holonomy manifolds
with conical singularities without the restriction to be compact.

   It should be mentioned that the study of explicit
metrics with exceptional holonomy has also importance in the context
of dualities of string theory and M-theory
\cite{Vafa}-\cite{Kachru}. Also generalizations of the work of
Acharya and Witten on singular $G_2$ spaces were investigated
recently in \cite{Berglund}, and over complete spaces with torus
symmetry and admitting only orbifold singularities in
\cite{Angelova} and \cite{Angelito}. Moreover compact $G_2$-holonomy
spaces with orbifold singularities are believed to arise as
quotients of a certain conical hyperkahler manifold in D=8 by one of
its isometries \cite{Witten2}, thus providing a new link between
special holonomy manifolds and hypergeometry. The range of
applications of this topic is very wide; some of them can be found
in \cite{Partuche}-\cite{Acho}.
\\

{\bf The purpose of this dissertation}
\\

    This dissertation is based on the works of the author during his
studies in order to get the degree of Candidate in Sciences
\cite{yo},\cite{yo2}, \cite{yo3} and \cite{yo4} and is related to
applications of reduced holonomy manifolds and hyperkahler torsion
geometry to string and M-theory. Our aim is the following:
\\

a) To show that in four dimensions weak hyperkahler torsion
structures are the same than hypercomplex structures, and also
equivalent to certain self-dual structures considered by Plebanski
and Finley \cite{yo4}.
\\

b) To find the most general form for a weak hyperkahler torsion
metric in four dimensions and discuss some particular examples
\cite{yo4}.
\\

c) To construct the hyperkahler target space metric corresponding to
several matter hypermultiplets of type IIA superstring compactified
on a Calabi-Yau threefold, even taking into account the D-instanton
contributions, by imposing the physical requirements due to Vafa and
Ooguri to a generic $4n$ dimensional hyperkahler metric \cite{yo3}.
\\

d) To construct a family of toric hyperkahler spaces with
tri-holomorphic killing vectors in eight dimensions. To lift the
result to new eleven dimensional backgrounds not related to D-brane
solutions. To find new IIA backgrounds by reduction along one of the
isometries and new IIB backgrounds by use of dualities \cite{yo2}.
\\

e) To construct a family of toric $G_2$ holonomy manifold and
$SU(3)$ torsion structures with two commuting isometries \cite{yo}.
To lift the result to new eleven dimensional backgrounds. To find
IIA backgrounds by reduction along one of the isometries and IIB
backgrounds by use of dualities \cite{yo2}.
\\

{\bf The structure of this dissertation}
\\

  In the preliminary material we explain in detail the elementary features about hyperkahler
geometry, hypercomplex structures, quaternion Kahler spaces, $G_2$
and $Spin(7)$ holonomy manifolds. The results of this section are
rather mathematical and not original at all, but the presentation is
new. We consider of great importance to make clear certain
mathematical features in simple language. In section 2.1.1 we
present the most important features about quaternion Kahler and
hyperkahler geometry in dimension higher than four. In section 2.1.2
we show that in four dimensions any quaternion Kahler space is
self-dual Einstein. In section 2.2 we give definition of an
hypercomplex structures. In this context we review in section 2.3
four dimensional hyperkahler geometry, in particular the
Ashtekar-Jacobson-Smolin and Plebanski formulations. This part is
importance for the points a) and b) mentioned above. Hyperkahler
spaces with isometries are presented in section 2.4, in particular
it is reviewed the Boyer-Finley classification of the possible
Killing vectors for this geometry. It is also remarked a geometrical
interpretation for the integrability of the axial $SU(\infty)$ Toda
equation that naturally appears in this context. In section 2.4.3 it
is presented a one to one link between self-dual conformal
structures and Einstein-Weyl structures, which is known as the
Jones-Tod correspondence. We also present in section 2.4.4 the most
general self-dual structures with two commuting isometries that are
non conformal to an hyperkahler space namely, the Joyce spaces. The
other part of section 2.4 present the subcases of the Joyce spaces
that are Einstein, and therefore quaternion Kahler. This are the
Calderbank-Pedersen spaces, which play an important role in the
present dissertation, in particular for the points d) and e).

   Section 2.5 concerns with certain aspects of higher dimensional
quaternion Kahler and hyperkahler geometry. In section 2.5.1 are
presented the most general $4n$ dimensional hyperkahler spaces with
$n$ commuting tri-holomorphic isometries, which is a higher
dimensional generalization of the Gibbons-Hawking metrics. The
metrics are written in the momentum map system, which is the most
suitable form for the purposes of this dissertation. This result
will be used in the point c) and d). We also present in section
2.5.3 the Swann extension, which is a construction of a higher
dimensional quaternion kahler space starting with other with less
dimensions. This part concerns mainly with the point d).

   Section 2.6 presents some features about the groups $G_2$ and $Spin(7)$
and the relation with the octonion algebra. We show the link between
$G_2$ and $Spin(7)$ holonomy manifolds, the presence of globally
defined covariantly constant spinor fields and octonionic
self-duality. We present the Bryant-Salamon construction, which
extend any quaternion kahler manifold in four dimensional to a seven
dimensional one with holonomy included in $G_2$ and to an eight
dimensional one with holonomy in $Spin(7)$. We also present spaces
with weak $G_2$ holonomy as the cones of certain $Spin(7)$ spaces,
and half-flat six dimensional manifolds as the cones of certain
$G_2$ spaces. We emphasize the relation between $G_2$ holonomy
manifolds and compactifications of $M$ theory preserving certain
amount of supersymmetries. This part is related to the point e).

  The remaining sections are devoted to solve the tasks a)-e) enumerated above.
The preliminary material plays an important role in order to
understand this section. In the conclusions we mention the main
results of the present work. We have included an appendix that
explain the relations between hyperkahler torsion geometry and
supersymmetric sigma models. At the end it is included the
bibliography and the publications of the author.

\newpage

\section{Preliminary material}

\subsection{Quaternion Kahler and hyperkahler spaces}

   By definition, a quaternion Kahler manifold is an euclidean $4n$
dimensional manifold with holonomy group $\Gamma$ included into the
Lie group $Sp(n)\times Sp(1)\subset SO(4n)$ \cite{Berger}. This
statement is non trivial if $D>4$, but in $D=4$ we have the
isomorphism $SO(4)\simeq SU(2)_L\times SU(2)_R \simeq Sp(1)\times
Sp(1)$ related to the orthogonal decomposition of spinors in left
and right states of chirality, and so the statement $\Gamma\subseteq
Sp(1)\times Sp(1)$ is trivial and should be modified. The first
subsection of this section concerns with spaces with dimension at
least $8$, the case $D=4$ will be treated separately. We will show
that a quaternionic Kahler metric has the following properties
\cite{Ishihara}, \cite{Wolf}.
\\

1) There exists three automorphism $J^i$ ($i=1$ ,$2$, $3$) of the
tangent space $TM_x$ at a given point $x$ such that $J^{i} \cdot
J^{j} = -\delta_{ij} + \epsilon_{ijk}J^{k}$, and for which the
metric $g$ is quaternion hermitian, that is \be\lb{hermoso}
g(X,Y)=g(J^i X, J^i Y), \ee being $X$ and $Y$ arbitrary vector
fields.
\\

2) The structures $J^i$ satisfy the fundamental relation
\be\lb{rela2} \nabla_{X}J^{i}=\epsilon_{ijk}J^{j}\omega_{-}^{k}, \ee
with $\nabla_{X}$ the Levi-Civita connection of the manifold and
$\omega_{-}^{i}$ its $Sp(1)$ part. As a consequence of hermiticity
of $g$ stated above the tensor
$\overline{J}^{i}_{ab}=(J^{i})_{a}^{c}g_{cb}$ is antisymmetric and
therefore there exists an associated 2-form
$$
\overline{J}^i=\overline{J}^{i}_{ab} e^a \wedge e^b
$$
for which
\be\lb{basta}
d\overline{J}^i=\epsilon_{ijk}\overline{J}^{j}\wedge\omega_{-}^{k},
\ee
being $d$ the usual exterior derivative.
\\

3) Corresponding to the $Sp(1)$ connection we can define the 2-form
$$
F^i=d\omega_{-}^i+\epsilon_{ijk}\omega_{-}^j \wedge \omega_{-}^k.
$$
Then for a quaternion Kahler manifold
\be\lb{lamas} R^i_{-}=\Lambda
\overline{J}^i, \ee \be\lb{rela} F^i=\Lambda' \overline{J}^i. \ee
being $\Lambda$ and $\Lambda'$ certain constants. The tensor
$R^a_{-}$ is the $Sp(1)$ part of the curvature. The last two
conditions implies that $g$ is Einstein with non zero cosmological
constant $R_{ij}\sim g_{ij}$.
\\

4) In a quaternion Kahler space the globally defined $(0,4)$ and
$(2,2)$ tensors
$$
\Theta=\overline{J}^1 \wedge \overline{J}^1 + \overline{J}^2 \wedge
\overline{J}^2 + \overline{J}^3 \wedge \overline{J}^3,
$$
$$
\Xi= J^1 \otimes J^1 + J^2 \otimes J^2 + J^3 \otimes J^3
$$
are covariantly constant with respect to the usual Levi Civita
connection.
\\

5) Any quaternion Kahler space is orientable.
\\

6) If there exists a local frame in which $\omega_{-}^i$ is equal to
zero, then the metric is called hyperkahler and is Kahler with
respect to any of the complex structures. The metric $g$ is in this
case Ricci-flat and the holonomy is reduced to a subgroup of
$Sp(n)$.
\\

7) In four dimensions quaternion Kahler spaces are the same than
Einstein spaces and with self-dual Weyl tensor.
\\

  This section is devoted to explain this concepts in more detail. By taken
granted properties 1-7 the reader can go directly to the next
section.

\subsubsection{Quaternionic Kahler spaces in dimension higher than four}

   It is well known that the generators $J^i$ of the Lie algebra
$sp(1)$ of $Sp(1)\simeq SU(2)$ have the multiplication rule
\be\lb{cape} J^{i}\cdot J^{j}=-\delta^{ij} I + \epsilon_{ijk}J^{k},
\ee which implies the $so(3)\simeq su(2)$ commutation rule
\be\lb{commutatus} [J^i, J^j]= \epsilon_{ijk}J^k. \ee We see that
$J^i J^i=-I$ and therefore $J^i$ will be called almost complex
structures. An useful $4n \times 4n$ representation is
$$
J^{1}=\left(\begin{array}{cccc}
  0 & -I_{n \times n} r& 0 & 0 \\
  I_{n \times n} & 0 & 0 & 0 \\
  0 & 0 & 0 & -I_{n \times n} \\
  0 & 0 & I_{n \times n} & 0
\end{array}\right),\;\;\;\;
J^{2}=\left(\begin{array}{cccc}
  0 & 0 & -I_{n \times n} & 0 \\
  0 & 0 & 0 & I_{n \times n} \\
   I_{n \times n} & 0 & 0 & 0 \\
  0 & -I_{n \times n} & 0 & 0
\end{array}\right)
$$
\be\lb{reprodui} J^{3}=J^{1}J^{2}=\left(\begin{array}{cccc}
  0 & 0 & 0 & -I_{n \times n} \\
  0 & 0 & -I_{n \times n} & 0 \\
  0 & I_{n \times n} & 0 & 0 \\
  I_{n \times n} & 0 & 0 & 0
\end{array}\right).
\ee
The group $SO(4n)$ is a Lie group and this means in particular
that for any $SO(4n)$ tensor $A^{a}_{b}$ the commutator $[A, J^{i}]$
will take also values in $SO(4n)$. We will say that $A$ belong to
the subgroup $Sp(n)$ of $SO(4n)$ if and only if
\be\lb{commutatus2}
[A, J^{i}]=0.
\ee
Condition (\ref{commutatus2}) together with
(\ref{commutatus}) implies that a tensor $B^{a}_{b}$ belongs to the
subgroup $Sp(n)\times Sp(1)$ if and only if
$$
[B,J^{i}]=\epsilon_{ijk}J^{j}B_{-}^{k},
$$
being $B_{-}^{k}$ the components of $B$ in the basis $J^{k}$. Both
conditions are independent of the representation.

   We will write a metric over a $4n$ dimensional manifold $M$
as $g=\delta_{ab}e^{a}\otimes e^{b}$, being ${e^{a}}$ the $4n$-bein
basis for which $g$ is diagonal. Let us define the triplet of
$(1,1)$ tensors \be\lb{bobo} J^{i}=(J^{i})^a_b e_{a}\otimes e^{b},
\ee defined by the matrices (\ref{reprodui}). If the holonomy is in
$Sp(n)\times Sp(1)$, then from the beginning $\omega^a_b$ will take
values on its lie algebra $sp(n)\oplus sp(1)$. As we saw above, this
implies that \be\lb{alban} [\omega,
J^{i}]=\epsilon_{ijk}J^{j}\omega_{-}^{k}. \ee As usual, the
connection $\omega^{a}_{b}$ is defined through
$$
\nabla_{X} e^{a}=-\omega^{a}_{b}(X)e^{b},
$$
together with the Levi-Civita conditions $\nabla g=0$ and
$T(X,Y)=0$. Using the chain rule $\nabla (A \otimes B)=(\nabla A)
\otimes B+ A \otimes (\nabla B)$ for tensorial products show us that
in the einbein basis \be\lb{alban2} [\omega, J^{i}]=\nabla_{X}J^{i}.
\ee Comparing (\ref{alban}) and (\ref{alban2}) we see that
quaternionic Kahler manifold are defined by the relation $$
\nabla_{X}J^{i}=\epsilon_{ijk}J^{j}\omega_{-}^{k},
$$
which is independent on the election of the frame $e^a$. This proves
that (\ref{rela2}) describe quaternion Kahler metrics
\cite{Ishihara} in dimension higher than four.

    The basis $e^a$ for a metric $g$ is defined up to an $SO(4n)$
rotation. Under this $SO(4n)$ transformation the tensors
(\ref{bobo}) are also transformed, but it can be shown that the
multiplication (\ref{cape}) is unaffected. In other words, given the
tensors $J^{i}$ one can construct a new set of complex structures
\be\lb{globos} J'^i= C^i_{j}J^{j},\qquad J'^{i}\cdot
J'^{j}=-\delta^{ij} I + \epsilon_{ijk}J'^{k}\qquad
\Longleftrightarrow\qquad C^i_{k} C^{k}_j=\delta^i_j \ee This can be
paraphrased by saying that a quaternionic Kahler manifold has a
bundle $V$ of complex structures parameterized by the sphere $S^2$.
Using the textbook properties of $\nabla$ it can be seen that
(\ref{rela2}) is unaltered under such rotations.

   Let us define three new tensors
$(\overline{J}^{i})_{ab}$ by
$(\overline{J}^{i})_{ab}=(J^{i})_{a}^{c}\delta_{cb}$. From
(\ref{reprodui}) it follows that
$$
(J^i)^a_b=-(J^i)^b_a \qquad\Longleftrightarrow\qquad
(\overline{J}^i)_{ab}=-(\overline{J}^i)_{ba}
$$
This show that $(\overline{J}^{i})_{ab}$ are the components of the
two-forms $\overline{J}^{i}$ defined by \be\lb{compo}
\overline{J}^{i}=(\overline{J}^{i})_{ab} e^{a}\wedge e^{b}. \ee The
forms (\ref{compo}) are known as the hyperkahler forms. From
(\ref{rela2}) it is obtained that
$$
\nabla_{X}J^{i}=\epsilon_{ijk}J^{j}\omega_{-}^{k}\qquad
\Longrightarrow \qquad d\overline{J}^{i}=\epsilon_{ijk}\omega_{-}^j
\wedge \overline{J}^{k},
$$ being $d$ the usual exterior derivative.
The last implication proves relation (\ref{basta}).

  If we change the frame $e^a$ to a new one $x_{\mu}$ then the definition
$(\overline{J}^{i})_{ab}=(J^{i})_{a}^{c}\delta_{cb}$ should be
modified by the covariant one
$(\overline{J}^{i})_{\alpha\beta}=(J^{i})_{\alpha}^{\gamma}g_{\gamma\beta}$.
Here the greek index indicates the components in the new basis and
$g_{\gamma\beta}$ are the corresponding components of the metric.
Therefore
$$
(\overline{J}^i)_{ab}=-(\overline{J}^i)_{ba}
\qquad\Longleftrightarrow\qquad
(J^{i})_{\alpha}^{\gamma}g_{\gamma\beta}=(J^{i})_{\beta}^{\gamma}g_{\gamma\alpha}
$$
The last relation is equivalent to $$ g(J^{i} X,Y)=g(X,
J^{i}Y)\qquad\Longleftrightarrow \qquad g(X,Y)=g(J^{i} X, J^{i}Y)$$
for arbitrary vector fields $X$ and $Y$ in $TM$. Then the metric $g$
will be always quaternion hermitian with respect to the complex
structures. Relation (\ref{hermoso}) is also invariant under the
automorphism of the complex structures.

    In general, if in a given manifold there exist three complex structures
satisfying (\ref{cape}), and we take intersecting coordinate
neighborhoods $U$ and $U'$, then we have two associated basis $J^i$
and $J'^i$. Both basis should be related by an $SO(3)$
transformation in order to satisfy (\ref{cape}). This means that any
quaternion Kahler space is orientable \cite{Ishihara}. Consider now
the fundamental 4-form \be\lb{lafunda} \Theta=\overline{J}^1 \wedge
\overline{J}^1 + \overline{J}^2 \wedge \overline{J}^2 +
\overline{J}^3 \wedge \overline{J}^3, \ee and the globally defined
$(2,2)$ tensor \be\lb{22} \Xi= J^1 \otimes J^1 + J^2 \otimes J^2 +
J^3 \otimes J^3. \ee By means of the formula (\ref{globos}) it
follows that both tensors (\ref{lafunda}) and (\ref{22}) are
globally defined on the manifold M. For a quaternionic Kahler
manifold it is obtained directly from (\ref{rela2}) and
(\ref{basta}) that \cite{Ishihara}
$$
\nabla \Theta=0,\;\;\;\;\;\;\; \nabla \Xi=0.
$$
In $D=8$ for a quaternion Kahler manifold $d\Theta=0$ and if the
manifold is of dimension at least $12$ then $d\Theta$ determines
completely $\nabla \Theta$. In particular $d\Theta=0$ implies
$\nabla \Theta=0$ \cite{Swann}.

      One of the most important consequences of (\ref{rela2}) is
that quaternionic Kahler spaces are always Einstein with
cosmological constant  \cite{Wolf}. The proof is briefly as follows.
From the definition of the curvature tensor $R(X,Y)=[\nabla_{X},
\nabla_{Y}]-\nabla_{[X,Y]}$ together with (\ref{rela2}) it follows
in the einbein basis that \be\lb{una}
R_{ijm}^{l}(J^{a})^m_{k}-R_{ijk}^{m}(J^{a})^l_{m}=
\epsilon_{abc}(F^{b})_{ij}(J^{c})^{l}_{k}. \ee where $R_{ijm}^{l}$
are the components of the curvature tensor and the two form $F^{a}$
was defined as \footnote{In the physical literature sometimes the
three components $\omega_{-}$ are referred as an $SU(2)$ vector
potential and $F^a$ as the corresponding strength tensor}
$$
F^a=d\omega_{-}^{a}+\epsilon_{abc}\omega_{-}^{b}\wedge
\omega_{-}^{c}.
$$
We can rewrite (\ref{una}) as a commutator
$$
[R(X,Y), J^a]=\epsilon_{abc}F^{b}J^{c},
$$
being $X$ and $Y$ arbitrary vector fields. Multiplying (\ref{una})
by $(J^{a})_l^{s}$ and contracting indices, and then multiplying by
$(J^{b})^{k}_{l}$ and using the identity \footnote{Which is clearly
true in the representation (\ref{reprodui})}
$$
(J^{a})_l^{s}(J^{b})_s^{l}=4n \delta^{ab},
$$
gives the formula \be\lb{labuena}
F^{a}_{ij}=\frac{1}{2n}R_{ijk}^{l}(J^{a})_{l}^{k}. \ee Inserting
(\ref{labuena}) into (\ref{una}) yields
$$
R_{ijk}^{l}(J^{a})_{l}^{k}=\frac{2n}{2+n}R_{im}(J^{a})_{j}^{m},
$$
which can also be expressed as \be\lb{lame} R^i_{-}=\frac{2n}{2+n}R
\overline{J}^i, \ee being $R$ is the scalar curvature and $R^i_{-}$
are the $Sp(1)$ components of the curvature tensor. The second
Bianchi identities together with (\ref{lame}) shows that $R$ is
constant and thus $R_{ij}\sim g_{ij}$ \cite{Wolf}. Thus, in any
dimension, quaternionic Kahler spaces are always Einstein with non
zero cosmological constant $\lambda$.

        Because $R$ is a constant we see from (\ref{lame}) that
$$
R^i_{-}=\Lambda \overline{J}^i,
$$ being $\Lambda$
certain constant. We also have from (\ref{labuena}) that
$$ F^i=\Lambda' \overline{J}^i,
$$
being $\Lambda'$ another constant. This condition are the same as
(\ref{lamas}) and (\ref{rela}) given above. In the limit
$\lambda\rightarrow 0$ the constants $\Lambda$ and $\Lambda'$ goes
simultaneously to zero.

       If there exists a rotation of the local frame for which
$\omega_{-}=0$ then the complex structures are locally covariantly
constant, that is \be\lb{hypi} \nabla_{X}J^{i}=0. \ee In this case
$R^i_{-}=F^i=0$ thus the space has self-dual curvature, which
implies Ricci flatness. This spaces are called hyperkahler, and
(\ref{hypi}) shows that they are Kahler with respect of any of the
complex structures. Condition (\ref{hypi}) implies that the holonomy
is in $Sp(n)$ and that \be\lb{nocon} d\overline{J}^a= 0 \ee together
with the annulation of the Niejenhuis tensor given by \be\lb{Nie}
N(X,Y)=[X,Y]+J[X,JY]+J[JX,Y]-[JX,JY], \ee \cite{Boyer}. A complex
structure for which $N(X,Y)=0$ is called integrable. We will explain
this relations in more detail in section 4.

\subsubsection{Quaternion Kahler manifolds in dimension four}

      As we saw starting the previous section, in four dimensions
the statement that the holonomy is $\Gamma \subseteq Sp(n)\times
Sp(1)$ is trivial due to the isomorphism $SO(4)\simeq SU(2)_L\times
SU(2)_R \simeq Sp(1)\times Sp(1)$. We will modify this definition
and we will say that a four dimensional manifold $M$ is quaternionic
Kahler if (\ref{lamas}) holds. This condition is not trivial, we
will show below that quaternion Kahler spaces in $d=4$ are Einstein
(as in the higher dimensional case) and with self-dual Weyl tensor.

     Let us consider a four dimensional metric
$g=\delta_{ab}e^a \otimes e^b$ and the connection $\omega^a_b$ given
by the first Cartan equation
$$
de^a + \omega^{a}_{b}\wedge e^b = 0,\;\;\;\; \omega_{ib}^a = -
\omega_{bi}^a.
$$
The notation $SU(2)_{\pm}$ denote the $SU(2)_L$ and $SU(2)_R$ groups
respectively. The $SU(2)_{\pm}$ components of the spin connection
are explicitly \be\lb{secon} \omega^{a}_{\pm}=\omega^a_{0}\pm
\epsilon_{abc}\omega^b_c. \ee The curvature tensor is given by the
second Cartan equation
$$
R^a_b=d\omega^a_b + \omega^a_s \wedge \omega^s_b = R^a_{b,st}e^s
\wedge e^t
$$
and the $SU(2)$ parts are \be\lb{secon2} R^{a}_{\pm}=R^a_{0}\pm
\epsilon_{abc}R^b_c. \ee The Ricci tensor is defined in the diagonal
basis by $R_{ij}=R^a_{i,aj}$ and the scalar curvature is $R_{ii}=R$.

    Instead of use the basis ${e^a \wedge e^b}$ we can use the basis
$\overline{J}^a_{\pm}=e^0 \wedge e^a \pm \epsilon _{abc} e^b \wedge
e^c$. Then it follows that $\overline{J}^a_{\pm}$ are separately
complex structure with definite self-duality properties, that is
$$
\ast \overline{J}^a_{\pm}=\pm \overline{J}^a_{\pm}.
$$
In this basis \be\lb{riemself} R^{a}_{+}=A_{ab}\overline{J}^{b}_{+}
+ B_{ab}\overline{J}^{b}_{-},\;\;\;
R^{a}_{-}=B_{ab}^{t}\overline{J}^{b}_{+} +
C_{ab}\overline{J}^{b}_{-} \ee where the matrices $A$ and $C$ are
symmetric. The components of the Ricci tensor are \be\lb{sericc}
R_{00}=Tr(A+B),\;\;\;
R_{0a}=\frac{\epsilon_{abc}}{2}(B_{bc}^t-B_{bc}),\;\;\;
R_{ab}=Tr(A-B)\delta_{ab} + B_{ab} + B_{ab}^t, \ee and the scalar
curvature is \be\lb{esca} R=4Tr(A)=4Tr(C). \ee It is clearly seen
from (\ref{sericc}) that the Einstein condition $R_{ij}=\Lambda
\delta_{ij}$ is equivalent to $B=0$ and $Tr(A)=Tr(C)=\Lambda$.

   The components of the Weyl tensor in the diagonal basis are given by
\be\lb{Weylo} W^{a}_{bcd}=
R^{a}_{bcd}-\frac{1}{2}(\delta_{ac}R_{bd}-\delta_{ad}R_{bc}
+\delta_{bd}R_{ac}-\delta_{bc}R_{ad})+\frac{R}{6}(\delta_{ac}\delta_{bd}
-\delta_{ad}\delta_{bc}). \ee The tensor $W$ is invariant under a
conformal transformation $g \rightarrow \Omega^2 g$ and the
associated two form is
$$
W^a_b=W^{a}_{bcd}e^c\wedge e^d.
$$
An explicit calculation shows that the $SU(2)_{\pm}$ of $W$ are
$$
W^{a}_{+}=W^a_{0}+ \epsilon_{abc}W^b_c=
(A_{ab}-\frac{1}{3}Tr(A)\delta_{ab})\overline{J}^b_{+},
$$
$$
W^{a}_{-}=W^a_{0}- \epsilon_{abc}W^b_c=
(C_{ab}-\frac{1}{3}Tr(C)\delta_{ab})\overline{J}^b_{-}.
$$
From this expressions we see that to say that an Einstein space is
self dual (i.e, $W^a_{-}=0$) is equivalent to \be\lb{lacond}
C_{ab}=\frac{\Lambda}{3}\delta_{ab}\;\;\;\Longleftrightarrow\;\;\;
R^{a}_{-}=\frac{\Lambda}{3}\overline{J}^a_{-}. \ee The second
(\ref{lacond}) is the same as (\ref{lamas}) in four dimensions. Thus
we conclude then that in $D=4$ quaternionic Kahler is the same as
self-dual Einstein.

  The self-duality condition $W_{-}=0$ is conformal invariant and
then if a metric $g$ is self-dual, then the family $[g]$ constructed
by $g$ through conformal transformations $g\rightarrow \Omega^2 g$
is also self-dual. But Einstein condition is not conformal invariant
in general.

\newpage

\subsection{Hypercomplex structures}

     As we have seen, the problem of classifying
the possible quaternion Kahler spaces appearing in $d=4$ is
equivalent to find all the Einstein spaces with self-dual Weyl
tensor. In this section we will focus our attention in four
dimensional self-dual spaces. Let us consider a metric $g$ and
denote with $[g]$ the family of metrics obtained from $g$ by
arbitrary conformal transformation $g\rightarrow\Omega^2g$. Because
the Weyl tensor is conformally invariant it follows that the
self-duality of $g$ implies the self-duality of $[g]$. The converse
is of this statement is also true.

     An important example of self-dual families are the hypercomplex structures
\cite{Boyer}, \cite{Finley}, that we will introduce in this section.
We will show in the following sections that hypercomplex structures
and weak torsion hyperkahler structures are the same concept in four
dimensions.

\subsubsection{Basic concepts}

    As it follows from equation (\ref{basta}), for hyperkahler spaces
there exist a frame for which \be\lb{noconf} d\overline{J}^i=0, \ee
that is, for which the Kahler triplet is \emph{closed}. Conversely,
it was shown in \cite{Plebanski} that any space in $d=4$ satisfying
(\ref{noconf}) is automatically hyperkahler. The curvature tensor
$R^a_{bcd}$ of an hyperkahler space is automatically self-dual, and
the almost complex structures are integrable, thus complex
structure.

    For a given hyperkahler structure $(g, \overline{J}^i)$
one can define an new structure $(\widetilde{g},
\widetilde{\overline{J}}^i)$ being $\widetilde{g}=\Omega^2 g$ and
$\widetilde{\overline{J}}^i=\Omega^2 \overline{J}^i$. This new
metric is quaternion hermitian with respect of the
$\widetilde{\overline{J}}^i$. But from (\ref{noconf}) one obtains
that \be\lb{noconf2} d\widetilde{\overline{J}}^i + 2 d
\log(\Omega)\wedge\widetilde{\overline{J}}^i=0. \ee This means that
$(\widetilde{g}, \widetilde{\overline{J}}^i)$ is not hyperkahler. A
conformally invariant generalization of the hyperkahler condition
(\ref{noconf}) is \be\lb{conf} d\widetilde{\overline{J}}^i
+\alpha\wedge\widetilde{\overline{J}}^i=0, \ee being $\alpha$ an
arbitrary 1-form. Under conformal transformation we have $\alpha + 2
d \log(\Omega)$ and (\ref{conf}) is unaltered \cite{Finley}. This
property define a conformal family $[g]$ with all the element
sharing the property (\ref{conf}).

     If $\alpha$ is a gradient ($\alpha=\nabla \phi$, being $\phi$ certain function
depending on all the coordinates), then it will exist a
representative $g$ of $[g]$ with self-dual curvature and therefore
hyperkahler. From the expression (\ref{Weylo}) of the Weyl tensor it
is possible to check that if $R$ is self-dual, then $W$ is also
self-dual. Thus the family $[g]$ corresponding to an hyperkahler
metric will have self-dual Weyl tensor (\ref{Weylo}). But the
explicit calculation of the $W_{abcd}$ for a metric satisfying
(\ref{conf}) shows that $\alpha$ disappear in its calculation.
Therefore the condition to be gradient is not necessary to ensure
$W_{-}=0$ and condition (\ref{conf}) defines a self-dual structure
in general \cite{Finley}. We will denote it as $([g],
[\overline{J}^i])$.

     A family of structures $([g],
[\overline{J}^i])$ satisfying (\ref{conf}) are the hypercomplex
structures in four dimensions \cite{Boyer}. For a given
representative $(g, \overline{J}^i)$ of an hypercomplex structure,
the metric $g$ is quaternion hermitian with respect to the
$\overline{J}^i$ and the Niejenhuis tensor \be\lb{rio}
N(X,Y)=[X,Y]+J[X,JY]+J[JX,Y]-[JX,JY] \ee corresponding to every
$\overline{J}^i$ vanish for every pair of vector fields $X,Y$. Such
an almost complex structures are called integrable, and they turn
into complex structures. The metric $g$ will be complex in this
case. The definition of hypercomplex structures is consistent due to
the fact that the integrability condition is conformally invariant.
In four dimensions we can select the self-dual complex structures
\be\lb{sibil} J^i=e_4 \otimes e^i - e_i \otimes e^4 + \epsilon
_{ijk} e_j \otimes e^k \ee up to an $SU(2)$ transformation. The
action of (\ref{sibil}) over the tangent space $TM_x$ is defined by
\be\lb{acshon}
\begin{array}{rclrclrclrcl}
J^1 ( e_1 )& = & e_2, &J^1( e_2 )& = &- e_1, &J^1 ( e_3 )& = & e_4,
&J^1 ( e_4 )& = &- e_3,
\\
J^2  ( e_1 )& = & e_3, &J^2 ( e_2 )& = &- e_4, &J^2 ( e_3 )& = &-
e_1, &J^2 ( e_4 )& = &  e_2,
\\
J^3 ( e_1 )& = & e_4, &J^3 ( e_2 )& = & e_3, &J^3 ( e_3 )& = &- e_2,
&J^3 ( e_4 )& = &- e_1.
\end{array}
\ee Then the annulation of the tensor $N^i(X,Y)$ will be equivalent
to the conditions
$$
[e_1,e_2]+[e_3,e_4]= - A_2 e_1 + A_1 e_2 + A_3 e_4 - A_4 e_3
$$
\be\lb{genasht} [e_1,e_3]+[e_4,e_2]= - A_3 e_1 + A_4 e_2 + A_1 e_3 -
A_2 e_4 \ee
$$
[e_1,e_4]+[e_2,e_3]= - A_4 e_1 - A_3 e_2 + A_2 e_3 + A_1 e_4
$$
for some set of functions ${A_1, A_2, A_3, A_4}$ \cite{Strachan}.
This is a direct consequence of (\ref{acshon}) and (\ref{rio}). We
immediately recognize that in the case $A_a=0$ we recover the
Ashtekar-Jacobson-Smolin quadratic equations for self-dual spaces
\cite{Ashtekar}. If we consider the conformal transformation $g
\rightarrow \Omega^2 g$, then it is seen that (\ref{genasht}) is
also satisfied with $A_a \rightarrow A_a + 2 e_a \log(\Omega)$. This
means that condition (\ref{genasht}) define a conformal family of
metrics $[g]$ and therefore the definition of hypercomplex
structures is consistent.

    Hypercomplex condition implies, but is not implied by, that
the family $[g]$ is self-dual \cite{Strachan}-\cite{Boyer}. This
follows from the fact that, as we will show now, condition
(\ref{genasht}) implies (\ref{conf}). Let us consider the connection
$\omega$ given by
$$
de^a + \omega^a_{b}\wedge e^b = 0.
$$
The antisymmetric part $\omega^a_{[bc]}$ is related to the structure
functions defined by the Lie bracket $[e_a,e_b]=c_{ab}^{c}e_c$ by
$$
\omega^a_{[bc]} = \frac{1}{2} c_{ab}^{c}\,.
$$
It is convenient to define the one-form
$$
\alpha = A-\chi
$$
where \be\lb{conectar} A = A_a e^a ,\;\;\;\; \chi =c_{ab}^{b} e^a,
\ee and the self-dual two-form $\overline{J}^1 = e^1 \wedge e^2+ e^3
\wedge e^4$. Then
$$
d \overline{J}^1 = d(e^1 \wedge e^2+ e^3 \wedge e^4)
$$
$$
=-\omega^1_{[ab]} e^a \wedge e^b \wedge e^2 +\omega^2_{[ab]} e^a
\wedge e^b \wedge e^1 -\omega^3_{[ab]} e^a \wedge e^b \wedge e^4
+\omega^4_{[ab]} e^a \wedge e^b \wedge e^3
$$
$$
= -\frac{1}{2} c^{1}_{ab} e^a \wedge e^b \wedge e^2 +\frac{1}{2}
c^{2}_{ab} e^a \wedge e^b \wedge e^1 -\frac{1}{2} c^{3}_{ab} e^a
\wedge e^b \wedge e^4 +\frac{1}{2} c^{4}_{ab} e^a \wedge e^b \wedge
e^3
$$
$$
= e^1\wedge e^2\wedge (c_{ab}^{a} e^b + A_3 e^3 + A_4 e^4)
+e^3\wedge e^4\wedge (c_{ab}^{a} e^b + A_1 e^1 + A_2 e^2)
$$
$$
=(e^1 \wedge e^2 + e^3 \wedge e^4) \wedge ( A-\chi)
$$
and therefore \be\lb{arri} d \overline{J}^1= \overline{J}^1\wedge
(A-\chi)=\overline{J}^1\wedge \alpha \ee This is equivalent to
(\ref{conf}). The same formula holds for $\overline{J}^2$ and
$\overline{J}^3$. Therefore we reach to the conclusion that
hypercomplex structures satisfies (\ref{conf}) and therefore are
self-dual \cite{Strachan}. It is straightforward to prove that under
$g \rightarrow \Omega^2 g$ the forms $\alpha$, $A$ and $\chi$
transform as
$$
\alpha_a \rightarrow \alpha_a + 2 e_a \log(\Omega)
$$
\be\lb{conformalito} A_a \rightarrow A_a - 2 e_a \log(\Omega) \ee
$$
\chi_a \rightarrow \chi_a - 3 e_a \log(\Omega).
$$
Hypercomplex structures are not the only self-dual structures in
four dimensions. Counterexamples are for instance the Joyce spaces
that will be introduced in another section.

\subsubsection{Examples of hypercomplex structures}

     Let us look now for hypercomplex structures $[g]$ over a manifold $M$
with two commuting $U(1)$ Killing vectors. The representatives $[g]$
of such structures will be of Gowdy form \be\lb{form} g = g_{ab}dx^a
dx^b + g_{\alpha\beta}dx^{\alpha} dx^{\beta}. \ee The latin and
greek indices takes values 1 and 2. Both $g_{ab}$ and
$g_{\alpha\beta}$ are supposed to be independent of the coordinates
$x^{\alpha}=(\theta, \varphi)$. Then the Killing vectors are
$\partial/\partial\theta$ and $\partial/\partial\varphi$ and are
commuting, so there is a $U(1)\times U(1)$ action on the manifold.

    By Gauss theorem there exists a local scale transformation
$g \rightarrow \Omega^2 g$ which reduce (\ref{form}) to
\be\lb{form2} \widetilde{g} = \Omega^2(d\rho^2+ d\eta^2) +
\widetilde{g}_{\alpha\beta}dx^{\alpha} dx^{\beta}, \ee being
$\widetilde{g}_{\alpha\beta}$ function of the new coordinates
$(\rho, \eta)$. We can express this metric as \be\lb{simio}
\widetilde{g}=\Omega^2(d\rho^2+ d\eta^2) +(\overline{A}_0 d\theta +
\overline{A}_1 d\varphi)^2 + (\overline{B}_0 d\theta +
\overline{B}_1 d\varphi)^2, \ee where $\overline{A}_i$ and
$\overline{B}_i$ are unknown functions of $(\rho, \eta)$ and the
factor $\rho$ was introduced by convenience. But instead to work
with the local form (\ref{simio}) we will work with the following
equivalent expression \be\lb{simplif}
\widetilde{g}=\Omega^2(d\rho^2+ d\eta^2 )+\frac{(A_0 d\theta -  B_0
d\varphi)^2 + ( A_1 d\theta - B_1 d\varphi)^2}{(A_{0}
B_{1}-A_{1}B_{0})^2}. \ee Anzatz (\ref{simplif}) looks more
complicated than (\ref{simio}), but has the advantage that the dual
basis $e_i$ takes the simple form \be\lb{einan}
e_1=\Omega\partial_{\rho},\;\;\; e_2=\Omega\partial_{\eta} \ee
\be\lb{einan2} e_3= B_1
\partial_{\theta} + A_1\partial_{\varphi},\;\;\; e_4= B_0
\partial_{\theta} + A_0 \partial_{\varphi}.
\ee
Inserting (\ref{einan}) and (\ref{einan2}) into the quadratic
system (\ref{genasht}) gives $\Omega=1$ and the Cauchy Riemman
equations \be\lb{caucho} (A_0)_{\rho}=(A_{1})_{\eta},\;\;\;\;\;
(A_0)_{\eta}=-(A_{1})_{\rho} \ee and the same equations for $B_0$
and $B_1$. This means that the metric (\ref{simplif}) is described
in terms of two holomorphic functions \be\lb{complex} F(z)= A_0 + i
A_1 ,\;\;\;\;\;\; G(z)= B_0 + i B_1,
\ee
being $z= \rho + i \eta$.

     We can find the same result in another way. Let us consider four
functions $f_1,...,f_4$ and $g_1,...,g_4$ depending on the
coordinates $x^1=\rho$ and $x^2=\eta$ and let us define the vector
fields
$$
e_a= f_a \partial_{\theta} + g_a \partial_{\varphi}+
\partial_{x^a} \;\;\;\;\;\; (a=1,2),
$$
$$
e_a= f_a \partial_{\theta} + g_a
\partial_{\varphi},\;\;\;\;\;\;(a=3,4).
$$
Clearly $e_a$ are the most general vector fields for a metric with
two commuting isometries up to a conformal scaling. Introducing this
expressions into (\ref{cuadra}) gives the system of equations
$$
(f_3)_{\eta}-(f_4)_{\rho}=0,\;\;\;\;\;(g_3)_{\eta}-(g_4)_{\rho}=0,
$$
\be\lb{sistem}
(f_2)_{\eta}+(f_1)_{\rho}=0,\;\;\;\;\;(g_2)_{\eta}+(g_1)_{\rho}=0,
\ee
$$
(f_1)_{\eta}-(f_2)_{\rho}=0,\;\;\;\;\;(g_1)_{\eta}-(g_2)_{\rho}=0.
$$
From the first we see that $f_3=(H)_{\rho}$ and $f_4=(H)_{\eta}$,
being $H(\rho, \eta)$ certain function. Therefore we can make the
coordinate change $(\rho, \eta, \theta, \varphi)\rightarrow (\rho-H,
\eta, \theta, \varphi)$ and eliminate $f_3$ and $f_4$. The same
holds for $g_3$ and $g_4$ and therefore we are dealing with the case
described in (\ref{einan2}). The corresponding hyperkahler metric is
then conformal to \be\lb{simplo} \widetilde{g}=d\rho^2+ d\eta^2
+\frac{(A_0 d\theta -  B_0 d\varphi)^2 + ( A_1 d\theta - B_1
d\varphi)^2}{(A_{0} B_{1}-A_{1}B_{0})^2}, \ee which is the same
metric as above with $\Omega=1$. We have redefined $e_1=A_0$,
$e_2=B_0$, $f_1=A_1$, $f_2=B_2$ in (\ref{simplo}).

  There exists another family of hypercomplex structures that
can be constructed in terms of holomorphic functions. It is direct
to check that Ashtekar equations (\ref{cuadra}) can be cast in the
following complex form \be\lb{compash} [e_1+ i e_2, e_1- i
e_2]-[e_3+ i e_4, e_3- i e_4]=0,\;\;\; [e_1+ i e_2, e_3- i e_4]=0
\ee Let $M$ be a complex surface with holomorphic coordinates $(z_1,
z_2)$ and let us define four vector fields $e_i$ as
$$
e_1 + i e_2=f_1\frac{\partial}{\partial z_1} +
f_2\frac{\partial}{\partial z_2}
$$
$$
e_3 + i e_4=f_3\frac{\partial}{\partial z_1} +
f_4\frac{\partial}{\partial z_2}
$$
being $f_j$ a complex function on $M$. Then (\ref{compash}) implies
that $\partial f_j/\partial \overline{z}_k=0$ and therefore we can
construct an hypercomplex structure using four arbitrary holomorphic
functions or two holomorphic vector fields \cite{Joyce}.

\subsection{Hyperkahler spaces in four dimensions}

   In a well known work \cite{Ashtekar} Ashtekar, Jacobson and Smolin
introduced a formulation for self-dual manifolds in which they
reduced the problem to solve certain quadratic equations for the
dual vector fields and a volume form preserving condition. Here we
review their construction in the context of hypercomplex structures.
\\

{\bf Proposition 3}{ \it Consider an oriented manifold $M$ and four
vector fields $e_1$, $e_2$, $e_3$ and $e_4$ forming an oriented
basis for $TM$ at each point. Let us suppose that the fields
satisfies the quadratic equations \be\lb{cuadra} [e_1, e_2]+[e_3,
e_4]= 0, \;\;\;\; [e_1, e_3]+[e_4, e_2]= 0, \;\;\;\; [e_1,
e_4]+[e_2, e_3]= 0, \ee and the volume preserving condition ${\cal
L}_{e_a}\Theta=0$ for some 4-form $\Theta$. Then the vectors $e_a$
are conformal to an orthonormal frame of a self-dual Ricci-flat
metric.}
\\

Before to explain Proposition 3 let us note that
$$
[e_1, e_2]+[e_3, e_4]= 0, \;\;\;\; [e_1, e_3]+[e_4, e_2]= 0,
\;\;\;\; [e_1, e_4]+[e_2, e_3]= 0,
$$
is the system (\ref{cuadra}) but with the $A_a$'s equal to zero.
Then (2.43) implies that $\chi=-\alpha$. Then if $\chi$ were a
gradient, the form $\alpha$ also would be a gradient and there will
exist a conformal change taking the condition
$$
d\overline{J}^i = \alpha \wedge \overline{J}^i
$$
to $d\overline{J}^i =0$. The corresponding metric will be conformal
to an hyperkahler one, which is automatically self-dual Ricci-flat.
As we will see, this is essentially the content of the volume
preserving condition of proposition 3.

   In general the covariant divergence of an arbitrary vector field
$V$ is obtained by the formula \be\lb{divo} {\cal L}_{V}{\Theta} =
(\nabla \cdot V)\Theta. \ee The divergence of a tetrad
$\overline{e}_a$ is given by \be\lb{divage} \nabla \cdot
\overline{e}_a=\overline{c}^{b}_{ab}, \ee being
$\overline{c}^{c}_{ab}$ defined by
$$
[\overline{e}_b,
\overline{e}_c]=\overline{c}_{bc}^{a}\overline{e}_a.
$$
Let us define a factor $\Omega$ through \be\lb{car}
\Omega^2=\Theta(e_1, e_2, e_3, e_4) \ee and take the derivative of
(\ref{car}) along $\overline{e}_a=\Omega^{-1}e_a$. Using the
condition ${\cal L}_{e_a}\Theta=0$ together with (\ref{divo}) and
(\ref{divage}) gives
$$
\overline{c}^{b}_{ab}=-\overline{e}_a (\log \Omega)\qquad
\Longleftrightarrow \qquad \chi=-\overline{e}_a (\log
\Omega)\overline{e}^a.
$$
The last implication is just the definition (\ref{conectar}). The
transformation rule (\ref{conformalito}) implies that under
$g\rightarrow \Omega^{-2} g$ we have \be\lb{finoli} \chi_a
\rightarrow \chi_a + 3 \overline{e}_a \log(\Omega)\qquad
\Longleftrightarrow \qquad
  \chi= 2 \overline{e}_a(\log\Omega)e^a.
\ee If (\ref{cuadra}) also hold we have \be\lb{barry}
\chi=-\alpha\qquad\Longleftrightarrow\qquad \alpha_a=- 2
\overline{e}_a (\log\Omega)=- 2 e_a (\frac{1}{\Omega}) \ee and
therefore $\alpha$ is a gradient. Then $g$ is conformal to an
hyperkahler metric. The converse of this theorem is also true, more
details of this assertions can be found in the original references
\cite{Ashtekar}.

    It is also possible to cast into this formalism
the Plebanski formulation for hyperkahler spaces \cite{Plebanski}.
By defining the complex vector fields
$$
u=e_1 + i e_2,\;\;\;\; v= e_1 - i e_2,\;\;\;\; w=e_3 + i
e_4,\;\;\;\; x= e_3 - i e_4,
$$
the system (\ref{cuadra}) takes the form \be\lb{complejin} [u,
w]=0,\;\;\;\;\;\; [v, x]=0,\;\;\;\;\;\;[u, v] + [w, x] = 0. \ee It
follows from the first (\ref{complejin}) that we can introduce
coordinates $(t, y)$ on each of these surfaces such that $$
u=\partial_t,\;\;\;\;\;\; \qquad {w} = {\partial}_y.
$$
Let us select the vector fields
$$
v = a_y {\partial}_x - b_y {\partial}_z, \qquad {x} = - a_t
{\partial}_x + b_t {\partial}_z.
$$
being $a$ and $b$ unknown functions. Equations (\ref{complejin})
give us
\begin{equation}
\{a, a_x\} = \{b, a_z\}, \qquad \{a, b_x\} = \{b, b_z\},
\label{heaven}
\end{equation}
where we have defined the Poisson Bracket by
\[
\{f, g\} = f_t g_y - f_y g_t.
\]
Let us suppose that the vector fields $e_i$ are divergence free with
respect to the volume form $\Theta = dt \wedge dx \wedge dy \wedge
dz$. Then it exists a function $\Omega$ such that $a = K_z$ and $b =
K_y$. By integrating equations (\ref{heaven}) once, and scaling the
coordinates $(y, z)$, one finds that $K$ satisfies the equation
\be\lb{heavenlo} \{ K_y, K_z \} = 1, \ee which is the First Heavenly
equation \cite{Plebanski}. This equation can be expressed as
$$
\left|\begin{array}{cc}
  K_{yt} & K_{yx} \\
  K_{zt} & K_{zx} \\
\end{array}\right|=1.
$$
The corresponding hyperkahler metric is
\[g = 4 \left( b_t a_y - a_t b_y \right)^{-1} \left[ dt \otimes
\left( a_t dz + b_t dx \right) + dy \otimes \left( a_y dz + b_y dx
\right) \right],\] or in terms of $K$, \be\lb{heaven} g= K_{ty}dt
dy+K_{tz} dt dz +K_{xy} dx dy + K_{xz} dx dz. \ee The metrics
(\ref{heaven}) are known as the Plebanski heavens \cite{Plebanski}
and are the most general hyperkahler examples. An hyperkahler space
is also Kahler, and the key function $K$ plays the role of the
Kahler potential.

     Let us consider metrics with one isometry and introduce a the dual
basis $e_a=f_a\partial/\partial x_1 +
\partial/\partial x_a$. Then (\ref{cuadra}) reduce to the system
$$
\frac{\partial f_1}{\partial x_2}-\frac{\partial f_3}{\partial
x_4}+\frac{\partial f_4}{\partial x_3}=0,\;\; \frac{\partial
f_1}{\partial x_3}-\frac{\partial f_4}{\partial x_2}+\frac{\partial
f_2}{\partial x_4}=0,\;\; \frac{\partial f_1}{\partial
x_4}-\frac{\partial f_2}{\partial x_3}+\frac{\partial f_3}{\partial
x_2}=0.
$$
Let us assume that $\partial/\partial x^1$ is a Killing vector and
let us denote $x_1=t$. This implies that the functions $f_{i}$ are
independent of $t$ and this system reduces to \be\lb{Gibb-Hawk}
\nabla V=\nabla \times A\;\;\;\;\Longleftrightarrow\;\;\;\;\;\;
dU=\ast dA \ee which implies that $\Delta V=0$ and we have defined
$V=f_1$ and $A=(f_2, f_3, f_4)$. The corresponding metric is
\be\lb{ashgib} g=V^{-1}(dt-A)^2 + V dx \cdot dx, \ee and it can be
checked that is indeed Ricci-flat.

      A particular solution of (\ref{Gibb-Hawk}) is
$$
(V(x))^{-1}=a+\sum^{n}_{i=1}2m_i|r-r_i|^{-1}
$$
being $a$ and $m_i$ constants and $r_i$ the position of certain
points on the manifold. If we select $m_i=M$ and $t$ periodic in the
range $0\leq t \leq \frac{8\pi M}{n}$ then the singularities in
$r=r_i$ are removable. Such spaces have finite energy and therefore
are gravitational instantons. The metrics corresponding $a=n=1$ are
known as the Taub-Nut metrics. If $a=0$ the metrics are known as
multi-instantons and if $n=2$ we obtain the Eguchi-Hanson instanton
\cite{Eguchi}.

    Metrics (\ref{ashgib}) are known as Gibbons-Hawking metrics
\cite{GibbHawko}, although they also were considered independently
by Kloster, Son and Das \cite{Kloster}. They are the four
dimensional self-dual examples that are Ricci-flat with a
$U(1)$-isometry for which condition ${\cal L}_{\partial_t}J^a=0$ is
satisfied. Such isometries are called tri-holomorphic. Another
characterization of (\ref{ashgib}) is that they have self-dual
Killing vector $K_{i}$, i.e, the tensor $K_{ij}=\nabla_i K_j$ is
self-dual \cite{Boyer}, \cite{Gegenberg}. We will return to this
point in the next subsection.

\newpage

\subsection{Quaternion Kahler and hyperkahler metrics in $d=4$ with
at least one isometry}

    This section present the most general local form of
a metric belonging to a self-dual conformal structure and at least
one Killing vector. The spaces in consideration are four
dimensional. Although this results are not new, we present them in
an slightly different way than the standard one. We show the
relation between self-dual structures with a Killing vector and
Einstein-Weyl ones, which satisfy a generalization of the Einstein
equations including conformal transformations, i.e, Einstein-Weyl
equations are invariant under the group $CO(n)=R_{+}\times SO(n)$
\cite{Hitchin}, \cite{JonTod}. A geometric interpretation for the
integrability of the non linear axial $SU(\infty)$ Toda equation
will be also presented. The argument is that the axial Toda equation
describe an hyperkahler space of the Gibbons-Hawking type (which are
entirely described in terms of a linear equation) in a coordinate
system that hides its linear nature.

  We should remark that $U(1)\times U(1)$ quaternion Kahler
metrics are important for the goals of the present work. By
extending them to spaces of $G_2$, $Spin(7)$ holonomy and higher
dimensional hyperkahler ones, we constructed certain classical
solutions of M-theory preserving some amount of supersymmetries. The
presence of isometries allows to find type IIA and IIB backgrounds
by use of dualities. We will discuss it in the next sections.

\subsubsection{Hyperkahler metrics with Killing vectors that are not self-dual}

    The most general form of an hyperkahler
metric is given by (\ref{heaven}) and defined in terms of a Kahler
potential $K$ satisfying the non linear equation (\ref{heavenlo}).
It will be convenient for the following to introduce two complex
coordinates $v,w$ and write the heavenly metrics (\ref{heaven}) in
complex form as \be\lb{heavenal} g= K_{vw} dv dw + K_{v\overline{w}}
dv d\overline{w} + K_{\overline{v}w} d\overline{v}dw +
K_{\overline{v}\overline{w}} d\overline{v}d\overline{w} \ee where
bar indicate complex conjugation. Because Plebanski metrics are
general it follows that the Gibbons-Hawking metrics (\ref{ashgib})
should arise as a subcase of (\ref{heavenal}). Such reduction goes
as follows. Select the key function $K$ as \be\lb{ah} K(v,
w,\overline{v},\overline{w})=K(i(\overline{w}-w), v, \overline{v})
\ee If one define $u=i(\overline{w}-w)$, then equation
(\ref{heavenlo}) becomes \be\lb{boomchak}
K_{uu}K_{v\overline{v}}-K_{uv}K_{u\overline{v}}=1. \ee Consider the
new independent variables $x, v, \overline{v}$ with $x=K_{u}$. By
setting \be\lb{defono} H(x,v,\overline{v})=xu-K \ee the heavenly
equation (\ref{boomchak}) reduce to \be\lb{lapoco}
H_{xx}+H_{yy}+H_{zz}=0 \ee being $y=v+\overline{v}$ and
$iz=v-\overline{v}$. The metric (\ref{heavenal}) in this coordinate
system take the form \be\lb{gegodas2} g=H_{xx} (dx^2 + dy^2 + dz^2)
+ H_{xx}^{-1}(dt + H_{xy}dz - H_{xz}dy)^2, \ee being $i
t=\overline{v}-v$. By selecting $H_{x}=G$ we can rewrite
(\ref{gegodas2}) in the more familiar form \be\lb{gegodas3} g=G_{x}
(dx^2 + dy^2 + dz^2) + G_{x}^{-1}(dt + G_{y}dz - G_{z}dy)^2, \ee By
identifying $G_{x}=V$ and $A=G_{y}dz - G_{z}dy$ we find that $\nabla
\times A=\nabla V$. Then (\ref{gegodas3}) are the Gibbons-Hawking
metrics. The Killing vector for such metric is $\partial/\partial t$
and it is called translational in the literature \cite{Boyer}.

    There exist another type of hyperkahler metrics with one Killing vector,
and related to certain limit of a Toda equation. Let us select the
key function $K$ as \be\lb{ah} K(v,
w,\overline{v},\overline{w})=K(\log |w|^2, v, \overline{v}) \ee By
choosing $F=\log |w|^2$ equation (\ref{heavenlo}) becomes
\be\lb{boomchak}
K_{uu}K_{v\overline{v}}-K_{uv}K_{u\overline{v}}=e^F. \ee Consider
the new independent variables $t, v, \overline{v}$ with
$\tau=K_{F}$. If we set \be\lb{defono} H(\tau,v,\overline{v})=\tau
F-K \ee then equation (\ref{boomchak}) is converted into
 \be\lb{lapoco}
H_{xx}+H_{yy}+(e^{H_{\tau}})_{\tau}=0 \ee with $x=v+\overline{v}$
and $iy=v-\overline{v}$. Defining $u=H_{\tau}$ and $z=\tau$ and
differentiating (\ref{lapoco}) with respect to $z$ gives
\be\lb{Toda} (e^u)_{zz} + u_{yy} + u_{xx}=0. \ee Equation
(\ref{Toda}) is known as the continuum limit of the $sl(n)$ Toda
equation. The metric (\ref{heavenal}) is expressed in terms of the
new coordinates as \be\lb{gegodas} g=u_z
[e^{u}(dx^2+dy^2)+dz^2]+u_{z}^{-1}[dt+(u_{x}dy-u_{y}dx)]^2, \ee
being $u$ a solution of the $SU(\infty)$ Toda equation (\ref{Toda}).
We have defined the coordinate $i t=\log (\frac{w}{\overline{w}})$
and the Killing vector is $\partial/\partial
t=\partial/\partial(\arg w)$. This explain why this Killing vectors
are called rotational in the literature.

       A result due to Boyer and Finley \cite{Boyero} show that both
(\ref{gegodas}) and (\ref{ashgib}) exhaust all the possible metrics
with at least one isometry. The first are the most general ones with
self-dual Killing vector, the second are those with a non self-dual
isometry. Metrics (\ref{ashgib}) are related to a linear equation
and therefore are more easy to construct. Instead the continuum Toda
equation (\ref{Toda}) is integrable but not many explicit solutions
are known. One interesting solution is those that is factorized by
$u=f(x,y)+g(z)$, then there appears three solutions given by
$$
e^{u_1}=\frac{1}{2}(z^2-a^2)[1+\frac{1}{8}(x^2+y^2)]^{-2},
\;\;\;\;\;\;\; z^2\geq a^2
$$
$$
e^{u_2}=\frac{1}{2}(z^2+a^2)[1+\frac{1}{8}(x^2+y^2)]^{-2},
\;\;\;\;\;\;\; z^2\geq a^2
$$
$$
e^{u_3}=\frac{1}{2}(a^2-z^2)[1+\frac{1}{8}(x^2+y^2)]^{-2},
\;\;\;\;\;\;\; z^2\leq a^2
$$
The first of this solutions corresponds to the Eguchi-Hanson metric.
This can be visualized defining $\sigma=1+\frac{1}{8}(x^2+y^2)$ and
the coordinate system $(r, \theta, \varphi)$ given by
$$
x\sigma^{-1}=\sqrt{2}\cos(\varphi)\sin(\theta),
$$
$$
y\sigma^{-1}=\sqrt{2}\sin(\varphi)\sin(\theta),
$$
$$
4z=r^2,
$$
then the metric (\ref{gegodas}) becomes the standard Eguchi-Hanson,
namely \be\lb{egola}
g=(1-\frac{\alpha^4}{r^4})^{-1}dr^2+\frac{r^2}{4}(d\theta^2+
\sin(\theta)^2d\varphi^2)+\frac{r^2}{4}
(1-\frac{\alpha^4}{r^4})(\cos(\theta)d\varphi+dt)^2 \ee being
$\alpha^4=16 a^2$. It is seen that there is another Killing vector
$\partial/\partial \varphi$. There exists a coordinate change that
bring this metric to the form of Hawking multi-instanton solution.
Therefore $\partial/\partial \varphi$ is self-dual but
$\partial/\partial t$ is not. The metrics corresponding to the other
solutions $u_2$ and $u_3$ have singular points for the scalar
curvature which cannot be excluded from the space time. Therefore
this spaces are not gravitational instantons \cite{Gegenberg}.

\subsubsection{Integrability of the axial continuum Toda equation}

    The Eguchi-Hanson metric is an example possessing simultaneously a self-dual
Killing vector and a not self-dual one (\ref{egola}). Such examples
are described simultaneously by the non linear equation (\ref{Toda})
for certain election of coordinates, and by the linear equation
(\ref{Gibb-Hawk}) for other coordinates. The transformation from one
of such coordinates into the other is a mapping between the axial
$SU(\infty)$ Toda equation and the so called Ward monopole equation,
which is linear. Therefore although the continuum Toda equation is
not linear, in the axial case it is completely equivalent to a
linear one. By axial Toda we mean the equation \be\lb{atoda}
(e^u)_{zz} + u_{xx}=0. \ee The geometrical origin of the equivalence
stated above is that the axial Toda equation describe
Gibbons-Hawking metrics in a coordinate system which hide its linear
nature \cite{Ward}.

     To see this statement more concretely let us take a Gibbons-Hawking metric
(\ref{ashgib}) and write the flat $3$-dimensional part in
cylindrical polar coordinates $(\eta,\rho,\varphi)$. The result is
$$
g_f = dx^2 + dy^2 + dz^2 = d\eta^2 + d\rho^2 + \rho^2 d\varphi^2.
$$
Let us also suppose that the generator of the axial symmetry
$\partial/\partial\varphi$ is also a Killing vector, then $V(\rho,
\eta)$ in (\ref{ashgib}) is independent on the $\varphi$ coordinate
and the equation (\ref{ashgib}) becomes the Ward monopole equation,
namely \be\lb{axialward} \rho^{-1}(\rho V_\rho)_\rho+
V_{\eta\eta}=0. \ee Note that if $\widetilde{V}$ is a solution of
(\ref{axialward}), so is $\widetilde{V}_\eta$, and
$\widetilde{V}_\eta$ determines $\widetilde{V}$ up to the addition
of $C_1\log(C_2\rho)$ for some constants $C_1$ and $C_2$. This
provides a way of integrating the equation $d{*dV}=0$ to give
$*dV=dA$. If we choose $V=\widetilde{V}_\eta$, then we can take
$A=\rho \widetilde{V}_\rho\,d\phi$. The corresponding hyperkahler
metric is \be\lb{queseyo1}
g=V_\eta(d\eta^2+d\rho^2+\rho^2d\phi^2)+V_\eta^{-1}(d\theta +\rho
V_\rho\,d\phi)^2, \ee where have reexpressed $\widetilde{V}$ as $V$
in order to simplify the notation.

    The metric (\ref{queseyo1}) has $U(1)\times U(1)$ isometry and
the two commuting Killing vectors are $\partial/\partial \varphi$
and $\partial/\partial \theta$. By construction $\partial/\partial
\theta$ is self-dual. But $\partial/\partial \varphi$ is not
self-dual in general. Therefore there exist a coordinate system
$(x,y,z, \varphi)$ for which (\ref{queseyo1}) take the form
(\ref{gegodas}). We can rewrite (\ref{queseyo1}) by completing
squares as
$$
g=V_\eta\biggl(d\rho^2+d\eta^2
+\frac1{V_\eta^2+V_\rho^2}d\theta^2\biggr)
+\frac{\rho^2(V_\eta^2+V_\rho^2)}{V_\eta}
\biggl(d\varphi+\frac{V_\rho}{\rho(V_\eta^2+V_\rho^2)}d\theta\biggr)^2.
$$
or equivalently as
$$
g=\frac{V_\eta}{\rho^2(V_\eta^2+V_\rho^2)}[\rho^2(V_\eta^2+V_\rho^2)(d\rho^2+d\eta^2)
+\rho^2d\psi^2]
$$
\be\lb{queseyo2} +\frac{\rho^2(V_\eta^2+V_\rho^2)}{V_\eta}
\biggl(d\varphi+\frac{V_\rho}{\rho(V_\eta^2+V_\rho^2)}d\theta\biggr)^2.
\ee It is extremely important to recall that (\ref{queseyo1}) and
(\ref{queseyo2}) are the \emph{same} metric. Comparison between
(\ref{gegodas}) and (\ref{atoda}) shows that \be\lb{monopolo}
u_{z}=\frac{V_\eta}{\rho^2(V_\eta^2+V_\rho^2)} \ee and that there it
should exists a coordinate system $(x,y,z)$ for which
\be\lb{identifo} \rho^2(V_\eta^2+V_\rho^2)(d\rho^2+d\eta^2)
+\rho^2d\theta^2=e^{u}(dx^2+dy^2)+dz^2 \ee with $u$ satisfying
(\ref{atoda}). The system $(x,y,z)$ is obtained as follows. There
are two exact differentials constructed with $V$, namely
$$
dV= V_{\rho}d\rho + V_{\eta}d\eta,
$$
$$
dG= \rho V_{\rho}d\eta - \rho V_{\eta}d\rho.
$$
The equation $d^2V=0$ is identically satisfied and $d^2G=0$ is a
consequence of the equation (\ref{axialward}), therefore $dV$ and
$dG$ are truly differentials. The function $G$ is defined by $V$
through the Backlund transformation
$$
\rho V_{\rho}= G_\eta,\;\;\;\;\;\; \rho V_{\eta}=-G_{\rho}.
$$
It is elementary to show that
$$
\rho^2(V_\eta^2+V_\rho^2)(d\rho^2+d\eta^2)
+\rho^2d\theta^2=\rho^2(dV^2+d\psi^2)+dG^2.
$$
Therefore
$$
\rho^2(dV^2+d\theta^2)+dG^2=e^{u}(dx^2+dy^2)+dz^2
$$
This means that $y=\theta$ and a Toda solution is defined through
the equation $e^u=\rho^2$. We have obtained the following
proposition \cite{Ward}.
\\

{\bf Proposition 4}{ \it Any solution $V$ of the equation
$V_{\eta\eta}+\rho^{-1}(\rho V_{\rho})_{\rho}=0$ defines locally the
coordinate system $(x, z)$ \be\lb{Wardchan} x=V_{\eta},\;\;\;\;\;\;
z = \rho V_{\rho}, \ee in terms of $(\rho, \eta)$ and conversely
(\ref{Wardchan}) defines implicitly $(\rho, \eta)$ as functions of
$(x,z)$. Then the function $u(x,z)=log(\rho^2)$ is a solution of the
axially symmetric Toda equation \be\lb{axialtoda} (e^u)_{zz} +
u_{xx}=0. \ee This procedure can be inverted in order to find a Ward
monopole $V$ starting with a given Toda solution.}
\\

Proposition 4 gives a method to find solutions of a non linear
equation (the continuum Toda one) starting with a solution of a
linear one (the Ward equation) and shows the equivalence that we
were looking for. Proposition 4 can be checked by an elementary (but
lengthy) chain rule exercise. It is difficult in practice to find
explicit solutions of (\ref{axialtoda}) and usually proposition 4
gives implicit solutions.

\subsubsection{Einstein-Weyl structures and hyperkahler metrics}

     An important feature about the three dimensional part
\be\lb{conformal} \gamma=e^{u}(dx^2+dy^2)+dz^2, \ee is that it is an
Einstein-Weyl structure. Einstein-Weyl condition is a generalization
of the ordinary Einstein condition in which the Lorenz group $SO(4)$
is extended to the conformal group $CO(4)=R_{+}\times SO(4)$
\cite{Hitchin} \footnote{Usually the conformal group is defined by
the angle-preserving (conformal) transformations. We are not using
this definition here. We are considering only $SO(4)$
transformations and dilatations}. The Einstein equations are
generalized in this picture in order to include invariance under
coordinate rotations plus dilatations. In this section some
important facts about them are sketched following \cite{Hitchin} and
\cite{JonTod}.

   It is known that for a given space $\overline{W}$ endowed with a
metric $g_{\mu\nu}$ the Levi-Civita connection $\nabla$ is uniquely
defined by \be\lb{Levi-Civita} \nabla g=0,\;\;\;T(\nabla)=0 \ee
where $T(\nabla)$ is the torsion. A Weyl-structure is defined by the
manifold $\overline{W}$ together with:
\\

(a) A class of conformal metrics $[g]$, whose elements are related
by the conformal scaling (or gauge transformation) \be\lb{congauge}
g_{\mu\nu}\rightarrow \Omega^2 g_{\mu\nu}, \ee together with
$SO(n)$-coordinate transformations. $\Omega^2$ is an smooth,
positive real function over $\overline{W}$.
\\

(b) A torsion-free connection $D_{\mu}$ which acts over a
representative $g_{ab}$ of the conformal class $[g]$ as
\be\lb{Weyldef} D_{\mu} g_{\nu\alpha}=\omega_{\mu}g_{\nu\alpha}, \ee
for certain functions $\omega_{\mu}$ defining an one form $\omega$.
Then it is said that $D$ preserves $[g]$.
\\

    The conformal group in $n$-dimensions is
$CO(n)=R_{+}\times SO(n)$ and includes rotations plus general scale
transformations. The structure $[g]$ is called $CO(n)$-structure
over $\overline{W}$. The connection $D$ is uniquely determined by
(\ref{Weyldef}) in terms of $\omega$ and $g$. This can be seen in a
coordinate basis $\partial_k$ in which the system (\ref{Weyldef})
takes the form
$$
g_{\mu\nu,\alpha}=g_{\beta\nu}\Upsilon^{\beta}_{\mu\alpha} +
g_{\mu\beta}\Upsilon^{\beta}_{\nu\alpha} +\omega_{\mu}g_{\nu\alpha},
$$
where the symbols $\Upsilon^{\beta}_{\mu\alpha}$ denotes the
connection coefficients of $D$, which are symmetric in the lower
indices by the torsionless condition. Thus a series of steps
analogous to those needed to determine the Levi-Civita connection
shows that $\Upsilon^{\beta}_{\mu\alpha}$ is \be\lb{nablacon}
\Upsilon^{\beta}_{\mu\alpha}=\Gamma^{\beta}_{\mu\alpha}+\gamma^{\beta}_{\mu\alpha}
\ee where $\Gamma^{\beta}_{\mu\alpha}$ are the Christoffel symbols
and the add $\gamma^{\beta}_{\mu\alpha}$ is \be\lb{add}
2\gamma^{\beta}_{\mu\alpha}=(\delta^{\beta}_{\mu} \omega_{\alpha} +
\delta^{\beta}_{\alpha} \omega_\mu + g_{\mu\alpha} \omega^{\beta}).
\ee The form $\omega$ is not invariant under (\ref{congauge}), its
transformation law can be obtained from (\ref{Weyldef}) and
(\ref{add}) and is \be\lb{congauge} \omega_{\mu}\rightarrow
\omega_{\mu}+ 2\partial_{\mu}log(\Omega). \ee It follows from
(\ref{add}) and (\ref{congauge}) that the Levi-Civita of any $g$ of
$[g]$ preserves the conformal structure.

    If for a Weyl-structure the symmetric part of the
Ricci tensor $\widetilde{R}_{(\mu\nu)}$ constructed with $D_{\mu}$
satisfies \be\lb{Einweyl} \widetilde{R}_{(\mu\nu)}=\Lambda
g_{\mu\nu} \ee for certain $\Lambda$, then will be called
Einstein-Weyl. If in addition the antisymmetric part of
$\widetilde{R}_{[\mu\nu]}$ vanish there exists a gauge in which
(\ref{Einweyl}) reduces to the vacuum Einstein equation with
cosmological constant $\Lambda$. To see this it is needed to
calculate
$$
\widetilde{R}=D_{X}D_{Y}-D_{Y}D_{X}-D_{[X,Y]}
$$
using the formula (\ref{nablacon}) for $D$. The relation
$CO(4)$=$R_{+}\times SO(4)$ decompose $\widetilde{R}$ into a real
component $R_0$ and into an $SO(4)$-component $R$ that is equal to
the curvature tensor constructed with $\nabla$. After contracting
indices it is obtained
$$
\widetilde{R}_{\mu\nu}=R_{\mu\nu}+\nabla_{\mu}\omega_{\nu}-\frac{1}{2}\nabla_{\mu}\omega_{\nu}-\frac{1}{4}\omega_{\mu}
\omega_{\nu} +
g_{\mu\nu}(\frac{1}{2}\nabla_{\alpha}\omega_{\alpha}+\frac{1}{4}\omega_{\alpha}\omega^{\alpha}),
$$
where $R_{\mu\nu}$ is the Ricci tensor found with $\nabla$. The
antisymmetric part is originated by the $R_0$ component and is
determined in terms of $\omega$ as \be\lb{antim}
\widetilde{R}_{[\mu\nu]}=\frac{3}{2}\nabla_{[\mu}\omega_{\nu]}. \ee
If (\ref{antim}) is zero, then $\omega$ is the gradient of certain
function $\Psi$ over $\overline{W}$. The conformal rescaling
(\ref{congauge}) with $\Omega=-e^{\Psi}$ set $\omega=0$. This reduce
the symmetric part
$$
\widetilde{R}_{(\mu\nu)}=R_{\mu\nu}-\frac{1}{2}\nabla_{(\mu}\omega_{\nu)}-\frac{1}{4}\omega_{\mu}\omega_{\nu}
$$
to $R_{\mu\nu}$ and (\ref{Einweyl}) is the Einstein equation with
$\Lambda$, thus $[g]$ contains an Einstein metric.

   As we recalled in the last subsection, the metric
$$
\gamma=e^{u}(dx^2+dy^2)+dz^2,
$$
is the representative of an Einstein-Weyl structure with the factor
$\omega$ given by \be\lb{conformal2} \omega=-u_{z}dz. \ee Therefore
the Gegenberg and Das hyperkahler metrics (\ref{gegodas}) are
constructed with a three dimensional basis that is Einstein-Weyl. It
can also be seen that if we identify
$$
V=u_{z},\;\;\;\; A=u_{x}dy-u_{y}dx,
$$
then (\ref{gegodas}) take the form
$$
g = V [ e^{u}(dx^2+dy^2)+dz^2 ] + \frac{(dt+ A)^2}{V},
$$
and it is seen from (\ref{Toda}) that the pair $(V, A)$ satisfy
\be\lb{geno} \ast dA = dV + \frac{\omega}{2} V. \ee The Hodge star
$\ast$ is taken with respect to the Einstein-Weyl structure.
Equation (\ref{geno}) is the conformal generalization of the
Gibbons-Hawking one $\ast dA = dV$ and therefore it will be called
generalized monopole equation.

\subsubsection{Self-dual structures with one Killing vector}

   In this subsection we show that the generalized monopole equation
(\ref{geno}) is conformally invariant and that it describe all the
self-dual conformal structures with one conformal Killing vector.
Let us consider a metric with the form
$$
g = V h + \frac{(dt+ A)^2}{V},
$$
being $h$ an Einstein-Weyl structure and the functions $(V, A)$
satisfying (\ref{geno}). Then if we perform the conformal change
$$
\widetilde{h}=\Omega^{-2} h,
$$
we can express $g$ as \be\lb{formita}
 g =V h + \frac{(dt+ A)^2}{V}=\Omega \widetilde{g} \ee
where $\widetilde{V} = V \Omega$. We want to prove that if
(\ref{geno}) holds then
$$
\ast_{\widetilde{h}} dA =
d\widetilde{V}+\frac{\widetilde{\omega}}{2}\widetilde{V}
$$
being $\widetilde{\omega}$ the conformal factor associated to
$\widetilde{h}$. If we express the conformal transformation
(\ref{congauge}) for $\omega$ as
$$
\Omega \widetilde{\omega} = \Omega \omega - 2 d\Omega
$$
then it is easily seen that
$$
d\widetilde{V} + \frac{\widetilde{\omega}}{2}\widetilde{V} = \Omega
( dV + \frac{\omega}{2} V )
$$
and from (\ref{geno}) and the last expression we have \be\lb{sale}
d\widetilde{V} + \frac{\widetilde{\omega}}{2}\widetilde{V} = \Omega
\ast_{h} dA. \ee
 By another side we can put
$$
dA = A_{ab} e^a \wedge e^b \;\;\;\; \Longrightarrow\;\;\;\; \ast_{h}
dA = \epsilon_{abc} A_{ab} e^c,
$$
where we denote as $A_{ab}$ the coefficients of $dA$ in the basis
$e^a \wedge e^b$, being $e^a$ the einbein corresponding to $h$. Then
$$
\Omega \ast_{h} dA = \epsilon_{abc} A_{ab} \Omega e^c =
\epsilon_{abc} A_{ab} \Omega^2 \widetilde{e}^c
$$
being $\widetilde{e}^c$ the einbein corresponding to
$\widetilde{h}$, namely $\widetilde{e}^c=e^c/\Omega$. Therefore
$$
\ast_{\widetilde{h}}( \Omega *_{h} dA ) = A_{ab} \Omega^2
\widetilde{e}^a \wedge \widetilde{e}^b = A_{ab} e^a \wedge e^b = dA.
$$
Using the fact that $\ast^2=1$ from the last formula we obtain
\be\lb{thelast}
 \Omega \ast_{h} dA =
\ast_{\widetilde{h}} dA. \ee From (\ref{sale}) and (\ref{thelast})
we see that (\ref{geno}) implies
$$
\ast_{\widetilde{h}} dA =
d\widetilde{V}+\frac{\widetilde{\omega}}{2}\widetilde{V}
$$
which implies that (\ref{geno}) is conformally invariant.

  We have also seen that equation (\ref{geno}) also describe
metrics with self-dual curvature as in (\ref{gegodas}) or
(\ref{ashgib}). It is evident from the result of this subsection
that conformal family $[g]$ associated to (\ref{gegodas}) and
(\ref{ashgib}) have self-dual Weyl tensor and a conformal Killing
vector, and that all their elements are described with an equation
of the form (\ref{geno}). This raise the question whether or not
\emph{any} self-dual family with one conformal Killing vector has
the form
$$
g =V h + \frac{(dt+ A)^2}{V},
$$
being $h$ an \emph{arbitrary} Einstein-Weyl structure and $(V, A)$ a
pair satisfying (\ref{geno}). The answer to this question is
positive, it was proved by Jones and Tod and is stated in the
following proposition \cite{JonTod}.
\\

{\bf Proposition 5 }{ \it a) Consider an Einstein-Weyl structure
$[h]$ in D=3 and a representative $h$. Then the four dimensional
metric \be\lb{Converse} g =V h + \frac{(dt+ A)^2}{V} \ee is
self-dual with one Killing vector $\partial_t$ if the pair of
functions $(V, A)$ satisfies the generalized monopole equation
\be\lb{nose} dA= *_{h}(dV - V \omega). \ee The Hodge star $*_{h}$ is
taken with respect to $h_{\mu\nu}$ and $\omega$ is the conformal
factor corresponding to $h$.

      b) Conversely given a metric $g$ that is self-dual and has one conformal
Killing vector $K^a$ then a conformal transformation can be
performed in order that $K^a$ becomes a Killing vector $\partial_t$
and there exists a system of coordinates in which $g$ takes the form
(\ref{Converse}), being $h$ a representative of an Einstein-Weyl
structure. The factor $\omega$ will be obtained in this case through
(\ref{nose}).}
\\

Some comments are in order. The Gegenberg and Das metrics
(\ref{gegodas}) and (\ref{ashgib}) generates all the families
conformal to an hyperkahler metric. For all the space over a Toda
base we have
$$
dV - \omega V= V_x dx + V_y dy + (V_z + u_z V)dz = V_x dx + V_y dy+
e^{-u}(e^u V)_z dz,
$$
and (\ref{nose}) is then explicitly
$$
dA=*_{h}(dV - \omega V)= V_{x} dz \wedge dy + V_{y} dx \wedge dz +
(V e^{u})_{z} dy \wedge dx.
$$
Therefore the integrability condition for the existence of $A$ is
\be\lb{integra} (V e^u)_{zz} + V_{yy} + V_{xx}=0. \ee In the next
subsection we will show an application of this correspondence, the
construction of all the self-dual structures with two commuting
Killing vectors that are surface orthogonal. This are the Joyce
spaces.

\subsubsection{The Joyce spaces}

      Let us focus in self-dual families $[h]$ with two commuting $U(1)$
isometries. Consider an structure $[g]$ over $M$ with
representatives $g$ that locally takes the form (\ref{simplif}), or
equivalently \be\lb{simpli2} g=(A_{0}B_{1}-A_{1}B_{0})(d\rho^2+
d\eta^2) +\frac{(A_0 d\theta - B_0 d\varphi)^2 + ( A_1 d\theta - B_1
d\varphi)^2}{(A_{0} B_{1}-A_{1}B_{0})}. \ee We made by convenience a
conformal transformation in (\ref{simplif}) with factor
$\Omega=A_{0}B_{1}-A_{1}B_{0}$, we can always do this if we are
dealing with a self-dual structure due to the conformal invariance
the Weyl tensor. The unknown functions $A_i$ and $B_i$ are supposed
to be independent on the coordinates $(\theta, \varphi)$. By
construction $\partial/\partial\theta$ and
$\partial/\partial\varphi$ are commuting Killing vectors of the
metric (\ref{simpli2}).

     We will fix the functions $A_i$ and $B_i$ in order that (\ref{simplif})
will be self-dual. The Jones-Tod correspondence insure that there
should exists a coordinate system $(x,y,z,t)$ defined in terms of
the old one $(\rho,\eta, \theta,\varphi)$ for which (\ref{simpli2})
is conformal to \be\lb{Piza} g =V h + \frac{1}{V}(dt+ A)^2 \ee
according to (\ref{Converse}). To find this system of coordinates we
write (\ref{simpli2}) by completing squares as
$$
g =\frac{A_0 B_1 - A_1
B_0}{\rho^2(A_0^2+A_1^2)}((A_0^2+A_1^2)(d\rho^2+d\eta^2)+ \rho^2
d\varphi^2)
$$
\be\lb{joycemetric2} +\frac{A_0^2+A_1^2}{A_0 B_1 - A_1
B_0}(d\theta-\frac{(A_0 B_0 + A_1 B_1)d\varphi}{A_0^2+A_1^2})^2 \ee
and is seen that after scaling by $\rho$ and identifying $t=\theta$
that it takes the form (\ref{Converse}) with a metric $h$ and a
monopole $(V,A)$ given by \be\lb{einsaxial}
h=(A_0^2+A_1^2)(d\rho^2+d\eta^2)+ \rho^2 d\varphi^2 ,\;\;\;\;\;\;
V=\frac{A_1 B_0-A_0 B_1}{\rho(A_0^2 + A_1^2)}, \ee \be\lb{reform}
A=-\frac{(A_0 B_0 + A_1 B_1)}{A_0^2 + A_1^2}d\varphi. \ee Comparison
of the first (\ref{einsaxial}) with (\ref{identifo}) and Proposition
4 shows the identification
$$
A_{0}=\rho \widetilde{V}_{\eta},\;\;\;\;\; A_{1}=-\rho
\widetilde{V}_{\rho}
$$
being $\widetilde{V}$ a solution of the Ward monopole equation
(\ref{axialward}). Therefore $A_i$ should satisfy
$$
(A_0)_{\eta}-(A_1)_{\rho}=0,
$$
and from the identity
$\widetilde{V}_{\rho\eta}=\widetilde{V}_{\eta\rho}$ we have the
integrability condition
$$
(\frac{A_{0}}{\rho})_{\rho}= -(\frac{A_{1}}{\rho})_{\eta}
$$
or equivalently
$$
(A_0)_{\rho}+(A_1)_{\eta}=A_0/\rho.
$$
Clearly the same reasoning holds for the $B_i$ and therefore we have
obtained part of the following result due to Joyce \cite{Joyce}.
\\

{\bf Proposition 6 }{ \it Consider a set of functions $A_i$ and
$B_i$ with $i=1,2$, satisfying \be\lb{joyce1}
(A_0)_{\rho}+(A_1)_{\eta}=A_0/\rho, \ee \be\lb{joyce2}
(A_0)_{\eta}-(A_1)_{\rho}=0, \ee and the same equations for $B_i$.
Then the family $[g_j]$ associated to the metric $g_j$ given by
\be\lb{joycemetric} g_{j}=(A_{0} B_{1}-A_{1}B_{0})\frac{d\rho^2 +
d\eta^2}{\rho^2}+ \frac{(A_0 d\theta - B_0 d\varphi)^2 + (A_1
d\theta - B_1 d\varphi)^2}{(A_{0} B_{1}-A_{1}B_{0})}, \ee is
self-dual. All self-dual metrics with two commuting surface
orthogonal Killing vectors arise locally in this way.}
\\

  Equation (\ref{joyce2}) implies the existence of a potential function
$G$ such that \be\lb{potencial1} A_0=G_\rho;\;\;\;\;\;\;
A_1=G_{\eta}. \ee Then (\ref{joyce1}) implies that $G_{\rho\rho} +
G_{\eta\eta}=G_{\rho}/\rho$. Conversely, as we have seen above,
(\ref{joyce1}) implies that \be\lb{potencial1} A_0=-\rho
V_\eta;\;\;\;\;\;\; A_1=\rho V_{\rho}, \ee and (\ref{joyce2}) gives
the Ward monopole equation \cite{Ward} \be\lb{Wardy}
V_{\eta\eta}+\rho^{-1}(\rho V_{\rho})_{\rho}=0. \ee The relations
\be\lb{shu} G_\rho=-\rho V_{\eta};\;\;\;\;\;\;\ G_{\eta}=\rho
V_{\rho}, \ee constitute a Backlund transformation allowing to find
a monopole V starting with a known $G$ or viceversa. The functions
$B_i$ can be also expressed in terms of another potential functions
$G'$ and $V'$ satisfying the same equations than $V$ and $G$.

    It should be reminded that we have presented
till now two different types of self-dual families with two
commuting Killing vectors, namely (\ref{joycemetric}) and
(\ref{simplif}). The equations (\ref{joyce1}) and (\ref{joyce2}) are
different to the Cauchy-Riemann one (\ref{caucho}), and is simple to
check that there does not exist a coordinate change converting
(\ref{joyce1}) and (\ref{joyce2}) into (\ref{caucho}). Therefore
(\ref{joycemetric}) and (\ref{simplif}) are not related by a
coordinate transformation. The first are the most general self-dual
structures with two commuting Killing vector that are
surface-orthogonal \cite{Joyce}, \cite{Pedersen}. This mean that the
manifold $M$ corresponding to (\ref{joycemetric}) is of the form
$M=N\times T^2$, being $T^2$ the two dimensional torus. Instead the
families (\ref{simplif}) admits a $T^2$ action but has no compatible
product structure $M=N\times T^2$.

     There exist a theorem given in \cite{Pedersen} that shows that the Killing
vector of a quaternion Kahler space are surface orthogonal.
Therefore in order to find the toric quaternion Kahler spaces we
should impose the Einstein condition on (\ref{joycemetric}) and not
on (\ref{simplif}). This is the purpose of the next subsection.

\subsubsection{Identification of the quaternionic-Kahler
metrics with at least one isometry}

   We have shown in the previous subsection a notable one to one
correspondence between self-dual structures with one isometry and
3-dimensional Einstein-Weyl structures. A generalization of this
result to the Einstein self-dual metrics was obtained independently
in \cite{Przanowski}, \cite{LeBrun}. The result is stated in the
following proposition.
\\

{\bf Proposition 7 }{ \it For any self-dual Einstein metric $g$ with
one Killing vector in D=4 there exist a system of coordinates
$(x,y,z,t)$ for which takes the form \be\lb{Prza} g =
\frac{1}{z^2}[V (e^{u}(dx^2 + dy^2) + dz^2) + \frac{1}{V}(dt+ A)^2].
\ee The functions $(V, A, u)$ are independent of the variable $t$
and satisfies \be\lb{prosa1} (e^u)_{zz} + u_{yy} + u_{xx}=0, \ee
\be\lb{prosa2} dA=V_{x} dz \wedge dy + V_{y} dx \wedge dz + (V
e^{u})_{z} dy \wedge dx, \ee \be\lb{prosa3} V=2-zu_{z}. \ee
Conversely, any solution of (\ref{prosa1}), (\ref{prosa1}) and
(\ref{prosa3}) define by (\ref{Prza}) a self-dual Einstein metric.}
\\

  If the condition (\ref{prosa3})
is relaxed then proposition 7 reduces to the proposition 5 up to an
scaling by $1/z^2$. Then (\ref{prosa3}) is the condition to be
satisfied in order to have an Einstein metric. It is sufficient
because the integrability condition
$$
(V e^u)_{zz} + V_{yy} + V_{xx}=0,
$$
is always satisfied for $V=2-zu_z$. In other words, every
$SU(\infty)$ Toda solution define a self-dual metric by
(\ref{Prza}).

     Proposition 7 is describe the local form of a quaternion
Kahler space with one isometry. Quaternion Kahler spaces with two
commuting Killing are a subcase of (\ref{Prza}). Equation
(\ref{prosa1}) show us that the family $[g]$ associated to such
quaternion Kahler metrics are of the Toda type. Therefore in order
to find the quaternion Kahler metrics with $U(1)\times U(1)$
isometry we should apply proposition 7 to the metrics of proposition
6 and not to (\ref{simplif}).

     To determine which $g$ among the Joyce metrics (\ref{joycemetric})
satisfy $R_{ij}\sim g_{ij}$ we apply condition (\ref{prosa3}) to
them. For Joyce spaces we can calculate the factor $\omega=-u_z dz$
by use of $\ast dA=dV-\omega V$, (\ref{einsaxial}) and
(\ref{reform}). The result is
$$
u_z=\frac{A_0}{\rho(A_0^2+A_1^2)}.
$$
Then we have \be\lb{arou} 2-zu_z=\frac{\rho(A_0^2+A_1^2)-G
A_0}{\rho(A_0^2+A_1^2)}. \ee Insertion of the expression for $V$
(\ref{einsaxial}) into the Einstein condition $V=2-zu_{z}$ and
taking into account (\ref{arou}) gives \be\lb{wold} A_1 B_0 - A_0
B_1=\rho( A_0^2 + A_1^2) - G A_0. \ee From (\ref{wold}) it is
obtained that $B_0 = \rho A_1 + \xi_0$ and $B_1 = G - \rho A_0 +
\xi_1$ with $A_1 \xi_0 = A_0 \xi_1$. The functions $\xi_i$ are
determined by the requirement that $B_i$ also satisfy the Joyce
system (\ref{joyce1}) and (\ref{joyce2}), the result is $\xi_0=-\eta
A_0$ and $\xi_1=-\eta A_1$. Therefore the metric $g_{j}/\rho z^2$ is
Einstein if and only if \be\lb{Caldon1} A_0=G_\rho;\;\;\;\;\;\;
A_1=G_{\eta} \ee \be\lb{Caldon2} B_0=\eta G_{\rho}-\rho
G_{\eta};\;\;\;\;\;\; B_{1}=\rho G_{\rho} + \eta G_{\eta}-G. \ee
Defining $G=\sqrt{\rho}F$ it follows that F satisfies
$$
F_{\rho\rho} + F_{\eta\eta}=\frac{3F}{4\rho^2}.
$$
Then inserting (\ref{Caldon1}) and (\ref{Caldon2}) expressed in
terms of F into $g_{j}/\rho z^2$ and making the identification $z=G$
gives the following assertion due to Calderbank and Pedersen
\cite{Pedersen}.
\\

{\bf Proposition 8}{ \it For any Einstein-metric with self-dual Weyl
tensor and nonzero scalar curvature possessing two linearly
independent commuting Killing fields there exists a coordinate
system in which the metric $g$ has locally the form
$$
g=\frac{F^2-4\rho^2(F^2_{\rho}+F^2_{\eta})}{4F^2}\frac{d\rho^2+d\eta^2}{\rho^2}
$$
\be\lb{metric} +\frac{[(F-2\rho F_{\rho})\alpha-2\rho
F_{\eta}\beta]^2+[(F+2\rho F_{\rho})\beta - 2\rho F_{\eta}\alpha
]^2}{F^2[F^2-4\rho^2(F^2_{\rho}+F^2_{\eta})]}, \ee where
$\alpha=\sqrt{\rho}d\theta$ and $\beta=(d\varphi+\eta
d\theta)/\sqrt{\rho}$ and $F(\rho, \eta)$ is a solution of the
equation \be\lb{backly}
 F_{\rho\rho} + F_{\eta\eta}=\frac{3F}{4\rho^2}.
\ee

on some open subset of the half-space $\rho>0$. On the open set
defined by $F^2 > 4\rho^2(F^2_{\rho}+F^2_{\eta})$ the metric $g$ has
positive scalar curvature, whereas for $F^2 <
4\rho^2(F^2_{\rho}+F^2_{\eta})$ -$g$ is self-dual with negative
scalar curvature.}
\\

The Einstein condition $R_{ij}=\kappa g_{ij}$ is not invariant under
scale transformations, so Proposition 8 gives all the
quaternionic-Kahler metrics with $T^2$ isometry up to a constant
multiple. The problem to find them is reduced to find an F
satisfying the linear equation (\ref{backly}), that is, an
eigenfunction of the hyperbolic laplacian with eigenvalue $3/4$.

  The equation for $V$ (\ref{Wardy}) and has solutions of the
integral form
$$
V_1(\rho,\eta)=W(\eta, i\rho) + c.c,\;\;\;V_2(\rho,\eta)=W(i\eta,
\rho) + c.c
$$
\be\lb{intexp} W(\eta, \rho)=\frac{1}{2\pi}\int^{2\pi}_0 H(\rho
sen(\theta)+ \eta)d\theta \ee where $H(z)$ is an arbitrary function
of one variable \cite{Ward}. The Backlund relations (\ref{shu})
define $V$ in terms of $G$, and consequently in terms of $F$, and
viceversa. For instance, non trivial eigenfunctions $F$ can be
constructed selecting an arbitrary $H(z)$, performing the
integration (\ref{intexp}) and finding $G$ through (\ref{shu}), then
$F=G/\sqrt{\rho}$.

\subsubsection{Examples of toric quaternion Kahler spaces}

   The following are quaternionic metrics constructed starting with
an arbitrary function G(x).

A. The trivial four dimensional toric metric
$$
ds^2=d\rho^2+d\eta^2+\rho^2d\phi^2+d\psi^2
$$
corresponds to the monopole
$$
V=\eta
$$
(This AHF holds using $G(x)=x \log(x)$ in the integral expression
(\ref{intexp})). The equations (\ref{shu}) are in this case
$$
H_{\rho}=-\rho;\;\;\;\;  H_{\eta}=0,
$$
and the eigenfunction F is given by
$$
F=\frac{\rho^{3/2}}{2}.
$$
The insertion of the last eigenfunction in (\ref{metric}) gives the
following metric \be\lb{quaterna}
g=-\frac{2d\rho^2+2d\eta^2}{\rho^2}-\frac{\rho^2(1+3\rho)+\eta^2(9+7\rho)}{2\rho^5}d\phi^2
-\frac{8}{\rho^4}d\psi^2-\frac{16\eta}{\rho^4}d\phi d\psi. \ee The
inequality $F^2<4\rho^2(F^2_{\rho}+F^2_{\eta})$ holds for $\rho>0$.
Invoking the Talderbank-Pedersen theorem, we see that for $\rho>0$
the quaternionic kahler metric is -g. In this case $\kappa=-1$ and
the three 1-forms $A^i$ are
$$
A^1=\frac{2 d\eta}{\rho};\;\;\; A^2=\frac{2d\phi}{\rho};\;\;\;
A^3=\frac{2\eta}{\rho^2}d\phi+ \frac{2}{\rho^2}d\psi.
$$
The metric constructed in this example is defined for all the
positive values of $\rho$.

B. With the function $G(x)=\log(x)x^3$ it is found the monopole
$$
V=3\eta\rho^2-2\eta^3
$$
and the hyperkahler metric
$$
ds^2=(3\rho^2-6\eta^2)(d\rho^2+d\eta^2+\rho^2d\phi^2)
+\frac{1}{3\rho^2-6\eta^2}(d\psi+6\eta\rho^2 d\phi)^2.
$$
The Backglund transformed F results
$$
F=\frac{3}{4}\rho^{3/2}(4\eta^2-\rho^2)
$$
and the corresponding metric is \be\lb{quaternb}
g=g_{\rho\rho}(d\rho^2+d\eta^2)+g_{\phi\phi}d\phi^2+g_{\psi\psi}d\psi^2+2
g_{\phi\psi}d\phi d\psi, \ee where the components of the metric
tensor are
$$
g_{\rho\rho}=-\frac{4(8\eta^4+6\eta^2\rho^2+3\rho^4)}{(\rho^3-4\eta^2\rho)^2}
$$
$$
g_{\phi\phi}=-\frac{8\eta^4\rho^2(19+5\rho)+16\eta^6(9+7\rho)+\rho^6(1+35\rho)
+3\eta^2\rho^4(35+61\rho)}{9\rho^5(\rho^2-4\eta^2)^2(8\eta^4+6\eta^2\rho^2+3\rho^4)}
$$
$$
g_{\psi\psi}=-\frac{64(4\eta^4+\rho^4)}{9\rho^4(\rho^2-4\eta^2)^2(8\eta^4+6\eta^2\rho^2+3\rho^4)}
$$
$$
g_{\psi\phi}=-\frac{32(8\eta^5-\rho^2\eta^3+3\eta\rho^4)}{9\rho^4(\rho^2-4\eta^2)^2(8\eta^4+6\eta^2\rho^2+3\rho^4)}.
$$

As in the example A, $F^2<4\rho^2(F^2_{\rho}+F^2_{\eta})$ for
$\rho>0$, -g is a quaternionic kahler metric and $\kappa=-1$. The
three forms $A^i$ are given by
$$
A^1=\frac{8\eta}{\rho^2-4\eta^2}d\rho+\frac{4(\rho^2-2\eta^2)}{\rho(\rho^2-4\eta^2)}d\eta
$$
$$
A^2=\frac{4}{3\rho(\rho^2-4\eta^2)}d\phi; \;\;\;\
A^3=\frac{4\eta}{3\rho(\rho^2-4\eta^2)}d\phi+\frac{4}{3\rho(\rho^2-4\eta^2)}d\psi.
$$

The metric (\ref{quaternb}) is singular at $\rho\rightarrow 0$ and
at the lines $2|\rho|=|\eta|$.

C. The powers $G(x)=x^n$ and $G(x)=\log(x)x^{2n+1}$ can be
integrated out giving polynomial solutions of higher degree. For
instance $G(x)=Log(x)x^5$ gives
$$
V=3\eta \rho^2-2\eta^3;\;\;\;\;F=\frac{1}{\eta}(8\eta^4+ 40 \eta^2
\rho^2+15 \rho^4).
$$
The even powers $G(x)=x^{2n}\log(x)$ can be integrated too, but the
expressions of the metrics are more complicated by the appearance of
logarithm terms. For example, with $G(x)=\log(x)x^{2}$ it is
obtained
$$
V=6\eta^2-\rho^2-6\eta^2\sqrt{1+\frac{\rho^2}{\eta^2}}+2(\rho^2-2\eta^2)\log(\frac{2}{\eta+\sqrt{\eta^2+\rho^2}})
$$
$$
F= \frac{4}{3}\eta^3 - 4\eta \rho^2 - \frac{4}{3}\eta^2(\eta^2 +
\rho^2)+ \frac{8}{3} \rho^2(\eta^2 + \rho^2) + 4\eta \rho^2
\log(\frac{2}{\eta + \sqrt{\eta^2 + \rho^2)}}).
$$

The expression for the quaternionic metric and the hyperkahler one
corresponding to such solutions is very large and difficult to
simplify.

D. The function $G(x)=e^{x}$ gives
$$
V=e^{\pm i\eta} I_0(\rho)+ c.c
$$
where $I_n(\rho)$ denotes the modified Bessel function of the first
kind, which are solutions of the equation
$$
\rho^2 H''(\rho)+\rho H'(\rho)-(\rho^2+ n^2)H(\rho)=0.
$$
 The hyperkahler space that corresponds to this monopole is:
$$
ds^2=\rho I_1(\rho)(d\rho^2+d\eta^2+\rho^2d\phi^2)\cos(\eta)
$$
\be\lb{kahlermia} +\frac{1}{\rho I_1(\rho)\cos(\eta)}
\{d\psi+\rho[I_1(\rho)+\frac{\rho}{2}(I_0(\rho)+I_2(\rho))]\sin(\eta)d\phi\}^2.
\ee The Backglund transformed eigenfunction F is given by
$$
F= \sqrt{\rho}I_1(\rho)e^{\pm i\eta}+ c.c.
$$
and from this solutions it follows the quaternionic metric
$$
ds_{qk}=\frac{\Theta(\rho, \eta)}{4\rho
I_1(\rho)^2}(d\rho^2+d\eta^2) +\frac{[2\rho^{3/2}I_1(\rho)\beta
\cos(\eta)-\rho^{3/2} (I_0(\rho)+ I_2(\rho))\alpha
\sin(\eta)]^2}{\Phi(\rho,\eta)}
$$
\be\lb{metricmia} +\frac{[2\rho^{3/2}I_1(\rho)\alpha \cos(\eta) +
\sqrt{\rho}(\rho I_0(\rho)+ 2 I_1(\rho)+ \rho I_2(\rho)) \beta
\sin(\eta)]^2}{\Phi(\rho,\eta)}, \ee where it has been defined
$$
\Theta(\rho, \eta)=\rho I_0(\rho)^2 + 2 I_1(\rho) I_2(\rho)+ \rho
I_2(\rho)^2 + 4 \rho I_1(\rho)^2 (\frac{\cos(\eta)}{\sin(\eta)})^2 +
2I_0(\rho)(I_1(\rho)+ \rho I_2(\rho))
$$
and
$$
\Phi(\rho,\eta)=-\rho^2 I_1(\rho)^2 \sin(\eta)^2[4\rho^2 I_1(\rho)^2
\cos(\eta)^2- I_1(\rho)^2 \sin(\eta)^2+ (\rho I_0(\rho)+ I_1(\rho)+
\rho I_2(\rho))^2 \sin(\eta)^2].
$$
It will be quaternionic for the regions of the plane $(\rho, \eta)$
in which $F^2< 4\rho^2(F^2_{\rho}+F^2_{\eta})$. In those regions
$k=-1$. The three one forms $A^i$ are
$$
A^1=\frac{1}{2}[1 + \tan(\eta)+
\frac{\rho}{I_1(\rho)}\tan(\eta)(I_0(\rho)+I_2(\rho))]
\frac{d\eta}{\rho}-\tan(\eta)d\rho
$$
$$
A^2=\frac{\alpha}{\sqrt{\rho}I_1(\rho)\cos(\eta)};\;\;\;\;
A^3=\frac{\beta}{\sqrt{\rho}I_1(\rho)\cos(\eta)}.
$$
The radial component of the metric (\ref{metricmia}) shows that some
of the singularities are the zeros of $I_1(\rho)$.

    By completeness we will also discuss the spaces corresponding
to the $m$-pole solutions investigated in \cite{Dancer} and
\cite{Pedersen}. The dynamics of the M-theory on toric $G_2$ cones
constructed with $m$-pole spaces as base manifolds have been
analyzed in \cite{Angelova} and \cite{Angelito}. Such examples leads
to toric Einstein self-dual spaces of positive scalar curvature
which are complete (thus compact) and admiting only orbifold
singularities. As such, they seem to be the best candidates with
$U(1)\times U(1)$ isometry for which the physical analysis given in
\cite{Witten2} can be applied. In the first reference of
\cite{Pedersen} it has been described the moduli space corresponding
to the $3$-pole solutions and has been shown that they encode some
well known examples appearing in the physics, like the Bianchi type
spaces. It will be shown that the hyperkahler metrics corresponding
to the $3$-pole solution are those discussed in \cite{Calderbank2}
which gives rise to the $3$-dimensional Eguchi-Hanson like
Einstein-Weyl metrics after the quotient by one of the isometries.
The following exposition follows closely those given in the
references \cite{Pedersen}.

   The basic eigenfunctions F of (\ref{backly}) which we need to consider are
\be\lb{solu} F(\rho, \eta,
y)=\frac{\sqrt{(\rho)^2+(\eta-y)^2}}{\sqrt{\rho}} \ee where the
parameter $y$ takes arbitrary real values. Using the Backglund
transformation it is found the basic monopole \be\lb{soludos}
V(\eta, \rho, y)=-\log[\eta-y + \sqrt{\rho^2 + (\eta-y)^2}]. \ee

   Being the equations for F and V linear, for any set of
real numbers $w_i$ the functions \be\lb{superpos} F=\sum^{k+1}_{j=0}
w_i F(\rho, \eta, y_j). \ee \be\lb{super2} V=\sum^{k+1}_{j=0} w_i
V(\rho, \eta, y_j) \ee will be solutions too. For this reason the
$2$-pole functions given by
$$
F_1=\frac{1+\sqrt{\rho^2+\eta^2}}{\sqrt{\rho}};\;\;\;
F_2=\frac{\sqrt{(\rho)^2+(\eta+1)^2}}{\sqrt{\rho}}-
\frac{\sqrt{(\rho)^2+(\eta-1)^2}}{\sqrt{\rho}},
$$
are eigenfunctions of the hyperbolic laplacian. The first one gives
rise to the spherical metric, while the second one gives rise to the
hyperbolic metric
$$
ds^2=(1-r_1^2-r_2^2)^{-2}(dr_1^2+dr^2_2 + r_1^2d\theta_1^2 +
r_2^2d\theta_2^2).
$$
The relation between the coordinates $(r_1, r_2)$ and $(\rho, \eta)$
can be extracted from the relation
$$
(r_1 + i r_2)^2=\frac{\eta-1+i\rho}{\eta+1+i\rho}.
$$
The hyperkahler metrics corresponding to both cases are
\be\lb{hyper1} ds^2=-\frac{1}{\sqrt{\rho^2 +
\eta^2}}(d\rho^2+d\eta^2+\rho^2d\phi^2) -\sqrt{\rho^2 +
\eta^2}(d\psi+ \frac{\eta}{\sqrt{\rho^2 + \eta^2}}d\phi)^2, \ee and
$$
ds^2=
\frac{\sqrt{\rho^2+(\eta-1)^2}-\sqrt{\rho^2+(\eta+1)^2}}{\sqrt{\rho^2+(\eta+1)^2}\sqrt{\rho^2+(\eta-1)^2}}
(d\rho^2+d\eta^2+\rho^2d\phi^2)
$$
\be\lb{hyper2}
+\frac{\sqrt{\rho^2+(\eta+1)^2}\sqrt{\rho^2+(\eta-1)^2}}{\sqrt{\rho^2+(\eta-1)^2}-\sqrt{\rho^2+(\eta+1)^2}}
[d\psi+(\frac{\eta+1}{\sqrt{\rho^2+(\eta+1)^2}}-\frac{\eta-1}{\sqrt{\rho^2+(\eta-1)^2}})d\phi]^2.
\ee

    The general "$3$-pole" solutions are
$$
F=\frac{1}{\sqrt{\rho}}+\frac{b+c/m}{2}\frac{\sqrt{\rho^2+(\eta+m)^2}}{\sqrt{\rho}}
+\frac{b-c/m}{2}\frac{\sqrt{\rho^2+(\eta-m)^2}}{\sqrt{\rho}}.
$$
By definition $-m^2=\pm 1$, which means that $m$ can be imaginary or
real. The corresponding solutions are denominated type I and type II
respectively. It is convenient to introduce the Eguchi-Hanson like
coordinate system defined by
$$
\rho= \sqrt{R^2 \pm 1} \cos(\theta), \;\;\;\; \eta=R \sin(\theta),
$$
where $\theta$ takes values in the interval $(-\pi/2, \pi/2)$. In
this coordinates \be\lb{unade} \sqrt{\rho}F=1+b R+c \sin(\theta),
\ee \be\lb{dosde}
\rho^{-1}[\frac{1}{4}F^2-\rho^2(F_{\rho}^2+F_{\eta}^2)]= \frac{b(R
\mp b) + c(\sin(\theta) + c)}{R^2 \pm sin^2(\theta)}. \ee and the
family of self-dual metrics corresponding to the $3$-pole are
expressed as
$$
ds^2=\frac{b^2-c^2+(bR-cS)}{(1+b R+c
S)^2}(\frac{dR^2}{R^2-1}+\frac{dS^2}{1-S^2})
$$
$$
\frac{1}{(1+b R+ cS)^2(b^2-c^2+(bR-cS))(R^2-S^2)}
$$
$$
*((R^2-1)(1-S^2)((bR-cS)d\varphi + (cR-bS) d\psi)^2
$$
\be\lb{enclo} + ((b(R^2-1)S+c(1-S^2)R)d\varphi + (c(R^2-1)S +
b(1-S^2)R + (R^2-S^2)d\psi)^2) \ee It has been denoted $S=
\sin(\theta)$ here. The expression (\ref{enclo}) includes some well
known metrics. Let us focus in the type I case. The formulas
(\ref{unade}) and (\ref{dosde}) allows to determine the domain of
definition of the metric (\ref{metric}). When b is nonzero for a
given value of $\theta$, $F=0$ if $R=-(1+c \sin(\theta))/b$  and
$(\frac{1}{4}F^2-\rho^2(F_{\rho}^2+F_{\eta}^2))=0$ if $R=(b^2+c^2+c
\sin(\theta))/b$. The case $c=0$ correspond to a bi-axial Bianchi IX
metric \cite{Pedor}. The domains of definition are $(-\infty,
R_{\infty}), (R_{\infty}, R_{\pm})$, and $(R_{\pm}, \infty)$. In the
first two cases the curvature is negative, and in the last one
positive, and in the two last cases there is an unremovable
singularity at $R=R_{\pm}$. In the case $b=0$ for $c>1$ and $c<1$
the metric will be of Bianchi VIII type \cite{Caldo}. The case $c=1$
corresponds to the Bergmann metric on $CH^2$.

  For the type II case, the range of R is $(1,\infty)$ but the moduli space is more complex
that in the type I case. For the lines $b=\pm c$ it is obtained the
hyperbolic metric if $b<0$ and the spherical metric if $b>0$. If
$(b,c)=(1,0)$ it is obtained the Fubbini-Study metric on $CP^2$
whereas the points $(0,1), (-1,0)$ and $(0,-1)$ yield again the
Bergmann metric on $CH^2$. Along the lines joining $(1,0)$ with
others we have bi-axial Bianchi metric IX, while along the lines
between $(0,1), (-1,0)$ and $(0,-1)$ the metric is Bianchi VIII. A
more complete description is given in \cite{Pedersen}.

   The triplet of one forms corresponding to this family of metrics is
$$
A^1=A^1_{+} +A^1_{-},
$$
$$
A^2=\frac{\sqrt{(R^2 \pm 1)(1-S^2)}}{(1+bR+cS)}d\phi\;\;\;
A^3=\frac{d\psi+\eta d\phi}{(1+bR+cS)},
$$
where it has been defined
$$
A^1_{\pm}=A^1_{1\pm}+A^1_{2\pm}+A^1_{3\pm},
$$
with
$$
A^1_{1\pm}=\frac{(b \pm c/m)(S R \pm m)}{(1+bR+cS)\sqrt{(1 -
S^2)(R^2\pm 1)+ ( S R\pm m)^2}}*
$$
$$
(\frac{S\sqrt{R^2\pm1}}{2\sqrt{1-S^2}}dS + \frac{R
\sqrt{1-S^2}}{\sqrt{R^2\pm 1}} dR )
$$
and
$$
A^1_{2\pm}=\frac{(b \pm c/m)\sqrt{(1 - S^2)(R^2\pm 1) + ( S R\pm
m)^2}}{2(1+bR+cS)\sqrt{R^2 \pm 1}} (R\frac{dS}{\sqrt{1-S^2}}+ S dR)
$$
and
$$
A^1_{3\pm}=[\frac{(b \pm c/m)(R^2\pm 1)^{1/2}(1 -
S^2)^{1/2}}{2\sqrt{(1 - S^2)(R^2\pm 1)+ ( S R \pm m)^2}}- \frac{(b
\pm c/m)\sqrt{(1 - S^2)(R^2\pm 1)+ (S R \pm m)^2}}{4(1 -
S^2)^{1/2}(R^2\pm 1)^{1/2}}]*
$$
$$
\frac{\sqrt{(1 - S^2)(R^2\pm 1)}}{(1+bR+cS)\sqrt{R^2 \pm 1}}
(R\frac{dS}{\sqrt{1-S^2}}+ S dR)
$$

(The sign $\pm$ in $(R^2\pm 1)$ depends only on the metric in
consideration, it is $+$ for type I and $-$ for type II.)

  The Backglund transformed function V reads
$$
V=\log(\rho)+\frac{(b+c/m)}{2}\log[\frac{\eta-m+\sqrt{(\eta-m)^2+\rho^2}}{\rho}]
$$
$$
+\frac{(b-c/m)}{2}\log[\frac{\eta+m+\sqrt{(\eta+m)^2+\rho^2}}{\rho}].
$$
For the type I case this is the potential for an axially symmetric
circle of charge, while the type II case corresponds to two point
sources on the axis of symmetry. The hyperkahler metrics obtained
are encoded in the following expression
$$
ds^2=\frac{bR+c\sqrt{1-S^2}}{R^2\pm (1-S^2)}
(d\rho^2+d\eta^2+\rho^2d\phi^2)
$$
\be\lb{hyper2} +\frac{R^2 \pm (1-S^2)}{bR+c\sqrt{1-S^2}} [ d\psi
+\frac{R^2\pm(1-S^2)-b(R^2\pm 1)\sqrt{1-S^2} + c R S^2}{R^2 \pm
(1-S^2)}d\phi ]^2. \ee This manifolds have been investigated
recently in \cite{Calderbank2} and it has been shown that the
quotient of (\ref{hyper2}) with $\frac{\partial}{\partial \phi}$
gives the Eguchi-Hanson type Einstein-Weyl metrics in D=3.

    The continuum limit of the expressions (\ref{superpos}) and (\ref{super2}) are
\be\lb{continium} F(\rho, \eta)=\int w(y) F(\rho, \eta, y)dy. \ee
\be\lb{continium2} V(\rho, \eta)=\int w(y) V(\rho, \eta, y)dy. \ee

where $w(y)$ is a distribution with compact support in $R$. A choice
of $w(y)$ for which at least one of the integrals (\ref{continium})
and (\ref{continium2}) converges gives rise to an smooth solution.
For instance for $w(y)=y/(y^2+1)^2$ it is obtained the following
non-trivial monopole
$$
V(\rho, \eta)=\frac{\cos(\frac{1}{2}Arg(1-2i\eta-\eta^2-\rho^2))
\log(\frac{|1-i\eta-\sqrt{(1-i\eta)^2+\rho^2}|}{|1+i\eta-\sqrt{(1-i\eta)^2+\rho^2}|})}{\sqrt{|(1-i\eta)^2+\rho^2}|}
$$
$$
+\frac{\sin(\frac{1}{2}Arg(1-2i\eta-\eta^2-\rho^2))
Arg(\frac{1-i\eta-\sqrt{(1-i\eta)^2+\rho^2}}{1+i\eta-\sqrt{(1-i\eta)^2+\rho^2}})}{\sqrt{|(1-i\eta)^2+\rho^2}|},
$$
and from (\ref{queseyo1}) follows an hyperkahler metric. But
(\ref{continium}) is divergent for this distribution.

    To conclude this subsection it should be mentioned that higher $m$-pole solutions
have been considered in \cite{Angelova} and \cite{Angelito}, and
that quaternionic spaces with torus symmetry have been investigated
recently in \cite{Valent} using the harmonic space formalism.

\newpage

\subsection{Higher dimensional hypergeometry}

   The previous sections were concerned with quaternion Kahler and
hyperkahler manifolds in $d=4$. In this section we deal with higher
dimensional ones. We present the most general $4n$-dimensional
hyperkahler metrics with $n$ commuting tri-holomorphic Killing
vectors \cite{Poon}, namely the Pedersen-Poon metrics. This result
generalize the Gibbons-Hawking one (\ref{ashgib}) to $d=4n$. We also
present a construction allowing to extend an arbitrary quaternion
Kahler metric in $d=4n$ to other quaternion Kahler and an
hyperkahler ones in $d=4(n+1)$. This is the Swann construction
\cite{Swann}. As we will see, the Swann extension of the
Calderbank-Pedersen metrics (\ref{metric}) preserves the Killing
commuting Killing vectors and they are tri-holomorphic with respect
to the new metric. Therefore the extension is of a Pedersen-Poon
type. Once this is understood it is straightforward to construct
certain classical supergravity solutions. This is the subject of the
next section.

\subsubsection{Construction of $4n$ hyperkahler manifolds with $T^n$
tri-holomorphic isometry}

    Let us remind that in section 3 we have described all the four
dimensional hyperkahler spaces with at least one self-dual Killing
vector namely, the Gibbons-Hawking metrics. We have seen that for
such spaces there always exist a coordinate system for which the
metric take the form \be\lb{weare} g=V^{-1}(dt-\omega)^2 + V dx
\cdot dx. \ee The vector $K=\partial_t$ satisfies \be\lb{Killing}
{\cal L}_{K}g=0 \qquad \Longrightarrow \qquad K_{\mu;\nu}+
K_{\nu;\mu}=0 \ee and is therefore a Killing vector. The tensor
$K_{\mu\nu}=\nabla_{\mu} K_{\nu}$ satisfies \be\lb{sdkv} \ast_{g}
K_{\mu\nu}=K_{\mu\nu} \ee and for this reason $K$ is called
self-dual in the terminology of the references
\cite{Boyero}-\cite{Gegenberg}.

    The vector $K=\partial_{t}$ also preserve the hyperkahler triplet
$\overline{J}^k$ of (\ref{weare}). Explicitly $\overline{J}^k$ is
given by \be\lb{usluga} \overline{J}^k=(dt+A)\wedge
dx^k-\epsilon_{rs}^k U(dx^r \wedge dx^s), \ee  and from
(\ref{usluga}) its is direct to check that \be\lb{triho} {\cal
L}_{K} \overline{J}^i=0. \ee In general a vector preserving
$\overline{J}^i$ as in (\ref{triho}) is called tri-holomorphic. If
instead we have that \be\lb{triho} {\cal L}_{K} J^i=0, \ee then $K$
preserve the complex structures and is called tri-hamiltonian.

    It can be shown that any two of the conditions given above
for $K$ implies the third. This mean in particular that for the
metric (\ref{weare}) the vector $K=\partial_{t}$ is Killing,
tri-holomorphic and tri-hamiltonian. For the spaces (\ref{gegodas})
this condition does not hold, the Gibbons-Hawking metrics
(\ref{weare}) are the only one with this properties in $d=4$.

      The higher dimensional analogs of (\ref{weare}) to $d=4n$ dimensions
are the hyperkahler spaces with $n$ commuting tri-holomorphic
Killing vectors $\partial_{t_i}$. A plausible anzatz is obtained by
introducing a set on $3n$ coordinates $x^i=\{x^i_r,\;\;
i=1,..,n\;\;r=1,2,3\}$ and the metric \be\lb{prove}
\overline{g}=U_{ij}dx_i\cdot dx_j + U^{ij}(dt_i+A_i)(dt_j+A_j), \ee
where the symmetric matrix $U_{ij}$ and then its inverse $U^{ij}$
are independent of the coordinates $t^i$. The $n$ one forms $A_j$
have the form \be\lb{exp} A_i=dx^j \cdot \Sigma_{ij} \ee where the
matrix $\Sigma_{ij}$ also do not depend on the $t_i$ coordinates.
The hyperkahler triplet corresponding to (\ref{prove}) is given by
\be\lb{multidim} \overline{J}^k =(dt_r + A_r) \wedge dx^r_k -
U_{ij}(dx^i \times dx^j)_k \ee Here $\times$ denote the exterior
product of forms (for instance the 3-component of  $dx^i \times
dx^j$ is $dx^i_1 \times dx^j_2$). Then it is directly seen that
${\cal L}_{\partial_i} \overline{J}^k=0$, therefore the Killing
vectors $\partial_{t_i}$ are tri-holomorphic and tri-hamiltonian.
For $n=1$ (\ref{prove}) reduce to the Gibbons-Hawking metrics
(\ref{weare}).

    Let us now find the conditions should be satisfied for
(\ref{prove}) to be an hyperkahler metric. In order to use the
$4n$-bein formalism we can express the matrix $U$ as \be\lb{upto} U
= K^{T}K \ee for a non singular matrix $K$, which is defined by
(\ref{upto}) up to an $SO(n)$ rotation. Then we can select the
$4n$-bein as
$$
e_{a}=K_{aj}dx^j \qquad (i=1,..,3n)
$$
\be\lb{4nbein} e_{a}=(dt_j+A_j)K^{aj} \ee being $K^{aj}$ the inverse
of $K_{aj}$. Then if (\ref{prove}) is hyperkahler, then $N^i(X,Y)=0$
for every of the complex structures $J^i$ corresponding to
(\ref{reprodui}). In $4n$ dimensions the conditions (\ref{geno}) are
generalized to a system of equations with a $4n$ vector potential
$\widetilde{A}_1,..., \widetilde{A}_{4n}$ that we will not write for
simplicity. But introducing (\ref{4nbein}) into such system gives
finally the following system of equations
$$
F_{x_{k}^i x_{l}^j}=\epsilon_{klm}\nabla_{x_{m}^i}U_j,
$$
\be\lb{pedero} \nabla_{x_{m}^i}U_j=\nabla_{x_{m}^j}U_i, \ee
$$
U_i=(U_{i1},...., U_{in}),
$$
being $F_{x_{k}^i x_{l}^j}$ the strength tensor corresponding to the
"vector potential" $A_a$. If the last two conditions holds then
(\ref{prove}) is hyperkahler.

      It was also shown that (\ref{prove}) are unique. In other
words, for any $T^n$ hyperkahler metrics there exists a coordinate
system in which they can be cast in the form (\ref{prove})
\cite{Poon}-\cite{Rocho}. Such system is the momentum map system. In
general momentum maps are related to a compact Lie group $G$ acting
over an hyperkahler manifold $M$ by tri-holomorphic Killing vectors
$X$, i.e, Killing vectors satisfying
$$
\pounds_X J^k=0.
$$
The last condition implies that $X$ preserves the hyperkahler forms
$\overline{J}^k$, that is
$$
\pounds_X \overline{J}^k=0= i_X d\overline{J}^k+ d(i_X
\overline{J}^k).
$$
Here $i_X \overline{J}^k$ denotes the contraction of $X$ with the
hyperkahler forms. By supposition $M$ is hyperkahler, then
$d\overline{J}^k=0$ and
$$
d(i_X \overline{J}^k)=0.
$$
This mean that $i_X \overline{J}^k$ can be expressed locally as the
differential of certain function $x^X_k$, namely \be\lb{momap}
dx^X_k=i_X\overline{J}^k. \ee The functions $x^X_k$ are called
\emph{momentum maps} and are defined up to a constant.

    In the case (\ref{prove}) the isometries are $\partial/\partial
t_i$ and the hyperkahler form corresponding to (\ref{gengibbhawk})
is \cite{Gibbin} \be\lb{quakal}
\overline{J}^k=(dt_i+A_i)dx^i_k-U_{ij}(dx^i \wedge dx^j)_k. \ee From
the last expression it follows that \be\lb{moroder} dx^{i}_k =
i_{t_i}\overline{J}^k \ee and therefore $(x^1_i, ...,x^n_i)$ are
really the momentum maps of the isometries. All the results of this
section are stated in the following proposition \cite{Poon}.
\\

{\bf Proposition 9 }{ \it For any hyperkahler metric in $D=4n$ with
$n$ commuting tri-holomorphic $U(1)$ isometries there exists a
coordinate system in which takes the form \be\lb{gengibbhawk}
\overline{g}=U_{ij}dx^i\cdot dx^j+ U^{ij}(dt_i+A_i)(dt_j+A_j), \ee
where $(U_{ij}, A_i)$ are solutions of the generalized monopole
equation
$$
F_{x_{\mu}^i
x_{\nu}^j}=\epsilon_{\mu\nu\lambda}\nabla_{x_{\lambda}^i}U_j,
$$
\be\lb{genmonop}
\nabla_{x_{\lambda}^i}U_j=\nabla_{x_{\lambda}^j}U_i, \ee
$$
U_i=(U_{i1},.., U_{in}),
$$
and the coordinates $(x^1_i,..., x^n_i)$ with $i=1, 2, 3$ are the
momentum maps of the tri-holomorphic vector fields
$\partial/\partial t_i$.}
\\

  The spaces in the form(\ref{gengibbhawk}) are known as the Pedersen-Poon
metrics. Equations (\ref{genmonop}) are known as monopole equations
by interpreting $F_{x_{\mu}^i x_{\nu}^j}$ as a vector potential and
$U_i$ as certain Higgs fields. Some Pedersen-Poon examples will be
constructed in the following section.

\subsubsection{Quaternion Kahler spaces in quaternion notation}

      It is convenient to introduce a notation
for quaternion Kahler manifolds in terms of quaternions, that has
the advantage to be more compact \cite{Swann}. It is fundamental to
observe that the holonomy of a quaternion Kahler space is in
$Sp(n)\times Sp(1)$ and therefore from the very beginning the
connection $\omega$ will take values in the algebra $sp(n)\oplus
sp(1)$ with splitting $\omega=\omega_{+} + \omega_{-}$. The Lie
algebra $sp(n)$ of $Sp(n)$ can be expressed in terms of quaternion
valued matrices $A$ with the property $A+\overline{A}^t=0$.

   Let us consider now two arbitrary
\textbf{H}-valued 1-forms
$$
\mu=\mu_{0}+ \mu_{1}I +\mu_{2}J +\mu_{3}K,\;\;\;\;\; \nu=\nu_{0}+
\nu_{1}I+\nu_{2}J+\nu_{3}K,
$$
being $I$, $J$ and $K$ unit quaternions. We define the quaternionic
wedge product as \be\lb{quaternionil} \mu \wedge_q
\nu=(\mu_0\wedge\nu_1-\mu_2\wedge\nu_3)I
+(\mu_0\wedge\nu_2-\mu_3\wedge\nu_1)J+(\mu_0\wedge\nu_3-\mu_1\wedge\nu_2)K
\ee
$$
+\mu_0\wedge\nu_0 + \mu_1\wedge\nu_1+ \mu_2\wedge\nu_2 +
\mu_3\wedge\nu_3 ,
$$
and in particular for two pure quaternionic forms we have
\be\lb{quaternioni} \overline{\mu} \wedge_q
\mu=(\mu_0\wedge\mu_1-\mu_2\wedge\mu_3)I
+(\mu_0\wedge\mu_2-\mu_3\wedge\mu_1)J+(\mu_0\wedge\mu_3-\mu_1\wedge\mu_2)K,
\ee with components only in Im \textbf{H}. This product can be
extended to $\textbf{H}^n$ valued 1-forms, with
$\textbf{H}$-components given by \be\lb{laextendo}
\mu^i=\mu_{0}^{i}+ \mu_{1}^i I +\mu_{2}^i J +\mu_{3}^i K,\;\;\;\;\;
\nu^i=\nu_{0}^i + \nu_{1}^i I+\nu_{2}^i J+\nu_{3}^i K. \ee The wedge
product $\overline{\wedge}$ of the forms (\ref{laextendo}) is
defined as
$$
\mu \overline{\wedge} \nu= \mu^i \wedge_q \nu^i,
$$
and is clearly seen that $\mu \overline{\wedge} \nu$ is
$\textbf{H}$-valued.

   Consider now a four dimensional metric
$$
g = e^1 \otimes e^1 + e^2 \otimes e^2 + e^3 \otimes e^3 + e^4
\otimes e^4.
$$
with an orthonormal tetrad $e^i$. We can extend such tetrad to a
quaternionic valued one
$$
e = e_1 + e_2 I + e_3 J + e_4 K,\;\;\;\;\; \overline{e} = e_1 - e_2
I - e_3 J - e_4 K,
$$
for which the distance $g$ is expressed as $g = e \otimes
\overline{e}$. The hyperkahler triplet associated to $g$
$$
\overline{J}^1=e_4 \otimes e^1 - e_1 \otimes e^4 + e_2 \otimes e^3 -
e_3 \otimes e^2
$$
\be\lb{trip} \overline{J}^2=e_4 \otimes e^2 - e_2 \otimes e^4 + e_3
\otimes e^1 - e_1 \otimes e^3 \ee
$$
\overline{J}^3=e_4 \otimes e^3 - e_3 \otimes e^4 + e_1 \otimes e^2 -
e_2 \otimes e^1
$$
is extended in this notation to a quaternion valued $(1,1)$ tensor
\be\lb{baba} \overline{J}=\overline{J}^1 I + \overline{J}^2 J +
\overline{J}^3 K. \ee
 From the definition (\ref{quaternioni}) it
follows that \be\lb{triqua} \overline{J}= e \wedge_q \overline{e}.
\ee In $4n$ dimensions we select the $\textbf{H}^n$-valued einbein
$e$ defined by \be\lb{laura} e^{a}=e^{a}_0 + e^a_1 I + e^a_2 J +
e^a_3 K. \ee The metric is then $g = e^a \otimes \overline{e}^a$.
Taking into account the representation of $Sp(n)$ in terms of
quaternions we can write $\omega=\omega_{+} + \omega_{-}$ also as a
$\textbf{H}^n$ one-form. Then it can be shown that the first Cartan
equation translates into \be\lb{carton} de = - \omega
\overline{\wedge} e \ee and that the hyperkahler triplet is
\be\lb{citala} \overline{J}=e \overline{\wedge} \overline{e}. \ee
The quaternionic expression of the fundamental relations
(\ref{rela}) and (\ref{basta}) in D=$4n$ are \be\lb{gaugo}
d\omega_{-}-\omega_{-} \overline{\wedge} \omega_{-}=\Lambda
\overline{J}, \;\;\;\; d\overline{J}-\omega_{-} \overline{\wedge}
\overline{J}+ \overline{J}\overline{\wedge} \omega_{-}=0. \ee
Introducing the second (\ref{gaugo}) into the first to give
\be\lb{gaugo2} \Lambda d\overline{J} +
\omega_{-}\overline{\wedge}d\omega_{-} - d\omega_{-}
\overline{\wedge} \omega_{-}=0,\;\;\;
d\omega_{-}-\omega_{-}\overline{\wedge}\omega_{-}+\Lambda\overline{J}=0,
\ee expressing $\overline{J}$ and $d\overline{J}$ entirely in terms
of the $sp(1)$ connection $\omega_{-}$ and its differentials.

\subsubsection{The Swann extension}

     The Swann construction extend any quaternionic Kahler in
dimension $D=4n$ to another $4(n+1)$-dimensional quaternion Kahler
one \cite{Swann}. Before to present it, let us consider an $4n$
dimensional hyperkahler manifold $M$ with metric
$\overline{g}=\delta_{ab} e^a \otimes e^b$. From the definition
$\overline{g}$ has holonomy in $Sp(n)$. We make the trivial
extension to a $4(n+1)$-dimensional metric given by \be\lb{extens}
g=\overline{g}+(du_0)^2+(du_1)^2 +(du_2)^2 + (du_3)^2, \ee being
$(u_0, u_1, u_2, u_3)$ four new coordinates. The metric
(\ref{extens}) is the direct sum of $\overline{g}$ plus a flat
metric. Then the holonomy $\Gamma$ of (\ref{extens}) lies in
$Sp(n)\subset Sp(n+1)$, and thus (\ref{extens}) is a trivial
extension of $\overline{g}$ to a higher dimensional hyperkahler one.

      Now let us generalize the trivial anzatz (\ref{extens}) by extending
a quaternionic Kahler metric $\overline{g}=\delta_{ab} e^a \otimes
e^b$ to another one in $D=4(n+1)$ of the form \be\lb{Swann}
g_s=g|u|^2 \overline{g} + f|du + u \omega_{-}|^2. \ee We have
introduced the quaternions
$$
u=u_0 + u_1 I + u_2 J + u_3 K ,\;\;\;\;\;\; \overline{u}= u_0 - u_1
I - u_2 J - u_3 K
$$
$$
\omega_{-}=\omega_{-}^1 I+\omega_{-}^2 J +\omega_{-}^3 K
$$
and the radius
$$
|u|^2=u\overline{u}=(u_0)^2+ (u_1)^2+ (u_2)^2+ (u_3)^2.
$$
being $I, J, K$ the unit quaternions. It is clear that if we
consider an hyperkahler base $\overline{g}$ then $\omega_{-}^i=0$
and we recover (\ref{extens}) from (\ref{Swann}) if the functions
$f(|u|^2)$ and $g(|u|^2)$ are simply constants. In general the
explicit form of the metric is \be\lb{explico}
g_s=g|u|^2\overline{g}+f[(du_0-u_i\omega_{-}^i)^2 +(du_i+
u_0\omega_{-}^i + \epsilon_{ijk}u_k\omega_{-}^k)^2] \ee although the
expression (\ref{Swann}) will be better for the following.

    If $g_s$ is quaternion Kahler space then $\Theta$ is closed
(see the discussion below (\ref{lafunda})) and this gives a system
of equation defining $f$ and $g$. The fundamental four form
(\ref{lafunda}) for $g_s$ is explicitly
$$
\Theta=fg(\alpha\wedge \overline{\alpha}\wedge u\overline{e}^t
\wedge e \overline{u} +u\overline{e}^t \wedge e \overline{u}\wedge
\alpha\wedge \overline{\alpha})
$$
\be\lb{lafundo} + g^2 |u|^4 \overline{e}^t \wedge e
\wedge\overline{e}^t\wedge e+ f^2 \alpha\wedge \overline{\alpha}
\wedge\alpha\wedge \overline{\alpha} \ee being $\alpha=du + u
\omega_{-}$. If we want to impose $d\Theta=0$ to (\ref{Swann}) we
need the identities \be\lb{ainou1} d(\overline{e}^t \wedge e \wedge
\overline{e}^t \wedge e)=0, \ee \be\lb{ainou2} dr^2=u
d\overline{u}+\overline{u}du=u\overline{\alpha}+\alpha\overline{u}.
\ee \be\lb{ainou3} d(u\overline{e}^t \wedge e \overline{u})=
\alpha\wedge\overline{e}^t \wedge e \overline{u} +u\overline{e}^t
\wedge e \wedge\overline{\alpha}, \ee \be\lb{ainou4} d(\alpha\wedge
\overline{\alpha}) =-c(u\overline{e}^t \wedge e \wedge
\overline{\alpha}+ \alpha\wedge \overline{e}^t \wedge e
\overline{u}). \ee An straightforward calculation by using
(\ref{ainou1})-(\ref{ainou4}) gives \be\lb{vule}
d\Theta=A\alpha\wedge \overline{\alpha}\wedge u\overline{e}^t \wedge
e \overline{u}+ A u\overline{e}^t \wedge e \overline{u}\wedge
\alpha\wedge \overline{\alpha} + B |u|^4 dr^2 \wedge \overline{e}^t
\wedge e \wedge \overline{e}^t \wedge e, \ee with the coefficients
$A$ and $B$ given explicitly by
$$
A=-\frac{3c}{|u|^2}f^2+ \frac{3}{|u|^2}fg+ (fg)_{|u|},
$$
$$
B=-\frac{2c}{|u|^2}fg+ \frac{2}{|u|^2}g^2+ 2g(g)_{|u|}.
$$
The closure of $\Theta$ implies $A=B=0$ and this gives two
differential equations for $f$ and $g$ with solution \be\lb{noli1}
f=\frac{a}{c(b |u|^2 + d)^2}, \ee \be\lb{noli2} g=\frac{1}{b |u|^2 +
d}, \ee with $a$, $b$, $c$ and $d$ constants. Introducing
(\ref{noli1})-(\ref{noli2}) into (\ref{explico}) gives the metric
\be\lb{abuela} g_s=\frac{|u|^2}{b|u|^2 + d}\overline{g}+\frac{a}{c(b
|u|^2+d)^2}[(du_0-u_i\omega_{-}^i)^2 +(du_i+ u_0\omega_{-}^i +
\epsilon_{ijk}u_k\omega_{-}^k)^2]. \ee The closure of $\Theta$ in
principle is not enough to prove that (\ref{Swann}) is quaternionic
Kahler. But it is possible to show that the other necessary and
sufficient conditions (\ref{lamas}) and (\ref{rela}) are indeed
satisfied for (\ref{abuela}). Therefore (\ref{abuela}) gives a
family of quaternion Kahler metrics, known as the Swann metrics. The
case $b=0$ corresponds to an hyperkahler metric.

\newpage

\subsection{Spaces with $G_2$ holonomy}

       The Lie group $G_2$ was considered by Berger
\cite{Berger} as one of the possible holonomy groups of a Riemmanian
Ricci-flat space. He also proved that $G_2$ holonomy is possible
only in seven dimensions. The existence of $G_2$ holonomy metrics
was proved rigorously in \cite{Bryant} 30 years after the Berger
work. In particular explicit non compact examples in \cite{Salamon}.

     This spaces have direct application to Kaluza-Klein
compactifications of M-theory. This is due to the fact that $G_2$
holonomy manifolds have globally defined one Killing spinor $\eta$
satisfying $D_i\eta=0$ and therefore as internal spaces they
preserve certain amount of supersymmetries after compactification.
In fact the presence of such spinor is the reason for the reduction
of the holonomy from $SO(7)$ to $G_2$.

     One of the main requirements in order to obtain a realistic
four dimensional theory after compactification is that the internal
space should be compact \cite{Duff}. Also the appearance of chiral
fermions in the effective theory is possible only if the internal
space has certain singularities \cite{Witten2}. Although the
existence of compact $G_2$ holonomy spaces was rigurously proved
recently by Joyce in \cite{Joyce3}, no explicit metrics over them
are known. Although this problem complicate the analysis of the
effective theory in $d=4$ all the physical phenomena near of the
singularity is local, that is, independent on the global properties
of the manifold \cite{Witten}. Non compact metrics possessing
conical singularities are therefore suitable for analyzing this kind
of phenomena.

    There also exist seven dimensional spaces that admit conformal Killing
spinors that is, satisfying $D_i \eta\sim \eta$. This spaces are
called weak $G_2$ holonomy spaces although their holonomy is
generically in $SO(7)$. Weak $G_2$ holonomy metrics also preserve
certain amount of supersymmetries and there are known compact
conical examples \cite{Bilal2}. In general weak $G_2$ holonomy
spaces are internal spaces with the presence of fluxes and can be
constructed as the conical part of cohomogeneity one $Spin(7)$
spaces.

    In this section we present certain features about $G_2$, weak
$G_2$ and $Spin(7)$ holonomy metrics. In particular we will show
that they are defined by certain self-duality conditions analog to
the four dimensional case but with the octonion constants replacing
the Kronecker symbols. The relation with octonion algebra is
explained in certain detail.

\subsubsection{The group $G_2$ and the octonions}

   In the following we make a brief description about the Lie group $G_2$.
It is known that $G_2 \subset SO(7)$ is one of the exceptional
simple Lie groups and that is simple connected. It's Lie algebra has
$14$ generators $G^{ab}$ and is the subgroup of $SO(7)$ that is
isomorphic to the automorphism group of the octonions $O$
\cite{Schafer}.

     The octonions constitutes the only non associative
division algebra and can be constructed as the doubling of the
quaternions $H$ (the associative ones are $R$, $C$ and $H$). An
arbitrary octonion $x \in O$ can be written as $x=x^0 + x^i e_i$,
where the set $e_i$ is a basis of $7$ unit octonions with the
multiplication rule \be\lb{mult} e_i . e_j = c_{ijk} e_k ;\;\;\; e_i
. 1 = 1 . e_i = e_i \ee and the components $x^i$ are real numbers.
The constants $c_{ijk}$ that define the multiplication table
(\ref{mult}) are totally antisymmetric and defined by \be\lb{const}
c_{123}=c_{246}=c_{435}=c_{367}=c_{651}=c_{572}=c_{714}=1, \ee up to
an index permutation. The constants corresponding to another set of
indices are zero. The conjugated octonion basis is
$\overline{e}_i=-e_i$ and it is clear that $\overline{x}=x^0 - x^i
e_i$.

    It is extremely important to recall that the octonions itself are
not an associative algebra. From (\ref{mult}) and (\ref{const}) it
is seen that $(e^3e^7)e^5 - e^3(e^7e^5) = -e^1$, which clearly shows
non associativity. For this reason the octonions itself cannot be
represented as a matrices with complex components and do not satisfy
the Jacobi identities. In other words, the algebra of $O$ is not a
Lie algebra.

      An automorphism is an isomorphism of an algebra A into itself. Under
an automorphism transformation the multiplication table
$$
x.y=z
$$
is left invariant, in other words
$$
T(x).T(y)=T(z) \;\;\;{\rm where} \;\; T \in {\rm Aut} A
$$
The following characterization of the group $G_2$ as the group of
automorphism of the octonion algebra $O$ is due to Zorn
\cite{Gursey}. Consider three arbitrary elements of $O$, for
instance, $e_1$, $e_2$ and $e_4$, and the transformation $\sigma$
given by
$$
\sigma(e_1)=e_1
$$
\be\lb{sigma} \sigma(e_2)=\cos(\alpha) e_2 + \sin(\alpha) e_3 \ee
$$
\sigma(e_4)=\cos(\beta) e_4 + \sin(\beta) e_7
$$
The images of the other elements are defined by the relations
$$
\sigma(e_2).\sigma(e_4)=\sigma(e_6),\qquad
\sigma(e_1).\sigma(e_2)=\sigma(e_3)
$$
\be\lb{ball} \sigma(e_1).\sigma(e_4)=\sigma(e_7),\qquad
\sigma(e_4).\sigma(e_3)=\sigma(e_5) \ee Explicitly we obtain that
$$
\sigma(e_1)=e_1
$$
$$
\left(%
\begin{array}{c}
  \sigma(e_2) \\
  \sigma(e_3) \\
\end{array}%
\right)=\left(%
\begin{array}{cc}
  \cos(\alpha) & \sin(\alpha) \\
  -\sin(\alpha) & \cos(\alpha) \\
\end{array}%
\right)\left(%
\begin{array}{c}
  e_2 \\
  e_3 \\
\end{array}%
\right)
$$
\be\lb{peto}
\left(%
\begin{array}{c}
  \sigma(e_4) \\
  \sigma(e_7) \\
\end{array}%
\right)=\left(%
\begin{array}{cc}
  \cos(\beta) & \sin(\beta) \\
  -\sin(\beta) & \cos(\beta) \\
\end{array}%
\right)\left(%
\begin{array}{c}
  e_4 \\
  e_7 \\
\end{array}%
\right) \ee
$$
\left(%
\begin{array}{c}
  \sigma(e_6) \\
  \sigma(e_5) \\
\end{array}%
\right)=\left(%
\begin{array}{cc}
  \cos(\alpha+\beta) & \sin(\alpha+\beta) \\
  -\sin(\alpha+\beta) & \cos(\alpha+\beta) \\
\end{array}%
\right)\left(%
\begin{array}{c}
  e_6 \\
  e_5 \\
\end{array}%
\right)
$$
Relation (\ref{peto}) can also be written as \be\lb{peto2}
\left(%
\begin{array}{c}
  \sigma(e_2) + i \sigma(e_3) \\
  \sigma(e_4) + i \sigma(e_7) \\
  \sigma(e_6) + i \sigma(e_5) \\
\end{array}%
\right)=e^{(a\lambda_3 + b\lambda_8)e_1}\left(%
\begin{array}{c}
  e_2 + i e_3 \\
  e_4 + i e_7 \\
  e_6 + i e_5 \\
\end{array}%
\right) \ee where $\lambda_3$ and $\lambda_8$ are the Gell-Mann
traceless matrices \be\lb{gell}
\lambda_3=\left(%
\begin{array}{ccc}
  1 & 0 & 0 \\
  0 & -1 & 0 \\
  0 & 0 & 0 \\
\end{array}%
\right),\qquad
\lambda_8=\left(%
\begin{array}{ccc}
  1 & 0 & 0 \\
  0 & 1 & 0 \\
  0 & 0 & -2 \\
\end{array}%
\right) \ee and $a$ and $b$ are related to the angles by
\be\lb{const} \alpha=a+\frac{b}{\sqrt{3}},\qquad
\beta=-a+\frac{b}{\sqrt{3}}. \ee Relations (\ref{ball}) and
(\ref{sigma}) implies that $\sigma(e_i)$ satisfy the same
multiplication rule than $e_i$. Conversely, every automorphism of
the octonions belong in this manner to at least one Cayley basis
(that is, a basis generated by three different elements). The
automorphism of the form given above are called canonical
automorphism. Each canonical automorphism has a fixed element and
three invariant planes
$$
\Psi(e_1)=\frac{1}{2}\left(%
\begin{array}{c}
  e_2 + i e_3 \\
  e_4 + i e_7 \\
  e_6 + i e_5 \\
\end{array}%
\right)=\frac{(1+i e_1)}{2}\left(%
\begin{array}{c}
  e_6 \\
  e_2 \\
  e_4 \\
\end{array}%
\right)
$$
$$
\Psi(e_2)=\frac{1}{2}\left(%
\begin{array}{c}
  e_4 + i e_6 \\
  e_3 + i e_1 \\
  e_5 + i e_7 \\
\end{array}%
\right)=\frac{(1+i e_2)}{2}\left(%
\begin{array}{c}
  e_4 \\
  e_3 \\
  e_5 \\
\end{array}%
\right)
$$
$$
\Psi(e_3)=\frac{1}{2}\left(%
\begin{array}{c}
  e_5 + i e_4 \\
  e_1 + i e_2 \\
  e_6 + i e_7 \\
\end{array}%
\right)=\frac{(1+i e_3)}{2}\left(%
\begin{array}{c}
  e_5 \\
  e_1 \\
  e_6 \\
\end{array}%
\right)
$$
\be\lb{beatit} \Psi(e_4)=\frac{1}{2}\left(%
\begin{array}{c}
  e_3 + i e_5 \\
  e_6 + i e_2 \\
  e_7 + i e_1 \\
\end{array}%
\right)=\frac{(1+i e_4)}{2}\left(%
\begin{array}{c}
  e_3 \\
  e_6 \\
  e_7 \\
\end{array}%
\right) \ee
$$
\Psi(e_5)=\frac{1}{2}\left(%
\begin{array}{c}
  e_1 + i e_6 \\
  e_4 + i e_3 \\
  e_7 + i e_2 \\
\end{array}%
\right)=\frac{(1 + i e_5)}{2}\left(%
\begin{array}{c}
  e_1 \\
  e_4 \\
  e_7 \\
\end{array}%
\right)
$$
$$
\Psi(e_6)=\frac{1}{2}\left(%
\begin{array}{c}
  e_2 + i e_4 \\
  e_5 + i e_1 \\
  e_7 + i e_3 \\
\end{array}%
\right)=\frac{(1+i e_6)}{2}\left(%
\begin{array}{c}
  e_2 \\
  e_5 \\
  e_7 \\
\end{array}%
\right)
$$
$$
\Psi(e_7)=\frac{1}{2}\left(%
\begin{array}{c}
  e_1 + i e_4 \\
  e_2 + i e_5 \\
  e_3 + i e_6 \\
\end{array}%
\right)=\frac{(1 + i e_7)}{2}\left(%
\begin{array}{c}
  e_1 \\
  e_2 \\
  e_3 \\
\end{array}%
\right)
$$
A canonical automorphism leaving invariant the element $e_A$ is
defined by \be\lb{defino}
\Psi'(e_A)=e^{(a_A\lambda_3 + b_A\lambda_8)e_A}\left(%
\begin{array}{c}
  e_B + i e_C \\
  e_D + i e_E \\
  e_F + i e_g \\
\end{array}%
\right)=\left(%
\begin{array}{c}
  e'_B + i e'_C \\
  e'_D + i e'_E \\
  e'_F + i e'_g \\
\end{array}%
\right) \ee (no Einstein convention is understood). Because we have
seven canonical transformations the group of automorphism of $O$ is
$14$ dimensional.

    Let us find the Lie algebra of the automorphisms. We will take the parameters $a_A$ and $b_A$.
Corresponding to $a_A$ we have
$$
\Psi'(e_A)=e^{a_A\lambda_3 e_A}\left(%
\begin{array}{c}
  e_B + i e_C \\
  e_D + i e_E \\
  e_F + i e_g \\
\end{array}%
\right)=\left(%
\begin{array}{c}
  e'_B + i e'_C \\
  e'_D + i e'_E \\
  e'_F + i e'_g \\
\end{array}%
\right)
$$
which gives
$$
\left(%
\begin{array}{c}
  e'_B \\
  e'_C \\
\end{array}%
\right)=\left(%
\begin{array}{cc}
  \cos(a_A) & \sin(a_A) \\
  -\sin(a_A) & \cos(a_A) \\
\end{array}%
\right)\left(%
\begin{array}{c}
  e_B  \\
  e_C \\
\end{array}%
\right)
$$
$$
\left(%
\begin{array}{c}
  e'_D \\
  e'_E \\
\end{array}%
\right)=\left(%
\begin{array}{cc}
  \cos(a_A) & -\sin(a_A) \\
  \sin(a_A) & \cos(a_A) \\
\end{array}%
\right)\left(%
\begin{array}{c}
  e_D  \\
  e_E \\
\end{array}%
\right)
$$
This mean that it is induced a rotation on the planes $(e_B, e_C)$
and $(e_D, e_E)$ with angles $a_A$ and -$a_A$. Therefore the
generator of this group action is \be\lb{gener1} J_{BC}-J_{DE} \ee
being $J_{BC}$ and $J_{DE}$ the anti-hermitian rotation generators.
In similar manner the generators for the transformation
corresponding to $b_A$ can be obtained, the result is \be\lb{gener2}
\frac{1}{\sqrt{3}}(J_{BC} + J_{DE} - 2J_{FG}). \ee The indices run
from 1 to 7 and therefore we have 14 generators. It is also direct
to check that they close under commutation and therefore they are a
subalgebra of $SO(7)$. This is the $G_2$ subalgebra. If we add
\be\lb{gener3} J_{BC} + J_{DE}+ J_{FG} \ee it can be seen that
(\ref{gener1}), (\ref{gener2}) and (\ref{gener3}) generate $SO(7)$
\cite{Gursey}.

     As we recalled before, the associator
\be\lb{associa} (a,b,c)=(a b)c- a (b c) \ee for three arbitrary
octonions $a$, $b$ and $c$ is not zero and therefore the algebra is
non associative. But (\ref{associa}) is fully antisymmetric and
therefore the octonion algebra is \emph{alternating}. Is therefore
obvious that
$$
(e_i, e_i, e_j)=0
$$
which implies that \be\lb{idento2} c_{acd}c_{bcd}=6\delta_{ab}. \ee
The octonions also obey the Moufang identity
\cite{Schafer}\be\lb{mofa} (x a)(x b)=x (a b) x. \ee which implies
that \be\lb{idento9} c_{paq}c_{qbs}c_{scp} = 3c_{abc}. \ee Let us
define the dual octonion constants \be\lb{psihatdef}
c_{ijkl}=\frac{1}{3!}\epsilon^{abcdefg}c_{efg}, \ee which are
explicitly \be\lb{const2}
c_{4567}=c_{2374}=c_{1357}=c_{1276}=c_{2356}=c_{1245}=c_{1346}=1.
\ee It can be checked the fundamental identity \be\lb{funcon}
c_{abcd}=(\delta_{bc}
\delta_{ad}-\delta_{ac}\delta_{bd})+c_{abe}c_{cde}. \ee The constant
$c_{abc}$ and $c_{abcd}$ can be assembled into a single object
$\psi^{abcd}$, with $a,b,\ldots = 1, \ldots 8$ defined by
\be\lb{identi} \psi_{abc8}=\widetilde{c}_{abc},\;\;\;\;\;\;
\psi_{abcd}=\widetilde{c}_{abcd} \ee Equation (\ref{psihatdef})
translates into the eight-dimensional self-duality of $\psi_{abcd}$,
that is \be\lb{octodual} \psi_{abcd}=\frac{1}{4!}
\epsilon^{abcdefgh}\psi_{efgh}. \ee If we consider the $8\times 8$
matrices $\gamma^a$ given by \be\lb{urra}
(\gamma^a)_{bc}=(\widetilde{c}_{abc}+\delta_{a[b}\delta_{c]8}) \ee
where the constants $\widetilde{c}_{abc}$ are equal to zero if one
of the index is equal to $8$ and to the octonion constants in other
case, then is seen that $\{\gamma^a, \gamma^b\}=4\delta^{ab}I$. This
means that $\gamma^a$ are a representation of the Dirac algebra in
seven dimensions and therefore \be\lb{repre}
(\gamma^{ab})_{cd}=\frac{1}{4}[\gamma^a,
\gamma^b]=(\widetilde{c}_{abcd}+
\widetilde{c}_{ab[c}\delta_{d]8}+\delta_{a[c}\delta_{d]b}) \ee are
also a representation of the Lie algebra of $SO(7)$. Consider the
spinor $\eta$ with components $\eta_a=\delta_{ab}$. Then by use of
the identities (\ref{idento9}) and (\ref{idento2}) it can be proved
that \be\lb{rero} c_{mnp}=i\overline{\eta}\gamma^{mnp}\eta,\qquad
\gamma^{mnp}=\gamma^{[m}\gamma^{n}\gamma^{p]}. \ee By use of Fierz
rearrangement identities in seven dimensions together with
(\ref{rero}) it is possible to show many identities relating
$c_{abc}$ and $c_{abcd}$ \cite{Nicolai}, \cite{Bilal}, they are the
following \be\lb{ident1} c_{abe}c_{cde} =
-c_{abcd}+\delta_{ac}\delta_{bd} - \delta_{ad}\delta_{bc} \ee
\be\lb{ident2} c_{acd}c_{bcd}=6 \delta_{ab} \ee \be\lb{ident3}
c_{abp}c_{pcde} = 3 c_{a[cd}\delta_{e]b} - 3c_{b[cd}c_{e]a} \ee
\be\lb{ident4} c_{abcp}c_{defp}= -3c_{ab[de}\delta_{f]c}
-2c_{def[a}\delta_{b]c} -3c_{ab[d}c_{ef]c} +
6\delta^{[d}_a\delta^e_b\delta^{f]}_c \ee \be\lb{ident5}
c_{abpq}c_{pqc} = -4c_{abc} \ee \be\lb{ident6} c_{abpq}c_{pqcd} =
-2c_{abcd} + 4(\delta_{ac}\delta_{bd} - \delta_{ad}\delta_{bc}) \ee
\be\lb{ident7} c_{apqr}c_{bpqr} = 24\delta_{ab} \ee
$$
c_{abp}c_{pcq}c_{qde} = c_{abd}\delta_{ce} - c_{abe}\delta_{cd} -
c_{ade}\delta_{bc} + c_{bde}\delta_{ac}
$$
\be\lb{ident8} - c_{acd}\delta^{be} + c_{ace}\delta_{bd} +
c_{bcd}\delta_{ae} - c_{bce}\delta_{ad} \ee \be\lb{ident9}
c_{paq}c_{qbs}c_{scp} = 3c_{abc} \ee Formulas
(\ref{ident1})-(\ref{ident9}) can also be derived from the basic
identity for the product of two $\psi$ \cite{Floratos}
\be\label{basicPsiid} \psi_{abcd}\psi^{efgd} = 6 \delta^{[e}_a
\delta^{f}_b \delta^{g]}_c -9 \psi_{[ab}{}^{[e f} \delta^{g]}_{c]}.
\ee Consider the operators \be\lb{defproj} (P_{14})_{ab}^{cd} =
\frac{2}{3}(\delta_{ab}^{cd} +\frac{1}{4}
c_{ab}^{cd}),\;\;\;\;\;\;\;\; (P_{7})_{ab}^{\ \ cd}
=\frac{1}{3}(\delta_{ab}^{cd} - \frac{1}{2}c_{ab}\,^{cd}) , \ee with
$\delta_{ab}^{cd} = \frac{1}{2}(\delta_a^c\delta_b^d -
\delta_a^d\delta_b^c)$. It is straightforward to see from
(\ref{ident3}) that $(P_{14})_{ab}^{cd}c_{cde} = 0$ and
$(P_{7})_{ab}^{cd}c_{cde} = c_{abe}$. Also from (\ref{ident6}) we
obtain
$$
(P_{14})_{ef}^{ab} (P_{7})_{ab}^{cd} S_{cd}=0,
$$
$$
(P_{14})_{ef}^{ab} (P_{7})_{ab}^{cd} S_{cd}=0,
$$
$$
[(P_{14})_{ab}^{cd}+(P_{7})_{ab}^{cd}] S_{cd} = S_{cd}
$$
and therefore (\ref{defproj}) are orthogonal projectors. This means
that for any tensor $S^{ab}$ that transforms in the adjoint $21$
representation of $SO(7)$ we have the decomposition \be\lb{sdua}
S_+^{ab} = \frac{2}{3}(S^{ab} + \frac{1}{4}c^{abcd}S^{cd}),
\;\;\;\;\;\;\; c^{abc}S_+^{bc}=0, \ee \be\lb{asdua} S_-^{ab} =
\frac{1}{3}(S^{ab} - \frac{1}{2}c^{abcd}S^{cd})=c^{abc}\zeta^c,
\;\;\;\;\;\;\; \zeta^c = \frac{1}{6}c^{cab}S^{ab}, \ee
\be\lb{sumita} S^{ab} = S_+^{ab}\oplus S_-^{ab}. \ee This is
analogous to the fact in $D=4$ that any tensor transforming under
the adjoint representation $SO(4)\sim SU(2)_L\times SU(2)_R$ is the
direct sum of its self-dual and anti-self-dual parts. We will say
that $S_{+}^{ab}$ and $S_{-}^{ab}$ are the self-dual and
antiself-dual parts of $S^{ab}$. From the facts stated above we see
that \be\lb{laaut} S_+^{ab}=\frac{1}{2}c_{abcd}S_+^{cd} \ee
\be\lb{laaut2} S_-^{ab}= -\frac{1}{4}c_{abcd}S_-^{cd} \ee that are
analogous to the well known self-duality condition $F^{ab}=\pm
\epsilon^{abcd} F^{cd}$ in $D=4$ but with the octonion constants
replacing the Kronecker symbols. By orthogonality of $P_{14}$ and
$P_{7}$ also holds the relation \be\lb{comple} S^{ab}U^{ab}=S_+^{ab}
U_+^{ab} + S_-^{ab} U_-^{ab}. \ee being $U^{ab}$ another $SO(7)$
tensor. The projectors (\ref{defproj}) induce an isomorphism of the
$SO(7)$ algebra given by \be\lb{G2} G^{ab}=\frac{2}{3}(\gamma^{ab} +
\frac{1}{4}c_{abcd}\gamma^{cd}), \ee \be\lb{comp}
C^{ab}=\frac{1}{3}(\gamma^{ab} - \frac{1}{2}c_{abcd}\gamma^{cd}).
\ee From (\ref{const}) and (\ref{const2}) we can see that there are
$14$ generators $G^{ab}$ and $7$ generators $C^{ab}$. $G^{ab}$ are a
representation of the Lie algebra of $G_2$.

     The group $G_2$ is the subgroup of $SO(7)$ has a one
dimensional invariant subspace generated by certain vector $\eta$.
This mean that $A \eta= \eta$ being $A$ an element of $G_2$. This
implies that $G^{ab}\eta=0$, being $G_{ab}$ the infinitesimal
generators of the group $G_2$. For the representation (\ref{G2})
this subspace is generated by an spinor with components
$\eta_a=\delta_{a8}$. This is the decomposition $8 \rightarrow 7 +
1$ of $SO(7)$. The form of $\eta$ depends on the representation, but
its existence of course not. $G_2$ is also the group which leaves
invariant the three-form \be\lb{3form} \Phi=\frac{1}{3!}c_{abc}e^a
\wedge e^b \wedge e^c \ee and the 4-form \be\lb{4form} \ast
\Phi=\frac{1}{3!}c_{abcd}e^a \wedge e^b \wedge e^c \wedge e^d \ee
being $e_1,..,e_7$ a basis for a seven dimensional space V. The
closure of (\ref{3form}) and (\ref{4form}) is one of the main facts
of $G_2$ holonomy manifolds, as we will see in the following.

\subsubsection{$G_2$ holonomy and self-duality}

     In the previous subsection we have defined the Lie group $G_2$ as
the subgroup of $SO(7)$ that has one and only one stable spinor
$\eta$, also as the automorphism group of the octonions, and as the
group that leave the 3-form (\ref{3form}) invariant. We have also
seen that the $G_2$ component of a tensor transforming under the
adjoint representation $21$ of $SO(7)$ is self-dual (\ref{sdua}). In
this section we will show that the analog holds for an space $M$
with holonomy $G_2$ that is, the presence of one Killing spinor
$\eta$ globally defined on it, or that there is defined a three-form
that is closed and co-closed over it. Moreover for a $G_2$ holonomy
space it exist a rotation of the frame which take to zero the
anti-self-dual part of the spin connection. Therefore the curvature
of a $G_2$ holonomy metric is self-dual.

      It is evident that the holonomy of an arbitrary seven
dimensional manifold $M$ is always included in $SO(7)$. From the
discussion of the previous subsection it follows that is reduced to
$G_2$ if and only if there exist only one Killing spinor $\eta$,
that is, an spinor globally defined over $M$ satisfying the
condition \be\lb{holG} D_{i}\eta=0, \ee being $D_{i}=I \partial_{i}
+ \omega^a_{i b} \gamma^{ab}$ the covariant derivative in spinor
representation. In other words the existence of a globally covariant
spinor allows the reduction $8 \rightarrow 7 + 1$ to $7$ of the
holonomy group.

    If we choose a the representation of $SO(7)$
given by (\ref{G2}) and (\ref{comp}), and the spinor $\eta$ such
that $\eta_a=\delta_{ab}$ it follows that
$(\gamma^{ab}\eta)_{c}=c_{abc}$. Then
$$
D_{i}\eta=-\frac{1}{4}\omega_{i ab}\gamma^{ab}\eta=
-\frac{1}{4}\omega_{i ab}c_{abc}=0.
$$
It can be checked by using (\ref{const}) and (\ref{const2}) that the
last condition is equivalent to \be\lb{7duality}
\omega^{a}_{b}=\frac{1}{2}c_{abcd}\omega^c_{d}. \ee A 7-dimensional
space for which $\omega^{a}_{b}$ satisfies (\ref{7duality}) is
called self-dual. We conclude that self-dual spaces in $D=7$ have
$G_2$ holonomy. It can be proved that this result is independent of
the representation of $SO(7)$ and that the converse is also true,
that is, for $G_2$ holonomy there exists always a rotation of the
frame for which the anti self-dual part $\omega_{-}$ goes locally to
zero \cite{Bilal}.

     $G_2$ manifolds are always Ricci-flat. In fact, if we apply $D_j$ to
(\ref{holG}) and anti-symmetrize we will obtain
$$
R^{a}_{bcd}\gamma^{ab}\eta=0
$$
where $R^a_b=R^{a}_{bcd}e^c\wedge e^d$ is the curvature tensor. Now
if we multiply by $\gamma^i$ this expression and using that
$R_{abcd}=R_{cdab}=-R_{abdc}$ and that $R^a_{[ebc]}=0$ we find that
$$
R_{ij}\gamma^j\eta=0 \;\;\;\Leftrightarrow\;\;\;R_{ij}\gamma^j=0
$$
But the matrices $\gamma^i$ are linearly independent and therefore
$R_{ij}=0$. Indeed, in any dimension, the presence of a covariantly
constant spinor implies Ricci-flatness by the reasoning given above.

     It is well known that in $D=4$ the self-duality of the spin connection
implies the self-duality of the curvature tensor. This is also true
in seven dimensions. In \cite{Bilal} a clear proof was given. It is
based in the following identity for the octonion constants
\be\lb{1id}
c_{abcp}c_{defp}=-3c_{ab[de}\delta_{e]b}-2c_{def[a}\delta_{b]c}+
6\delta^{[d}_{a}\delta^{e}_{b}\delta^{f]}_{c}-2c_{def[a}\delta_{b]c}.
\ee If (\ref{7duality}) holds it is clear that $d\omega^{a}_{b}$
will be self-dual. The self-duality of
$\omega^{a}_{c}\wedge\omega^{c}_{b}$ follows using that
$$
\frac{1}{2}c_{abcd}\omega^{c}_{e}\wedge\omega^{e}_{d}=
\frac{1}{4}c_{abcd}c_{edfg}\omega^{c}_{e}\wedge\omega^{f}_{g}
$$
\be\lb{prueba}
=-\frac{1}{2}c_{abfe}\omega^{f}_{c}\wedge\omega^{c}_{e}
+\frac{1}{4}c_{aefg}\omega^{b}_{e}\wedge\omega^{f}_{g}
-\frac{1}{4}c_{befg}\omega^{a}_{e}\wedge\omega^{f}_{g}
+\omega^{a}_{c}\wedge\omega^{c}_{b} \ee
$$
=-\frac{1}{2}c_{abcd}\omega^{c}_{e}\wedge\omega^{e}_{d}
+2\omega^{a}_{c}\wedge\omega^{c}_{b}\;.
$$
The identity (\ref{1id}) has been used here. From (\ref{prueba}) is
seen that
$$
\frac{1}{2}c_{abcd}\omega^{c}_{e}\wedge\omega^{e}_{d}=
\omega^{a}_{c}\wedge\omega^{c}_{b}
$$
which implies the self-duality of $R=d\omega+\omega\wedge\omega$.

   From (\ref{7duality}), the antisymmetry of (\ref{const2}) and the
Bianchi identity $R_{d[ebc]}=0$ it follows that
 \be\lb{flat}
R_{ab}=R_{acbc}=\frac{1}{2}c_{acde}R_{debc}=\frac{1}{2}c_{acde}R_{d[ebc]}=0.
\ee This is another proof that $G_2$ holonomy spaces are Ricci-flat.
For such case the Weyl tensor $W_{abcd}$ defined by \be\lb{weyl}
W_{abcd}=R_{abcd}+\frac{R}{(n-1)(n-2)}(g_{ac}g_{bd}-g_{ad}g_{bc})
-\frac{1}{(n-2)}(g_{ac}R_{bd}-g_{ad}R_{bc}-g_{bc}R_{ad}+g_{bd}R_{ac})
\ee will be equal to the Riemann tensor. The first one is traceless,
in consequence Ricci-flat manifolds have traceless curvature tensor.

    The condition to have holonomy $G_2$ is also equivalent to the existence
of a closed and co-closed three form \cite{Bryant}. The natural
candidates to consider are the $G_2$ equivariant $3$-form $\Phi$
(\ref{3form}) and its dual $\ast\Phi$ \cite{Bilal} \be\lb{24form}
\Phi=\frac{1}{3!}c_{abc}e^a\wedge e^b\wedge e^c \;\;\;\;
\ast\Phi=\frac{1}{4!}c_{abcd}e^a\wedge e^b\wedge e^c\wedge e^d. \ee
Taking into account the identity
$$
c_{abp}c_{pcde}=3c_{a[cd}\delta_{e]b}-2c_{b[cd}\delta_{e]a}
$$
it is obtained
$$
d\Phi=-\frac{1}{3!2}c_{abc}e^a\wedge e^b\wedge\omega^{cd}\wedge e^d=
-\frac{1}{3!}c_{abc}c_{cdef}e^a\wedge e^b\wedge \omega^{ef}\wedge
e^d
$$
$$
=\frac{1}{3!}c_{ade}e^a\wedge e^d \wedge \omega_{eb} \wedge
e^b=-2d\Phi\;.
$$
From here follows \be\lb{clauphi} d\Phi=0\;. \ee So, $\Phi$ is a
closed form. Similarly, using (\ref{1id}) follows that
$$
d\ast\Phi=-\frac{1}{4!6}c_{abcd}\omega_{ae}\wedge e^e \wedge e^b
\wedge e^c\wedge e^d
=-\frac{1}{4!12}c_{aefg}c_{abcd}\omega^{fg}\wedge e^e \wedge e^b
\wedge e^c \wedge e^d
$$
$$
=\frac{1}{4!3}c_{fbcd}\omega^{fe}\wedge e^e \wedge e^b\wedge
e^c\wedge e^d= -2d\ast\Phi\;,
$$
which implies the closure of $\ast\Phi$. This proves that $G_2$
holonomy implies the closure of $\Phi$ and $\ast\Phi$. But in
\cite{Bryant} it has been proved that the closure and co-closure of
$\Phi$ and the constance of $\eta$ give the same number of
independent equations and therefore, both conditions are indeed
equivalent.

\subsubsection{The Bryant-Salamon construction}

      We present in this subsection an extension of a four
dimensional quaternionic Kahler space to a seven dimensional one
with holonomy lying in $G_2$. This construction was performed by
Bryant and Salamon in \cite{Salamon} and considered recently in
\cite{Mahapatrah}. In this section we intend to present such
construction not in the original form \cite{Salamon}, but in a way
more clear for physicists.

   Consider a 7-manifold $M=R^3 \times M^4$
with metric \be\lb{anzal} g=(du^1)^2+(du^2)^2+(du^3)^2 + g_h \ee
being $g_h$ 4-dimensional metric doesn't depending of the tree
coordinates $u_i$. Condition (\ref{7duality}) reduce for
(\ref{anzal}) to
$$
\omega^{i}_{j}=\frac{1}{2}\epsilon_{ijkl}\omega^k_{l}
$$
being $\omega^{i}_{j}$ the spin connection of $g_h$. Therefore if
$g_h$ has self-dual connection or at least there exists locally a
frame in which the anti-self-dual part of the connection goes to
zero (and therefore $g_h$ is hyperkahler) then $g=g_{R^3}\oplus
g_{h}$ has holonomy included in $SU(2)\subset G_2$. We can
generalize the anzatz (\ref{anzal}) and consider instead
\be\lb{anzo} g=e^{2g}[(D_{-}u^1)^2+(D_{-}u^2)^2+(D_{-}u^3)^2]
+e^{2f}g_q , \ee where $g_q$ is a quaternionic Kahler space in four
dimension and we had replaced the usual differentials $du^i$ by the
covariant derivative \be\lb{covo}
D_{-}u^i=du^i+\epsilon^{ijk}\omega_{-}^j u^k. \ee As usual
$\omega_{-}^i$ is the anti-self-dual part of the spin connection.
The metric (\ref{anzal}) corresponds to the hyperkahler limit (in
which locally we can select $\omega_{-}^i=0$) with $f$ and $g$
reduced to constants. We select $f$ and $g$ as functions on the
radius $|u|=\sqrt{u_i^2}$. Then condition (\ref{7duality}) for
(\ref{anzo}) is a constraint to be satisfied by $f$ and $g$ the
resulting metric will have holonomy $\Gamma \subset G_2$.

    The explicit calculation of the connection $\omega$ gives
$$
\omega^{ij} = {f' \over |u|} \,D_{-} u^{[i} u^{j]} - \epsilon^{ijk}
    \omega_{-}^k \
$$
\be\lb{spinconn} \omega^{mn} =\omega^{mn} - {\kappa} e^{2(f-g)}
    \epsilon^{ijk} u^k J^j_{mn} D_{-}u^i \
\ee
$$
\omega^{mi} = {g' \over |u|} \, e^{g-f} \, e^m u^i - {\kappa}
    e^{f -g} \, \epsilon^{ijk} \, u^k J^j_{mn} e^n
$$
where $(...)'$ denotes the derivative with respect to $|u|$. The
self-duality condition (\ref{7duality}) gives the equations
$$
0 = c_{iab}\omega^{ab} = c_{ijk} \omega^{jk}
    + c_{imn} \omega^{mn} = \Big[{f' \over |u|} +
    \kappa\, e^{2(f-g)}\Big]
\epsilon^{ijk} \, D_{-}u^j u^k
$$
\be\lb{bronski} 0 = c_{mab}\omega^{ab} = c_{mni} \omega^{ni}
    = J^i_{mn} \omega^{ni} =
    \Big[ {g' \over |u|}  - \kappa \, e^{2(f-g)}\Big]
    e^{g-f}\, J^i_{mn} e^n \, u^i.
\ee Equations (\ref{bronski}) are invariant under $f \rightarrow f +
\lambda$, $g \rightarrow g + \lambda$ which give a constant
conformal scaling of the metric (\ref{anzatz}). Also we have the
second symmetry $f \rightarrow f - \lambda$, $u \rightarrow
e^{\lambda} \, u$ which leaves the metric invariant. Using these
symmetries the solution result \be\lb{solucion} e^{-4f} = 2 \kappa
|u|^2 + c \qquad , \qquad e^{4g} = 2 \kappa |u|^2 + c. \ee By
introducing (\ref{solucion}) into (\ref{anzatz}) we obtain the
Bryant-Salamon family of $G_2$ metrics \be\lb{anzatz}
g=\frac{1}{\sqrt{2\kappa |u|^2+c}}
(du^i+\epsilon^{ijk}\omega^j_{-}u^k)^2+\sqrt{2\kappa |u|^2+c}\; g_q.
\ee Clearly in the limit $\omega^j_{-}=0$ and $\kappa=0$ we recover
(\ref{anzal}).

    The Bryant-Salamon metrics (\ref{anzatz}) are non-compact, because the
coordinates $u_i$ can take any real value, and this represent a
problem when consider applications to compactification of
$M$-theory. Nevertheless they have the important property to be
asymptotically conical. To see that they are conical let us write
(\ref{anzatz}) introducing spherical coordinates
$$
u_1=|u|\cos(\theta),
$$
$$
u_2=|u|\sin(\theta)\cos(\varphi),
$$
$$
u_3=|u|\sin(\theta)\sin(\varphi),
$$
and defining the new radial coordinate
$$
r^2=\sqrt{2\kappa |u|^2+c}.
$$
Then it is obtained \be\lb{polar}
g=\frac{dr^2}{(1-\frac{4c}{r^4})}+\frac{r^2}{4\kappa}(1-\frac{4c}{r^4})g_{ab}
(dx^a+\xi^a_i \omega_{-}^i)(dx^b + \xi^b_j
\omega_{-}^j)+\frac{r^2}{2}g_q, \ee where $g_{ab}$ and $\xi$ are the
metric and the killing vectors of $S^2$ respectively. The
corresponding 7-bein is explicitly
$$
e_i=\sqrt{\frac{\kappa}{2}} r V_i
$$
$$
e_5=\frac{r}{2}(d\theta-\sin(\varphi)\omega_{-}^2+\cos(\varphi)\omega_{-}^3)
$$
$$
e_6=\frac{r}{2}(\sin(\theta)d\varphi+\sin(\theta)\omega_{-}^1
-\cos(\theta)\cos(\varphi)\omega_{-}^2-\cos(\theta)\sin(\varphi)\omega_{-}^3)
$$
$$
e_7=\frac{dr}{(1-\frac{4c}{r^4})^{1/2}}
$$
being $V_i$ the einbeins of the quaternion base $g_q$. If we
consider $r^2>>c$ then (\ref{polar}) is \be\lb{cono} g \approx dr^2
+ r^2\Omega, \ee being $\Omega$ the metric of a $6$ manifold
independent of the coordinate $r$. Thus the family (\ref{anzatz}) is
conical. The explicit form of $\Omega$ is \be\lb{weaksu} \Omega =
g_{ab} (dx^a+\xi^a_i \omega_{-}^i)(dx^b+\xi^b_j \omega_{-}^j)+ g_q.
\ee The metric (\ref{weaksu}) is of weak $SU(3)$ holonomy, we will
return to this point below.

      It is instructive to check also the existence of a closed and
co-closed 3-form. It is seen by (\ref{3form}) that \be\lb{laforma}
\Phi = \frac{1}{6} (1+|u|^2)^{-3/4}\, \epsilon_{ijk}\, D_{-}u_i
\wedge D_{-}u_j\wedge D_{-}u_k + 2(1+|u|^2)^{1/4}\, D_{-}u_i \wedge
J^i\, \ee and the dual $\ast \Phi$ is \be\lb{ladual} \ast\Phi=
4(1+|u|^2)\, \Theta + \epsilon_{ijk}\, D_{-}u_i\wedge D_{-}u_j\wedge
J^k. \ee The 4-form $\Theta$ is defined in (\ref{lafunda}) and in
four dimensions is equal to the volume form of $M$
$$
\Theta= \frac{1}{2} \overline{J}^1\wedge \overline{J}^1 =
\frac{1}{2}\overline{J}^2\wedge \overline{J}^2 =
\frac{1}{2}\overline{J}^3\wedge \overline{J}^3.
$$
The covariant derivative $D_{-}$ is given by (\ref{covo}) and it
follows that $(D_{-})^2 u_i = \epsilon_{ijk} F^j u_k$. For a
quaternionic base $d\Theta=0$ and from (\ref{basta}) it follows that
$D_{-}\overline{J}^a=0$. It can be shown by use of this relations
that $\Phi$ is closed and co-closed.

\subsubsection{Weak $G_2$ and Spin(7) holonomy spaces}

      As it follows from the previous sections
$G_2$ holonomy manifolds admits a globally defined Killing spinor
$\eta$. In this subsection we study eight dimensional spaces with
the same property. Also we study seven dimensional ones that admits
a conformal Killing spinor, that is, an spinor satisfying $D_j \eta
\sim \eta$. For the former the holonomy is included in $Spin(7)$.
The second have generically $SO(7)$ holonomy but are defined by
certain conditions that are analog of the $G_2$ case. For this
reason we call such them "weak $G_2$ holonomy manifolds"
\cite{Gray}.

    In seven dimension the presence of conformally Killing spinor $\eta$
namely \be\lb{confspin} D_{i} \eta = i\lambda \gamma_{i} \eta, \ee
is a generalization of $G_2$ holonomy condition (\ref{holG}), which
is obtained as the limit $\lambda \rightarrow 0$ of
(\ref{confspin}). If we act on (\ref{confspin}) by $D_{i}$ and
antisymmetrize it is obtained
$$
R^{ab}_{ij} \gamma_{ab} \eta= 8\lambda^2 \gamma_{ij} \eta.
$$
Now if we multiply the last expression by $\gamma^{k}$ and using
that $R_{abcd}=R_{cdab}=-R_{abdc}$ and that $R^a_{[ebc]}=0$ we get
$$
R_{kj}\gamma^{j}\eta=4(d-1) \widetilde{\lambda}^2\gamma_{k}\eta
$$
which implies \be\lb{g2einst} R_{ij}=4(d-1) \lambda^2 \delta_{ij}.
\ee Therefore if there exist a globally defined $\eta$ satisfying
(\ref{confspin}) the metric is Einstein $R_{ij}\sim g_{ij}$ but with
cosmological constant different than zero. Spaces satisfying
condition (\ref{confspin}) have holonomy that lies in $SO(7)$ and
just in the limit $\lambda\rightarrow 0$ we can state that the
holonomy is reduced to a subgroup of $G_2$. For $\lambda\neq 0$ the
space are weak $G_2$ holonomy ones \cite{Gray}, \cite{Bilal}.

    As before, let us select the spinor $\eta_i=\delta_{i8}$, then
$$
D_j \eta = i\lambda \gamma_j \eta
$$
together with (\ref{urra}) and (\ref{repre}) implies that
\be\lb{wg21} c_{abc}\omega^{ab}_j=-\frac{\lambda}{2}
\delta_{jc}\qquad \Longleftrightarrow \qquad c_{abc}\omega^{bc} =
-\frac{\lambda}{12} e^a. \ee The relation (\ref{wg21}) is equivalent
to \be\lb{wg23} \omega_{-}^{ab}= -\frac{\lambda}{12}\, c_{abc} e^c
\qquad \Longleftrightarrow \qquad \omega^{ab} =\frac{1}{2}c_{abcd}\,
\omega^{cd} - \frac{\lambda}{4}c_{abc} e^c. \ee Condition
(\ref{wg23}) is the weak self-duality condition in seven dimensions.
If for a manifold there exists a frame for which (\ref{wg23}) hold
then it will be weak $G_2$ holonomy type.

   The form $\Phi$ is defined as before in (\ref{24form}) and
using (\ref{wg21})-(\ref{wg23}) we obtain
$$
d\Phi = -\frac{1}{2}c_{abc} e^a\wedge e^b \wedge \omega^{cd}\wedge
e^d = -\frac{1}{4} c_{abc}c_{cdef} e^a \wedge e^b \wedge \omega^{ef}
\wedge e^d
$$
\be\lb{boom1}
 +\frac{\lambda}{8} c_{abc}c_{cde} e^a \wedge e^b
\wedge e^e \wedge e^d =-2 d\Phi + 3\lambda *\Phi. \ee where we used
(\ref{ident1}) in the last step. So we conclude that \be\lb{wg24}
d\Phi =\lambda \ast \Phi, \ee and the last condition implies that
$d\ast\Phi=0$. As expected, if the cosmological constant is zero
(\ref{wg24}) reduce to the closure and co-closure of $\Phi$ and the
holonomy is again in $G_2$.

     Weak $G_2$ metrics can be constructed as the angular part
of certain conical $Spin(7)$ holonomy spaces, that we will present
in the following. This is analogous to the fact that weak $SU(3)$
manifolds are the angular part of certain conical $G_2$ holonomy
manifolds. $Spin(7)$ spaces have a globally defined Killing spinor
$\eta$ as their seven dimensional Ricci flat counterpart.

     $Spin(7)$ is the subgroup of
$SO(8)$ preserving the 4-form \be\lb{4formil}
\hat{\Phi}=\psi_{abcd}e^a \wedge e^b \wedge e^c \wedge e^d \ee being
$\psi_{abcd}$ the self-dual constants associated to octonions
defined in (\ref{identi}). Because we are in the middle dimension,
we see that $\ast \Phi=\Phi$. Intuition could suggest that as in the
$G_2$ case, the holonomy will be reduced from $SO(8)$ to $Spin(7)$
if $d\hat{\Phi}=0$.

      Let us consider the $SO(8)$ $\gamma$-matrices
$\Gamma_a=(\Gamma_a,\Gamma_8)$, with $a=1,...,8$ \be
\Gamma_a=\left(\matrix{0&i\gamma_a\cr -i\gamma_a&0}\right)\, ,
~~~~~~ \Gamma_8= \left(\matrix{0&1\cr 1&0}\right)\, , ~~~~~~~
\{\Gamma_a,\Gamma_b\}=2\delta_{ab}\, , \ee that correspond to the
standard embedding of $SO(7)_v$ in $SO(8)$. The $SO(8)$ generators
$\Gamma^{ab}=\frac{1}{2}[\Gamma^a,\Gamma^b]$ satisfy the relations
\be \Gamma^{ab}=c^{abc}\Gamma^9\Gamma_{8c}\, ,~~~
c_{abcd}\Gamma^{ab}=-(4+2\Gamma^9)\Gamma_{cd}\, ,
 \label{8c}
\ee where $\Gamma^9=\left(\matrix{1&0\cr 0&-1}\right)$ is the
chirality matrix. Then it can be shown that the right handed part of
an spinor $\eta$, namely $\eta_+=(1-\Gamma^9)\eta$ is invariant
under the action of \be\lb{nou}
G^{ab}=\frac{3}{8}\left(\Gamma^{ab}+\frac{1}{6}{\psi^{ab}}_{cd}
\Gamma^{cd}\right)\, ,  \ee that is
 \be
G^{ab}\eta_+=0
%~~~~ \eta_\a=\delta_{8\a}
\, . \label{Gh} \ee We can define as before the associated
projectors \be\lb{defproj2} (P_{21})_{ab}^{cd} =
\frac{3}{8}(\delta_{ab}^{cd} +\frac{1}{6}
\psi_{ab}^{cd}),\;\;\;\;\;\;\;\; (P_{7})_{ab}^{\ \ cd}
=\frac{5}{8}(\delta_{ab}^{cd} - \frac{1}{10}\psi_{ab}\,^{cd}) , \ee
corresponding to the embedding of $Spin(7) \subset SO(8)$. Let us
call as before $A_+^{ab}$ and $A_{-}^{ab}$ the part annihilated by
$P_7$, or $P_{21}$ respectively of an $SO(8)$ tensor $A^{ab}$. As in
the seven dimensional case the $Spin(7)$ part $A_{+}^{ab}$ satisfies
a self duality condition related to the constants $\psi_{abcd}$
namely \be\lb{8duality}
A_{+}^{ab}=\frac{1}{2}\hat{c}_{abcd}A_{+}^{cd}. \ee Moreover, there
is a Killing spinor $\eta$ for which $A_{+}^{ab}\eta=0$.

    From the previous paragraph it follows that a
manifold $M$ has $Spin(7)$ holonomy if and only if there is a global
Killing spinor $\eta$ for which $D_i \eta=0$ or equivalently, if its
connection $\omega^{a}_{b}$ satisfies \be\lb{nueva8}
\omega_{b}^{a}=\frac{1}{2}\hat{c}_{abcd}\omega_{d}^{c}. \ee The last
condition is equivalent to the closure of $\hat{\Phi}$, namely
\be\lb{nuevaro} d\hat{\Phi}=0. \ee We omit the proof because they
are similar to those presented in the $G_2$ case. $Spin(7)$
manifolds are of course Ricci-flat as far as they admit a Killing
spinor $\eta$. If we select one direction, say $8$, we obtain from
(\ref{nueva8}) that \be\lb{seldir}
\omega^8_r=-\frac{1}{2}c_{rpq}\omega^{p}_{q}. \ee The rest of
(\ref{nueva8}), namely
$$
{\omega^p}_q=\frac{1}{2}c^p_{qrs}\omega^{r}_{s}-
c^p_{qr}\omega^{r8}\,
$$
are automatically satisfied if eq.(\ref{seldir}) holds.

     There exists a Bryant-Salamon extension of a quaternionic Kahler space
to one with $Spin(7)$ holonomy. It is possible to construct such
spaces in a way similar to the presented for the Swann spaces
(\ref{abuela}). This is because it should be imposed $d\hat{\Phi}=0$
for (\ref{4formil}), which is the analog of the Swann 4-form
(\ref{lafundo}) for the $Spin(7)$ case. We start with the anzatz
(\ref{Swann}) for the eight dimensional metric, namely
$$
g_s = g \overline{g} + f |du + u \omega_{-}|^2.
$$
The quaternion $u$ and the anti-self-dual $Im H$ valued 1-form
$\omega_{-}$ are
$$
u = u_0 + u_1 I + u_2 J + u_3 K ,\;\;\;\;\;\; \overline{u}= u_0 -
u_1 I - u_2 J - u_3 K
$$
$$
\omega_{-}=\omega_{-}^1 I+\omega_{-}^2 J +\omega_{-}^3 K
$$
$$
|u|^2=x\overline{x}=|u|^2=(u_0)^2+ (u_1)^2+ (u_2)^2+ (u_3)^2,
$$
being $I, J, K$ the unit quaternions. The functions $f$ and $g$
depends only on the radius $|u|$. Condition $d\hat{\Phi}=0$ gives a
differential equation for $f$ and $g$. The explicit expression of
$\Phi$ is obtained from (\ref{4formil}) and is
$$
\hat{\Phi}= 3 f g [\alpha\wedge \overline{\alpha}\wedge
\overline{e}^t \wedge e + \overline{e}^t \wedge e \wedge \alpha
\wedge \overline{\alpha}]
$$
\be\lb{expresate} + g^2  \overline{e}^t \wedge e
\wedge\overline{e}^t\wedge e + f^2 \alpha \wedge \overline{\alpha}
\wedge \alpha \wedge \overline{\alpha} \ee where $\alpha = du + u
\omega_{-}$. Then is straightforward by use of
(\ref{ainou1})-(\ref{ainou4}) to impose $d\Phi=0$ and find the
system defining $f$ and $g$. The final solution is
$$
f=\frac{1}{(2\kappa |u|^2+c)^{2/5}},
$$
$$
g=(2\kappa |u|^2+c)^{3/5}.
$$
with the corresponding metric \be\lb{sou} g_s=(2\kappa
|u|^2+c)^{3/5}\overline{g} + \frac{1}{(2\kappa
|u|^2+c)^{2/5}}|\alpha|^2. \ee Spaces defined by (\ref{sou}) are the
Bryant-Salamon $Spin(7)$ ones.

    Metrics (\ref{sou}) are non compact (because $|u|$ is not bounded),
and asymptotically conical. Introducing spherical coordinates for
$u$ gives
$$
u_1=|u| \cos(\theta),
$$
$$
u_2=|u| \sin(\theta)\sin(\varphi),
$$
$$
u_3=|u| \sin(\theta)\cos(\varphi)\sin(\phi),
$$
$$
u_4=|u| \sin(\theta)\cos(\varphi)\cos(\phi),
$$
and defining the radial variable
$$
r^2=\frac{9}{20}(2\kappa |u|^2 + c)^{3/5}
$$
we obtain the spherical form of the metric \be \lb{spo} g = {dr^2
\over \kappa (1- {c / r^{10/3}})} + {9 \over 100 \, \kappa} r^2
\left(1 - {c \over r^{10/3}}\right) \left(\sigma^i -
\omega_{-}^i\right)^2 + {9 \over 20} \, r^2 \,g_q \ee being
$\sigma^i$ the left-invariant one-forms on $SU(2)$
$$
\sigma_1=\cos(\varphi)d\theta + \sin(\varphi) \sin(\theta)d\phi
$$
$$
\sigma_2=-\sin(\varphi)d\theta + \cos(\varphi) \sin(\theta)d\phi
$$
$$
\sigma_3=d\varphi + \cos(\theta)d\phi.
$$
If we consider the limit $r>>c$ we find the behavior \be\lb{cono2} g
\approx dr^2 + r^2\Omega, \ee with $\Omega$ a seven dimensional
metric independent of the coordinate $r$, namely \be\lb{brg2}
\Omega= \left(\sigma^i - \omega_{-}^i\right)^2 +  g_q \ee We also
see that the subfamilies of (\ref{spo}) with $c=0$ are exactly
conical and their angular part is (\ref{brg2}). Such subfamilies are
known as cohomogeneity one metrics.

    There is a correspondence between cohomogeneity one $Spin(7)$
spaces and weak $G_2$ holonomy ones, and in particular we can prove
that the cone (\ref{brg2}) is of weak $G_2$ type. From (\ref{cono2})
it is obtained the seven-bein $\tilde e^a$ and the connection
$\tilde \omega^{ab}$ of an eight dimensional cohomogeneity one
space, namely
 \be \lb{noli} \tilde e^8
=dr \,, \qquad\qquad \tilde e^a = -{\lambda\over4}r e^a. \ee being
$e^a$ the einbein of the angular part. The first Cartan structure
$$
d\tilde e^a + \tilde\omega^{ab}\wedge\tilde e^b = 0 $$ implies that
\be\lb{soda} \omega^{ab} = \tilde\omega^{ab} \,, \qquad\qquad
\tilde\omega^{8a} = {\lambda\over4} e^a. \ee From the einbeins
(\ref{noli}) and (\ref{4formil}) we obtain \ba\lb{camal}
\widetilde{\Phi} = \left({\lambda r\over4}\right)^3\, dr\wedge \Phi
+ \left({\lambda r\over4}\right)^4\, *\Phi \,, \cr \label{w20} d
\widetilde{\Phi} = - \left({\lambda r\over4}\right)^3\, dr \wedge
\bigl(d\Phi-\lambda *\Phi\bigr) + \left({\lambda r\over4}\right)^4\,
d *\Phi \,, \ea where $\Phi$ and $*\Phi$ are the usual
seven-dimensional 3- and 4-form constructed from the $e^a$. We see
from (\ref{camal}) that $Spin(7)$ holonomy condition, namely
$d\widetilde{\Phi}=0$ is equivalent to weak $G_2$ holonomy of the
seven-dimensional base space, that is, to the condition
$$
d\Phi-\lambda *\Phi=0,\;\;\;\;\;\;\;\;\;\;d *\Phi=0.
$$
and the converse is also true. Moreover equations (\ref{noli}) are
exactly equivalent to
$$
\tilde\omega^{ab} = {1\over2}\psi_{abcd}\tilde \omega^{cd}.
$$
which is the $Spin(7)$ self-duality condition (\ref{nueva8}). Then
we conclude that there is a one to one correspondence between
co-homogeneity one $Spin(7)$ and weak $G_2$ metrics \cite{Hitcho}.

\subsection{$G_2$ holonomy spaces and M-theory compactifications}

     In this subsection we show that $G_2$ and weak $G_2$ holonomy
manifolds are internal spaces of M-theory that preserve at least one
supersymmetry after Kaluza-Klein reduction to four dimensions. The
bosonic lagrangian of the Sugra theory in $D=11$, that is the low
energy effective action of the M-theory is \be\lb{sugralag} {\cal
L}_{bos}= \frac{1}{2\kappa^2}[ e R -\frac{e}{48}F_{MNPQ}F^{MNPQ}-
\frac{1}{3!4!4!}\epsilon^{m_1... m_{11}} F_{m_1... m_4}F_{m_5...
m_8}A_{m_9m_{10}m_{11}}] \ee being $F_{MNPQ}
=24\,\partial_{[M}A_{NPQ]}$ and $R=R^{AB}_{MN}e^M_A \wedge e^N_B$.
We are interested in the bosonic part of the theory because we want
to compactify it to obtain a four dimensional theory with maximal
space-time symmetry \cite{Duff}. This means that the vacuum should
be invariant under $SO(1,4)$ or $SO(2,3)$ as the cosmological
constant is positive, negative or zero, corresponding to the de
Sitter, Minkowski or anti de Sitter spaces. The first requirement of
maximal symmetry is that the vacuum expectation value of the
Rarita-Schwinger fermion should vanish, therefore we set
$$
<\Psi_M>=0.
$$
The equations of motion in this case reduce to \be\lb{a2}
R_{MN}-\frac{1}{2}g_{MN}R=\frac{1}{3}[F_{MNPQ}F_{N}^{PQR}
-\frac{1}{8}g_{MN}F_{PQRS}F^{PQRS}] \ee \be\lb{a3} \nabla_m
F^{MNPQ}=-\frac{1}{576}\epsilon^{m_1..m_8MNPQ}F_{m_1...m_4}F_{m_5...m_8}
\ee Following the Kaluza-Klein procedure we consider the product
space $M_4\times M_7$ with the eleven dimensional einbein
\be\lb{factorizacion} e_m^a (X^n) = \left(\begin{array}{cc}
e_\mu^m(x^\nu) & 0 \\ 0 & e^a_i(x^j)
\end{array} \right), \ee
the greek coordinates will correspond to $M_4$ and the latin to
$M_7$. This means that the metric is factorized as \be\lb{a4}
g_{\alpha\beta}=g_{\alpha\beta}(x^{\mu}),\;\;\;\;\;g_{ab}=g_{ab}(x^m),
\;\;\;\;\\;\;\;\;\;g_{a\alpha}=0 \ee \be\lb{a5}
F_{\alpha\beta\gamma\delta}=F_{\alpha\beta\gamma\delta}(x^{\mu})
\;\;\;\;\; F_{abcd}=F_{abcd}(x^m)\;\;\;\;\;
F_{abc\alpha}(x^m)=F_{ab\alpha\beta}(x^m)=F_{a\alpha\beta\gamma}(x^m)=0
\ee The $x^{\alpha}$ independence of $F_{abcd}$ and the $x^a$
independence of $F_{\alpha\beta\gamma\delta}$ are a consequence of
the Bianchi identity \be\lb{a6}
\partial_{[M}F_{NPQR]}=0,
\ee together with (\ref{a5}). Let us introduce the following anzatz
\be\lb{factoback} F_{\mu\nu\rho\sigma} = A
\det(e_\mu^m)\epsilon_{\mu\nu\rho\sigma}, \;\;\; F_{ijkl} = B e_i^a
e_j^b e_k^c e_l^d\, c^{abcd}, \ee into the equations (\ref{a2}),
with $A$ and $B$ constants. If we set $B=0$ then we obtain the well
known Freund Rubin anzatz \cite{Freund}, but we will not assume this
for the moment. The Einstein equations are \be\lb{Einsfact}
R_{\mu\nu} = -\frac{1}{3}(A^2+\frac{7}{2} B^2)g_{\mu\nu}\;\;\;
R_{ij} = \frac{1}{6}(A^2+5 B^2) g_{ij} \ee and we see that for any
non trivial selection of $A$ and $B$, i.e., non trivial backgrounds
$F_{MNPQ}$ the four dimensional part is Einstein with negative
curvature. The equations of motion for $F_{MNPQ}$ reduces to
\be\lb{formal} B(d\Phi - A *\Phi) =0, \ee where $\Phi$ is the
three-form \be\lb{phi} \Phi=\frac{1}{3!4!}e_\mu^m
\det(e_i^a)\epsilon_{ijklmnp}F^{lmnp} \,dx^i\wedge dx^j\wedge dx^k=
e_i^a e_j^b e^c_k c^{abc}\,\,dx^i\wedge dx^j\wedge dx^k. \ee Now we
turn our attention to the criterion for unbroken supersymmetry in
the effective $D=4$ theory. If the vacuum is supersymmetric, then
condition $<\Psi_m>=0$ should be invariant under supersymmetry
transformations, that is \be\lb{susyvar} \delta\psi_M = {\cal
D}_M\epsilon -\frac{1}{288}F_{PQRS}
\left({\Gamma^{PQRS}}_M-8\Gamma^{QRS}\delta^P_M\right)\epsilon=0 \,,
\ee In particular, invariance (\ref{susyvar}) impose restrictions
over the internal manifold. Taking into account the $O(3,1)\times
O(7)$ decomposition of the gravitino and the gamma matrices given by
$$
\psi_M \longrightarrow \psi_\mu\otimes\hat\psi_\alpha,\;\;\;\;\;
\psi\otimes\hat\psi_{i,\alpha}
$$
$$
\Gamma^M \longrightarrow \gamma^\mu\otimes I,\;\;\;\;
\gamma_5\otimes\gamma^i ,
$$
we obtain that
$$
\delta(\psi_\mu\otimes\hat\psi_\alpha) = {\cal
D}_\mu(\epsilon\otimes\hat\epsilon_\alpha) +\frac{1}{12}\tilde
B(\gamma_\mu\epsilon)\otimes(\delta_{\alpha\beta} -8\delta_{\alpha
8}\delta_{\beta 8})\hat\epsilon_\beta
-{i\over6}A(\gamma_\mu\gamma_5\epsilon)\otimes\hat\epsilon_\alpha
$$
\be\lb{descomposicion1} =\check{\cal
D}_\mu(\epsilon\otimes\hat\epsilon_\alpha) \ee
$$
\delta(\psi\otimes\hat\psi_{i,\alpha}) = {\cal
D}_i(\epsilon\otimes\hat\epsilon_\alpha) +\frac{1}{12}(A-i\tilde
B\gamma_5)\epsilon\otimes(i\gamma_i\hat\epsilon)_\alpha
+{i\over3}\,\tilde B(\gamma_5\epsilon)\otimes(
e_{i\alpha}\hat\epsilon_8 +3\delta_{\alpha8}
e_{i\beta}\hat\epsilon_\beta)
$$
\be\lb{descomposicion2} = \check{\cal
D}_i(\epsilon\otimes\hat\epsilon_\alpha) \,. \ee Therefore in order
to preserve a supersymmetry (\ref{descomposicion1}) and
(\ref{descomposicion2}) should be zero. The compatibility conditions
for this is given by \be\lb{implicancia1} 0=[\check{\cal
D}_\mu,\check{\cal D}_\nu]\,\epsilon\otimes\hat\epsilon_\alpha \ee
Taking into account that $[D_{\mu}, D_{\nu}]\epsilon=R_{\mu\nu}^{mn}
\gamma_{mn}\epsilon$ we obtain from (\ref{implicancia1}) that
 \be\lb{implicancia2} {\cal
R}_{\mu\nu}\,(\epsilon\otimes\hat\epsilon_\alpha)
=-{1\over3}g_{\mu\nu}\left[
A^2\,(\epsilon\otimes\hat\epsilon_\alpha) +{1\over4}\tilde
B^2\,(\epsilon\otimes\hat\epsilon_\alpha
+48\delta_{\alpha8}\,\epsilon\otimes\hat\epsilon_8)\right] \,. \ee
Equation (\ref{implicancia2}) is compatible with the Einstein
equations (\ref{Einsfact}) only if $B=0$ and then it follows that
$$
\delta(\psi_\mu\otimes\hat\psi_\alpha) = {\cal
D}_\mu(\epsilon\otimes\hat\epsilon_\alpha) -
{1\over4}\sqrt{|R|\over3}\,
(i\gamma_\mu\gamma_5\epsilon\otimes\hat\epsilon_\alpha),
$$
being $R=R_{\mu\nu}g^{\mu\nu}$. Taking $B=0$ we get $R=5A^2/7$ and
that $\epsilon$ is a conformal Killing spinor, that is
\be\lb{implico} {\cal D}_i \hat\epsilon = -{1\over2}\sqrt{ R\over42}
(i\gamma_i\hat\epsilon) = - \frac{1}{12}A (i\gamma_i\hat\epsilon). \
\ee The variation (\ref{implico}) cancels (\ref{descomposicion2}) as
should be and the internal space is of weak $G_2$ holonomy. From
(\ref{confspin}) and (\ref{wg24}) we get that \be\lb{alfin} d\Phi =
-{2\over3} A *\Phi . \ee This mean that $B\neq 0$ is incompatible
with the requirement of supersymmetry. In other words, any non zero
background field $F_{abcd}$ broke susy. In absence of fluxes, i.e,
$A=0$ we have the product of Minkowski in four dimensions by a $G_2$
holonomy one, in other case we have the product of the $AdS_4$ by a
weak $G_2$ holonomy space.

\newpage

\section{Heterotic geometry without isometries}

     We present some properties of hyperkahler torsion
(or heterotic) geometry in four dimensions that makes it even more
tractable than its hyperkahler counterpart. We show that in $d=4$
hypercomplex structures and weak torsion hyperkahler geometries are
the same. We present two equivalent formalisms describing such
spaces, they are stated in the propositions of section 1. The first
is reduced to solve a non linear system for a doblet of potential
functions, which was found first Plebanski and Finley. The second is
equivalent to find the solutions of a quadratic
Ashtekar-Jacobson-Smolin like system, but without a volume
preserving condition. Is in this second sense that heterotic spaces
are simpler than usual hyperkahler ones. We also analyze strong the
strong version of this geometry. Certain examples are presented,
some of them are of the Callan-Harvey-Strominger type and others are
not. We discuss the benefits and disadvantages of both formulations
in detail.

\subsection{Introduction}

 Supersymmetric $\sigma$ models in two dimensions
with $N=2$ supersymmetry occur on Kahler manifolds while $N=4$
occurs on hyperkahler spaces \cite{Alvarez}. More general
supersymmetric $\sigma$ models can be constructed by including
Wess-Zumino-Witten type couplings in the action
\cite{Witten}-\cite{Sierra}. These couplings can be interpreted as
torsion potentials and the relevant geometry is a generalization of
the Kahler and hyperkahler ones with closed torsion. Such spaces are
known as strong Kahler and hyperkahler torsion (HKT) geometries. The
first examples of N=(4,4) SUSY sigma model with torsion (and the
corresponding HKT geometry) were found in \cite{ivokrivo1} and
developed further in \cite{ivokrivo2}-\cite{ivokrivo4} by using of
the harmonic superspace formalism.

        Hyperkahler torsion geometry is also an useful mathematical
tool in order to construct heterotic string models
\cite{Sen}-\cite{Hull}. In particular heterotic $(4,0)$
supersymmetric models are those that lead to strong hyperkahler
torsion geometry \cite{Bergshoeff}-\cite{Spindel}. If the torsion is
not closed we have a weak hyperkahler geometry and this case has
also physical significance \cite{Callan}-\cite{Gibpapa}.

     A direct way of classifying the possible
HKT spaces is to find the most general weak hyperkahler spaces and
after that to impose the strong condition, i.e, the closure of the
torsion. Callan, Harvey and Strominger noticed that under a
conformal transformation any usual hyperkahler geometry is mapped
into one with torsion \cite{Callan}; such examples are sometimes
called minimal in the literature \cite{Tod} and they are not the
most general \cite{Valent}. As there is a particular interest in
constructing spaces with at least one tri-holomorphic Killing vector
due to applications related to dualities \cite{Rocek}, there were
found heterotic extensions of the known hyperkahler spaces and, in
particular, the Eguchi-Hanson and Taub-Nut ones. The heterotic
Eguchi Hanson geometry was shown to be conformal to the usual one,
while heterotic Taub-Nut is a new geometry \cite{Valent}.

   One of the main properties of hyperkahler torsion spaces is the integrability of the
complex structures, that is, the annulation of their Nijenhuis
tensor. In four dimensions this implies that the Weyl tensor of such
manifolds is self-dual \cite{Strachan}-\cite{Finley}. The converse
of this statement is not true in general. On the other hand there
exist a one to one correspondence between four dimensional self-dual
structures with at least one isometry and 3-dimensional
Einstein-Weyl structures \cite{JonTod}. The Einstein-Weyl condition
is a generalization of the Einstein one to include conformal
transformations, and weak hyperkahler spaces should correspond with
certain special Einstein-Weyl metrics. In
\cite{Papadopoulos}-\cite{Tod} it was shown that the self-dual
spaces corresponding to the round three sphere and the Berger sphere
(which are Einstein-Weyl) are of heterotic type. Arguments related
to the harmonic superspace formalism suggest that indeed there are
more examples \cite{Ivanov}-\cite{Delduc}.

    The present work is related to the construction of weak heterotic geometries
in $d=4$ without Killing vectors. This problem is of interest also
because any weak space with isometries should arise as subcases of
those presented here. For the sake of clarity we resume the result
presented in this letter in the following two equivalent
propositions.
\\

{\bf Proposition 10}{ \it Consider a metric $g$ defined on a
manifold $M$ together with three complex structures $J^{i}$
satisfying the algebra $J^{i}\cdot
J^{j}=-\delta_{ij}+\epsilon_{ijk}J^{k}$ and for which the metric is
quaternion hermitian, i.e, $g(X,Y)=g(J^i X, J^i Y)$. Define the
conformal family of metrics $[g]$ consisting of all the metrics $g'$
related to $g$ by an arbitrary conformal transformation.
\\

a) Then we have the equivalence \be\lb{equivalencia} d\overline{J}^i
+ \alpha \wedge \overline{J}^i=0
 \qquad\Longleftrightarrow\qquad N^i(X,Y)=0,
 \ee
where $\overline{J}^i$ and $N^i(X,Y)$ are the Kahler form and the
Niejenhuis tensor associated to $J^{i}$, $d$ is the usual exterior
derivative and $\alpha$ is a 1-form. If any of (\ref{equivalencia})
hold for $g$, then (\ref{equivalencia}) is also satisfied for any
$g'$ of the conformal structure $[g]$, i.e, (\ref{equivalencia}) is
conformally invariant.
\\

 b) Any four dimensional weak hyperkahler torsion metric is equivalent
to one satisfying (\ref{equivalencia}) and there exists a local
coordinate system $(x, y, p, q)$ for which the metric take the form
\be\lb{metropo} g= (dx -\Phi_x dp + \Phi_x dq)\otimes dp +(dy +
\Psi_y dp -\Psi_x dq)\otimes dq, \ee up to a conformal
transformation $g\rightarrow \omega^2 g$. The potentials $\Psi$ and
$\Phi$ satisfy the non-linear system \be\lb{maestro} [\Phi_y
\partial_x
\partial_x + \Psi_x \partial_y \partial_y -
(\Phi_x + \Psi_y)\partial_x \partial_y+\partial_x \partial_p
+\partial_y \partial_q ]\left(%
\begin{array}{c}
  \Phi \\
  \Psi \\
\end{array}%
\right)=0. \ee

c) Conversely any metric (\ref{metropo}) defines a conformal family
$[g]$ in which all the elements $g'$ are weak hyperkahler torsion
metrics. The torsion $T$ corresponding to (\ref{metropo}) is given
by
$$
T=-\Xi_x dq\wedge (dy \wedge dx+\Phi_y dp\wedge dx- \Psi_y dy\wedge
dp)
$$
\be\lb{mono} +\Xi_y dp\wedge (dy\wedge dx +\Phi_x dy\wedge dq
-\Psi_x dq\wedge dx),\ee where $\Xi=\Phi_x-\Psi_y$.  Under the
conformal transformation $g \rightarrow \omega^2 g$ the torsion is
transformed as $ T\rightarrow T + \ast_{g} 2 d\log
(\omega)$\footnote{ The action of the Hodge star $\ast_{g}$ is
defined by
$$
\ast_{g}e^a=\epsilon_{abcd}e^b \wedge e^c \wedge e^d.
$$}}.
\\

Proposition 10 should not be considered as a generalization of the
Kahler formalism for weak HKT spaces. Although $(\Phi, \Psi)$ is a
doublet potential, the metric (\ref{metropo}) is not written in
complex coordinates. The use of holomorphic coordinates for such
spaces is described in detail in \cite{Valent}. The following is an
Ashtekar-Jacobson-Smolin like formulation for the same geometry.
\\

 {\bf Proposition 11}{ \it Consider a
representative $g=\delta_{ab}e^a \otimes e^b$ of a conformal family
$[g]$ defined on a manifold $M$ as at the beginning of Proposition
1, $e^a$ being tetrad 1-forms for which the metric is diagonal.
\\

 a) Then all the elements $g'$ of $[g]$ will be weak hyperkahler torsion
iff
$$
[e_1,e_2] + [e_3,e_4]= - A_2 e_1 + A_1 e_2 - A_4 e_3 + A_3 e_4
$$
\be\lb{genou} [e_1,e_3]+[e_4,e_2]= - A_3 e_1 + A_4 e_2 + A_1 e_3-
A_2 e_4 \ee
$$
[e_1,e_4]+[e_2,e_3]= - A_4 e_1 - A_3 e_2 + A_2 e_3 + A_1 e_4
$$
where $e_a$ is the dual tetrad of $e^a$ and $A_i$ are arbitrary
functions on $M$.
\\

b) Conversely any solution of (\ref{genou}) defines a conformal
family $[g]$ in which all the elements $g'$ are weak hyperkahler
torsion metrics. The torsion $T$ corresponding to (\ref{metropo}) is
given by \be\lb{maza} T=\ast_{g} (A_a-c^b_{ab})e^a \ee where
$c^c_{ab}$ are the structure functions defined by the Lie bracket
$[e_a,e_b]=c_{ab}^{c}e_c$. The transformation of the torsion under
$g \rightarrow \omega^2 g$ follows directly from Proposition 1.}
\\

  To conclude, we should mention that the
properties of hyperkahler torsion geometry in higher dimension were
considered for instance, in \cite{Poon1}-\cite{Ofer2}. In
particular, the quotient construction for HKT was achieved in
\cite{Poon3}. Also it was found that when a sigma model is coupled
to gravity the resulting target metric is a generalization of
quaternionic Kahler geometry including torsion \cite{Ofer}. This
result generalizes the classical one given by Witten and Bagger
\cite{Wite}, who originally did not include a Wess-Zumino term to
the action. A generalization of the Swann extension for quaternion
Kahler torsion spaces was achieved \cite{Poon2}. To the knowledge of
the authors the harmonic superspace description of quaternion Kahler
geometry was already obtained in \cite{Ivi}, \cite{Ogo} but the
extension to the torsion case is still an open problem.

   The present part is organized as follows.
In section 3.2 we define what is weak and strong heterotic geometry.
We show that the problem to finding weak examples is equivalent to
solving a conformal extension of the hyperkahler condition (namely,
the first (\ref{equivalencia}) given above) together with the
integrability condition for the complex structure. In section 3.3 we
present the consequences of (\ref{equivalencia}) following a work of
Plebanski and Finley \cite{Finley}. We found out that integrability
and the conformal extension of the hyperkahler condition are
equivalent in $d=4$, which is the point a) of Proposition 10.
Therefore weak heterotic geometry and hypercomplex structures are
exactly the same concept. We also show that strong representatives
of a given heterotic structure are determined by a conformal factor
satisfying an inhomogeneous Laplace equation. Just in the case when
the structure contains an hyperkahler metric it is possible to
eliminate the inhomogeneous part by a conformal transformation. We
discuss our results in the conclusions, together with possible
applications. For completeness, we show in the appendix how this
geometry arise in the context of supersymmetric sigma models.

\subsection{Hyperkahler torsion manifolds}

\subsubsection{Main properties}

   In this section we define what is hyperkahler torsion
geometry (for an explanation of its physical meaning see Appendix).
We deal with $4n$-dimensional Riemannian manifolds $M$ with metric
\be\lb{numeral} g=\delta_{ab}e^a\otimes e^b, \ee being $e^a$ a
tetrad basis for which the metric is diagonal, which is defined up
to an $SO(4n)$ rotation. Consider the $4n \times 4n$ matrices
$$
J^{1}=\left(\begin{array}{cccc}
  0 & -I_{n \times n} & 0 & 0 \\
  I_{n \times n} & 0 & 0 & 0 \\
  0 & 0 & 0 & -I_{n \times n} \\
  0 & 0 & I_{n \times n} & 0
\end{array}\right),\;\;\;\;
J^{2}=\left(\begin{array}{cccc}
  0 & 0 & -I_{n \times n} & 0 \\
  0 & 0 & 0 & I_{n \times n} \\
   I_{n \times n} & 0 & 0 & 0 \\
  0 & -I_{n \times n} & 0 & 0
\end{array}\right)
$$
\be\lb{reprodui} J^{3}=J^{1}J^{2}=\left(\begin{array}{cccc}
  0 & 0 & 0 & -I_{n \times n} \\
  0 & 0 & -I_{n \times n} & 0 \\
  0 & I_{n \times n} & 0 & 0 \\
  I_{n \times n} & 0 & 0 & 0
\end{array}\right).
\ee
 It is direct to check that (\ref{reprodui}) satisfy the quaternion algebra
\be\lb{almcomp} J^{i}\cdot J^{j}=-\delta_{ij}+\epsilon_{ijk}J^{k}.
\ee In particular from (\ref{almcomp}) it is seen that $J^i\cdot
J^i=-I$, a property that resemble the imaginary condition $i^2=-1$.
Associated to (\ref{reprodui}) we define the $(1,1)$ tensors
\be\lb{unaal} J^{i}=(J^{i})^{b}_{\;a} e_b \otimes e^a \ee which are
known as almost complex structures. Here $e_a$ the dual of the
1-form $e^a$. Let us introduce the triplet $\overline{J}^i$ of
$(0,2)$ tensors by \be\lb{susu} \overline{J}^i(X,Y)=g(X, J^i Y). \ee
It is direct to check that the metric (\ref{numeral}) is quaternion
hermitian with respect to any of the complex structures
(\ref{unaal}), that is \be\lb{defcon} g(J^i X, Y)=-g(X, J^i Y) \ee
for any X,Y in $T_x M$ (in this notation $J^i X$ denotes the
contraction of $J^i$ with $X$). By virtue of (\ref{defcon}) the
tensors (\ref{susu}) are skew-symmetric and define locally a triplet
of 2-forms which in the vielbein basis take the form \be\lb{dosal}
\overline{J}^i=(\overline{J}^i)_{ab} e^a \wedge e^b. \ee The 2-forms
(\ref{dosal}) are known as the hyperkahler triplet.
\\

\textbf{Definition}
\\

Heterotic geometry (or torsion hyperkahler geometry) is defined by
the following requirements.
\\

(1) \textit{Hypercomplex condition}. The almost complex structures
(\ref{sdcom}) should be integrable. This is equivalent to say the
Niejenhuis tensor \be\lb{nijui}
N^i(X,Y)=[X,Y]+J^{i}[X,J^{i}Y]+J^{i}[J^{i}X,Y]-[J^{i} X,J^{i} Y] \ee
associated to the complex structure $J^{i}$ is zero for every pair
of vector fields $X$ and $Y$ in $TM_x$.
\\

 (2) \textit{Existence of a torsion}. There exists a torsion tensor
$T^{\mu}_{\nu\alpha}$ defined in terms of the metric for which the
$T_{\mu\nu\alpha}=g_{\mu\xi}T^{\xi}_{\nu\alpha}$ is fully
skew-symmetric and define a three form (we write it in the tetrad
basis) \be\lb{tor} T= \frac{1}{3!}T_{\mu\nu\lambda}dx^{\mu}\wedge
dx^{\nu}\wedge dx^{\lambda}. \ee Here greek indices denote the
quantities related to a coordinate basis $x^{\mu}$\footnote{In the
vienbein basis relation (\ref{tor}) will be
$$
T= \frac{1}{3!}T_{abc}e^a\wedge e^b \wedge e^c.
$$}. If (\ref{tor}) is closed, i.e, $dT=0$, the geometry
will be called strong, in other case weak.
\\

(3) \textit{Covariant constancy of $J^{i}$}. Let us define a
derivative $D_{\mu}$ with the connection \be\lb{conoc}
\Upsilon^{\rho}_{\mu\nu}=\Gamma^{\rho}_{\mu\nu}-{1\over
2}T^{\rho}_{\mu\nu}, \ee being $\Gamma^{\rho}_{\mu\nu}$ the
Christofell symbols associated to the usual Levi-Civita connection.
By definition the structures $(J^i)_{a}^{b}$ of an hyperkahler
torsion space satisfy \be\lb{equo} D_{\mu}(J^i)_{\nu}^{\rho}=0 \ee
that is, they are covariantly constant with respect to $D_{\mu}$.
\\

     We are concerned through this work with $d=4$. In this case the explicit
form of the $(1,1)$ tensors (\ref{unaal}) is obtained by putting
$n=1$ into (\ref{reprodui}), the result is
$$
J^1 = -e_1 \otimes e^2 + e_2 \otimes e^1 -e_3\otimes e^4 +
e_4\otimes e^3
$$
\be\lb{sdcom} J^2 = -e_1 \otimes e^3 + e_3 \otimes e^1 - e_4\otimes
e^2 + e_2\otimes e^4 \ee
$$
J^3 = - e_1 \otimes e^4 + e_4 \otimes e^1 - e_2\otimes e^3 +
e_3\otimes e^2,
$$
and the corresponding hyperkahler triplet (\ref{dosal}) is
$$
\overline{J}^1 = e^2\wedge e^1 + e^4\wedge e^3
$$
\be\lb{sdcomo} \overline{J}^2 = e^3\wedge e^1 + e^2\wedge e^4  \ee
$$
\overline{J}^3 = e^4\wedge e^1 + e^3\wedge e^2,
$$
Matrices (\ref{reprodui}) with $n=1$ are a non unique $4\times 4$
representation the algebra (\ref{almcomp}). In fact due to the
$SO(4)$ freedom to select $e^a$ it follows that (\ref{sdcom}) are
defined up to an $SO(4)$ rotation. The multiplication rule
(\ref{almcomp}) is unchanged under such rotation. It can also be
checked from (\ref{defcon}) that the metric $g(X,Y)$ is quaternion
hermitian with respect to any $SO(4)$ rotated complex structures.

    As we have seen condition $N^{i}(X,Y)=0$ for every $J^{i}$ in (\ref{sdcom}) is equivalent to
solve the quadratic system (\ref{genou}) which is invariant under a
transformation
$$
e_a\rightarrow \omega e_a \; (g\rightarrow \omega^2 g),\;\;\;\;\;
A_a\rightarrow A_a + e_a \log \omega.
$$
This mean that weak hyperkahler spaces are some special kind of
hypercomplex structures. In the limit $A_i=0$ (\ref{genou}) reduce
to a system equivalent to the Ashtekar-Jacobson-Smolin one
\cite{Ashtekar} for hyperkahler spaces. Nevertheless one can not
conclude that (\ref{genou}) always describe spaces which are
conformally equivalent to hyperkahler ones, even in the case
$A_a=0$. \footnote{In order to have an hyperkahler manifold the
tetrad should also preserve certain volume form. If this condition
hold the solutions are also called minimal.} In the special cases in
which the resulting family is conformal hyperkahler we will call it
of the Callan-Harvey-Strominger type \cite{Callan}.

   It is important to recall that although the complex structures
are defined up to certain $SO(4)$ automorphisms of (\ref{almcomp}),
this does not affect the integrability condition $N^{i}(X,Y)=0$. A
$SO(4)$ rotation of the complex structures can be compensated by an
$SO(4)$ rotation of the frame leaving (\ref{acshon}) and therefore
$N^a(e_i,e_j)=0$ invariant. By the results that we will present
below it will be clear that such rotations also do not affect
conditions 2 and 3.

\subsubsection{Relation with the Plebanski-Finley conformal structures}

   The requirements 1, 2 and 3 of the definition of HKT geometry are not independent
because the last two implies the first. To see this clearly let us
write the expression for the Nijenhuis tensor in a coordinate basis
$x^{\mu}$, \be\lb{lacord} N^{\rho}_{\mu\nu}=
(J^i)_{\mu}^{\lambda}[\partial_{\lambda} (J^i)_{\nu}^{\rho}\ -\
\partial_{\nu} (J^i)_{\lambda}^{\rho}]\ -(\mu\ \leftrightarrow \
\nu)=0. \ee Then (\ref{lacord}) together with (\ref{equo}) implies
that \be\lb{coord} T_{\rho\mu\nu}
-(J^i)_{[\mu}^{\lambda}(J^i)_{\nu}^{\sigma}T_{\rho]\lambda\sigma}=0.
\ee It is simple to check that (\ref{coord}) is satisfied for any
skew-symmetric torsion. Let us consider the tetrad basis $e^a$ for
which the complex structures take the form (\ref{reprodui}), then
(\ref{coord}) is a direct consequence of the skew symmetry of
$T_{\mu\nu\alpha}$. Since $N^i(X,Y)$ is a tensor then it is zero in
any basis. Thus we have proved that the requirements 2 and 3 imply
first.

    We will show now that the task to find a weak HKT geometry
is in fact equivalent to obtain the solutions of \be\lb{miruta}
d\overline{J}^i + \alpha \wedge \overline{J}^i=0 \ee where $\alpha$
is an arbitrary 1-form. To prove this we note that the relation
$$
2D_{\mu} g_{\nu\alpha}=- g_{\mu\xi}T^{\xi}_{\nu\alpha} -
g_{\nu\xi}T^{\xi}_{\mu\alpha}
$$
together with the skew symmetry of
$T_{\mu\nu\alpha}=g_{\mu\xi}T^{\xi}_{\nu\alpha}$ yields directly
that \be\lb{yeah} D_{\mu} g_{\nu\alpha}=0. \ee Equation (\ref{yeah})
implies the equivalence \be\lb{poh}
D_{\mu}(J^i)^{\nu}_{\alpha}=0\qquad \Longleftrightarrow \qquad
D_{\mu}(\overline{J}^i)_{\nu\alpha}=0,\ee and therefore for an
hyperkahler torsion space
$$
D_{\mu}(\overline{J}^i)_{\nu\alpha}=0.
$$
From the identity \cite{Futa}
$$
D_{\mu}(\overline{J}^i)_{\nu\alpha}=D_{[\mu}(\overline{J}^i)_{\nu\alpha]}-3
D_{[\mu}(\overline{J}^i)_{\xi\rho]}(J^i)^{\xi}_{\nu}
(J^i)^{\rho}_{\alpha} + (\overline{J}^i)_{\mu\xi}
(N^i)^{\xi}_{\nu\alpha} ,
$$
is easily seen that \be\lb{cuba}
D_{\mu}(\overline{J}^i)_{\nu\alpha}=0\qquad\Longleftrightarrow\qquad
D_{[\mu}(\overline{J}^i)_{\nu\alpha]}=0,\qquad N^i(X,Y)=0. \ee where
the brackets denote the totally antisymmetric combination of
indices. But we have seen that $N^i(X,Y)=0$ is satisfied by our
hypotesis. An explicit calculation using (\ref{equo}),(\ref{conoc})
and the skew-symmetry of $T_{\mu\nu\alpha}$ shows that \be\lb{hap}
D_{[\mu}(\overline{J}^i)_{\nu\alpha]}=0\qquad
\Longleftrightarrow\qquad d\overline{J}^i +{1\over
2}T_{dbc}(J^i)_{a}^{d}e^{a}\wedge e^{b}\wedge e^{c}=0. \ee The
second (\ref{hap}) is the generalization of the hyperkahler
condition for torsion manifolds. If the torsion $T_{\mu\nu\alpha}$
is zero, then the second (\ref{hap}) implies that the hyperkahler
triplet is closed, which is a well known feature of hyperkahler
geometry. Consider now the explicit form of (\ref{hap}) for
$\overline{J}^1$. Then (\ref{equo}) and the first (\ref{reprodui})
gives us that
$$
d\overline{J}^1+T_{234}e^1\wedge e^3 \wedge e^4-T_{134}e^2\wedge e^3
\wedge e^4 +T_{312}e^4\wedge e^1 \wedge e^2-T_{412}e^3\wedge e^1
\wedge e^2
$$
$$
 =d\overline{J}^1+(T_{234}e^1+T_{134}e^2-T_{123}e^4+T_{124}e^3)\wedge (e^1\wedge
e^2+e^3\wedge e^4)=0
$$
and therefore
$$
d\overline{J}^1+(\ast_g T)\wedge \overline{J}^1=0.
$$
The same is obtained for $\overline{J}^2$ and $\overline{J}^3$,
namely \be\lb{ejem} d\overline{J}^i + (\ast_g T)\wedge
\overline{J}^i=0. \ee Let us define now the 1-form $\alpha$ given by
\be\lb{alfo} \alpha=\ast_{g} T \qquad \Longleftrightarrow \qquad
\ast_{g} \alpha=T. \ee Then we have from (\ref{ejem}) that
(\ref{hap}) is equivalent to \be\lb{pito} d\overline{J}^i +\alpha
\wedge \overline{J}^i=0. \ee Therefore formulas
(\ref{poh})-(\ref{hap}) implies that hyperkahler torsion geometry is
completely characterized by (\ref{miruta}) which is what we wanted
to show.

   Condition (\ref{pito}) is the conformally invariant extension of the
hyperkahler one and has been studied by Plebanski and Finley in the
seventies \cite{Finley}. A detailed calculation shows that
(\ref{pito}) is invariant under \be\lb{trofo2} e^a\rightarrow \omega
e^a\;(g\rightarrow \omega^2 g)\qquad
 \alpha\longrightarrow \alpha + 2 d\log
\omega, \ee in consequence it defines a conformal family $[g]$. The
transformation law (\ref{trofo2}) implies in particular that if
$\alpha$ is a gradient there exist a representative of $[g]$ for
which
$$
\alpha=0\qquad \Longleftrightarrow \qquad d\overline{J}^i=0.
$$
Clearly such families will be conformal to an hyperkahler space and
is of the Callan-Harvey-Strominger type. The main result of
Plebanski and Finley is that in general (\ref{pito}) implies that
the Weyl tensor of $[g]$ is self-dual \cite{Finley}. From
(\ref{alfo}) we deduce that under $g \rightarrow \omega^2 g$ it
follows that \be\lb{trofo3} T\longrightarrow T + \ast_{g} 2 d \log
(\omega), \ee which is a transformation found by Callan et all in
\cite{Callan}.

   Some more comments are in order. As we have seen the requirements 2 and 3
imply that $N^i(X,Y)=0$ and also imply (\ref{pito}). Therefore we
have \be\lb{imp1} N^i(X,Y)=0 \qquad \Longleftarrow \qquad
d\overline{J}^i + \alpha \wedge \overline{J}^i=0. \ee But is also
known that \cite{Strachan}-\cite{Boyer}
 \be\lb{imp}
N^i(X,Y)=0 \qquad \Longrightarrow \qquad d\overline{J}^i + \alpha
\wedge \overline{J}^i=0. \ee To see (\ref{imp}) consider the
connection $\omega$ given by the first structure equation
$$
de^a + \omega^a_{b}\wedge e^b = 0.
$$
It is well known that the antisymmetric part $\omega^a_{[bc]}$ is
related to the structure functions defined by the Lie bracket
$[e_a,e_b]=c_{ab}^{c}e_c$ by
$$
\omega^a_{[bc]} = \frac{1}{2} c_{bc}^{a}\,.
$$
Consider now the form
$$
\alpha = A-\chi
$$
where \be\lb{conectar} A = A_a e^a ,\;\;\;\; \chi =c_{ab}^{b} e^a.
\ee and the self-dual form $\overline{J}^1 = e^1 \wedge e^2+ e^3
\wedge e^4$. Then by use of (\ref{genou}) one obtain that
$$
d \overline{J}^1 = d(e^1 \wedge e^2+ e^3 \wedge e^4)
$$
$$
= -\frac{1}{2} c^{[1}_{ab}  e^{2]} \wedge e^a \wedge e^b
-\frac{1}{2} c^{[3}_{ab} e^{4]} \wedge e^a \wedge e^b
$$
$$
= e^1\wedge e^2\wedge (c_{ab}^{a} e^b + A_3 e^3 + A_4 e^4)
+e^3\wedge e^4\wedge (c_{ab}^{a} e^b + A_1 e^1 + A_2 e^2)
$$
$$
=(e^1 \wedge e^2 + e^3 \wedge e^4) \wedge ( A-\chi)
$$
and therefore
$$
d \overline{J}^1+ \overline{J}^1\wedge (\chi-A)=0.
$$
This is equivalent to (\ref{oso}) with $\alpha=\chi-A$. The same
formula holds for $\overline{J}^2$ and $\overline{J}^3$. Then we
conclude that \be\lb{imp3} N^i(X,Y)=0 \qquad \Longleftrightarrow
\qquad d\overline{J}^i + \alpha \wedge \overline{J}^i=0, \ee that
is, hypercomplex structures are \emph{the same} as the conformal
structures defined by (\ref{pito}).

\subsection{The general solution}

       We have seen in the previous section that to find an hyperkahler torsion geometry
is equivalent to find a metric which solve the conditions stated in
(\ref{miruta}). This task was solved by Plebanski and Finley and is
stated in the following proposition \cite{Finley}.
\\

{\bf Proposition 12} {\it For any four dimensional metric $g$ for
which
$$
d\overline{J}^i +\alpha \wedge \overline{J}^i=0
$$
hold there exist a local coordinate system $(x, y, p, q)$ for which
$g$ take the form \be\lb{metropo2} g = (dx -\Phi_x dp + \Phi_x
dq)\otimes dp +(dy + \Psi_y dp -\Psi_x dq)\otimes dq, \ee up to a
conformal transformation $g \rightarrow \omega^2 g$. The functions
$\Psi$ and $\Phi$ depends $(x, y, p, q)$ and satisfy the non-linear
system \be\lb{maestro2} [\Phi_y
\partial_x
\partial_x + \Psi_x \partial_y \partial_y -
(\Phi_x + \Psi_y)\partial_x \partial_y+\partial_x \partial_p
+\partial_y \partial_q ]\left(%
\begin{array}{c}
  \Phi \\
  \Psi \\
\end{array}%
\right)=0. \ee The converse of this assertion is also true.}
\\

For completeness we intend to give the proof with certain detail,
following the original reference \cite{Finley}. It is convenient to
introduce the Penrose notation and write the metric $g$ in the
following complex form \be\lb{compo} g=\delta_{ab}e^a\otimes e^b=E^1
\otimes E^2 + E^3 \otimes E^4. \ee We have defined the new complex
tetrad
$$
E^1=\frac{1}{\sqrt{2}}(e^1 + ie^2),\qquad E^2=\frac{1}{\sqrt{2}}(e^1
- ie^2)
$$
\be\lb{penu} E^3=\frac{1}{\sqrt{2}}(e^3 + ie^4),\qquad
E^4=\frac{1}{\sqrt{2}}(e^3 - ie^4). \ee In term of this basis the
complex structures (\ref{sdcomo}) are expressed as
$$
\overline{J}^1=2 E^4\wedge E^1,\qquad \overline{J}^3=2 E^3\wedge E^2
$$
\be\lb{penstru} \overline{J}^2=-E^1\wedge E^2 + E^3\wedge E^4. \ee
The advantage of this notation is that allow us to directly apply
the Frobenius theorem.
\\

{\bf Frobenius theorem} {\it In $n$-dimensions, if in a certain
domain $U$ there exist $r$ 1-forms $\beta^i$ ($i=1,..,r$) such that
$$
\Omega=\beta^1\wedge...\wedge \beta^r \neq 0
$$
and there exist a 1-form $\gamma$ such that \be\lb{frobo}
d\Omega=\gamma\wedge\Omega, \ee then there exists some functions
$f^i_j$ and $g^j$ on $U$ such that \be\lb{efe} \beta^i =
\sum_{j=1}^r f^i_j dg^j. \ee}

   The two conditions
\be\lb{oso} d\overline{J}^1 +\alpha \wedge \overline{J}^1=0, \qquad
d\overline{J}^3 +\alpha \wedge \overline{J}^3=0 \ee of (\ref{pito})
are of the form (\ref{frobo}) with $E^i$ are playing the role of the
1-forms $\beta^i$, $\overline{J}^1$ and $\overline{J}^3$ are the
analog of $\Omega$ and $\alpha$ play the role of $\gamma$. Then
Frobenius theorem implies the existence of scalar functions
$\widetilde{A},...,\widetilde{H}$ and $p,..,s$ such that
$$
E^1=\widetilde{A} dp + \widetilde{B} dq,
$$
$$
E^2=\widetilde{E} dr + \widetilde{F} ds,
$$
\be\lb{anlo} E^3=-\widetilde{G} dr - \widetilde{H} ds \ee
$$
E^4=-\widetilde{C} dp - \widetilde{D} dq.
$$
The functions $\widetilde{A},...,\widetilde{H}$ and  $p,..,s$ are
the analog of $f^i_j$ and $g^j$ in (\ref{efe}) respectively.

    Consider the two functions $\phi$ and $f$ defined by
$$
\widetilde{A}\widetilde{D}-\widetilde{B}\widetilde{C}=
(\phi^{-1}e^{f})^2,
$$
\be\lb{def} \widetilde{E}\widetilde{H}-\widetilde{F}\widetilde{G}=
(\phi^{-1}e^{-f})^2. \ee It is more natural to use the variables
$$
(A,B,C,D)=(\phi^{-1}e^{f})^{-1}(\widetilde{A}, \widetilde{B},
\widetilde{C}, \widetilde{D})
$$
$$
(E,F,G,H)=(\phi^{-1}e^{-f})^{-1}(\widetilde{E}, \widetilde{F},
\widetilde{G}, \widetilde{H})
$$
with the functions $A,..., H$ normalized as \be\lb{dinamic}
AD-BC=1,\qquad EH-GF=1. \ee Using (\ref{def}) we can express
(\ref{anlo}) as
$$
E^1=\phi^{-1}e^{f}(A dp + B dq) ,
$$
$$
E^2=\phi^{-1}e^{-f}(E dr + Fds),
$$
\be\lb{anl} E^3=-\phi^{-1}e^{-f}(G dr + H ds) ,\ee $$
 E^4=
-\phi^{-1}e^{f}(C dp + D dq),
$$
Since
$$
E^1\wedge E^2\wedge E^3\wedge E^4=\phi^{-4} dp \wedge dq \wedge dr
\wedge ds \neq 0
$$
one can consider the functions $p$, $q$, $r$ and $s$ as independent
coordinates.

    From (\ref{anl}), (\ref{penstru}) and (\ref{oso}) it is
obtained the following expression for $\alpha$ \be\lb{plin} \alpha=d
\log (\phi)+ f_p dp + f_q dq - f_r dr - f_s ds. \ee We now consider
the resting condition (\ref{pito}), namely \be\lb{mm}
d\overline{J}^2 +\alpha \wedge \overline{J}^2=0. \ee By use of
(\ref{penstru}) together with (\ref{anl}) we find that
$$
\overline{J}^2= \phi^{-4}[(AE+CG)dp\wedge dr+(AF+CH)dp\wedge ds
+(BE+DG) dq\wedge dr
$$
\be\lb{difo} +(BF+DH) dq \wedge ds]. \ee A direct calculation shows
that (\ref{mm}) together with (\ref{plin}) implies
$$
(AE+CG)2f_s-(AF+CH)2f_r=(AE+CG)_s-(AF+CH)_r,
$$
$$
(BE+DG)2f_s-(BF+DH)2f_r=(BE+DG)_s-(BF+DH)_r, $$ \be\lb{system}
(AE+CG)2f_q-(BE+DG)2f_p=-(AE+CG)_q + (BE+DG)_p,\ee
$$
(AF+CH)2f_q-(BF+DH)2f_p=-(AF+CH)_q+(BF+DH)_p.
$$
The first two (\ref{system}) shows that there exist certain
functions $x$ and $y$ such that
$$
AE+CG=e^{2f}x_r,
$$
$$
BE+DG=e^{2f}y_r,
$$
\be\lb{ping2} AF+CH=e^{2f}x_s, \ee
$$
BF+DH=e^{2f}y_s.
$$
By multiplying the last two (\ref{system}) by $e^{2f}$ and using
(\ref{ping2}) we also obtain that \be\lb{subso}
(e^{4f}x_r)_q=(e^{4f}y_r)_p,\qquad (e^{4f}x_s)_q=(e^{4f}y_s)_p. \ee
The solution of (\ref{ping2}) and (\ref{subso}) are
$$
G=A(e^{2f}y_r)-B(e^{2f}x_r)
$$
$$
H=A(e^{2f}y_s)-B(e^{2f}x_s)
$$
\be\lb{solo} E=D(e^{2f}x_r)-C(e^{2f}y_r) \ee
$$
F=D(e^{2f}x_s)-C(e^{2f}y_s)
$$
The normalization (\ref{dinamic}) and (\ref{solo}) implies that
$$
J=\left|%
\begin{array}{cc}
  x_r &  x_s \\
  y_r &  y_s \\
\end{array}%
\right|=e^{-4f}\neq 0,
$$
and therefore $x$ and $y$ can be used as independent coordinates
instead of $r$ and $s$. Moreover, the functions $A,..,D$ are not
determined by the equations but just constrained by the first
(\ref{dinamic}). This reflects that the tetrad and the vielbein are
defined up to an $SO(4)$. This rotations leaves condition $AD-BC=1$
invariant. Therefore without lost of generality we can select
$A=D=1$ and $B=C=0$. The resulting tetrad is now
$$
E^1=(\phi e^{-f})^{-1}dp,
$$
$$
E^2=(\phi e^{-f})^{-1}(dx + K dp + L dq),
$$
\be\lb{ij} E^3=-(\phi e^{-f})^{-1}(dy + M dp + N dq), \ee
$$
E^4=-(\phi e^{-f})^{-1}dq,
$$
being $K=-x_p$, $L=-x_q$, $M=-y_p$ and $N=-y_q$. The dual basis of
(\ref{ij}) is
$$
E_2=\phi e^{-f}
\partial_x
$$
$$
E_1=\phi e^{-f}(\partial_p - K \partial_x - M \partial_y) ,
$$
\be\lb{duo} E_4=-\phi e^{-f}(\partial_q - L
\partial_x - N\partial_y), \ee
$$
E_3=-\phi e^{-f}\partial_y.
$$
Because we are working with a conformal structure is easy to check
that we can make eliminate $\phi e^{-f}$ in (\ref{duo}) by a
conformal transformation. From (\ref{oso}) it is calculated that
\be\lb{fingo} 2\alpha=(L-M)_x dq-(L-M)_y dp. \ee Also from
(\ref{mm}) and (\ref{fingo}) it follows that
$$
\partial_x K +\partial_y L=0, \qquad \partial_x M +\partial_y  N=0,
$$
\be\lb{cuento}  (\partial_p - K \partial_x - M \partial_y) L
-(\partial_q - L
\partial_x - N\partial_y) K =0,\ee
$$
 (\partial_p - K \partial_x - M \partial_y)
N-(\partial_q - L
\partial_x - N\partial_y) M =0
$$
The first two (\ref{cuento}) implies the existence of two functions
$\Phi$ and $\Psi$ such that \be\lb{pico} K=-\Phi_x, \qquad L=\Phi_x,
\qquad M=\Psi_y, \qquad N=-\Psi_x, \ee and the first two
(\ref{cuento}) together with (\ref{pico}) gives the following
non-linear equations for the doublet $(\Phi, \Psi)$ \be\lb{master}
[\Phi_y
\partial_x
\partial_x + \Psi_x \partial_y \partial_y -
(\Phi_x+\Psi_y)\partial_x \partial_y+\partial_x \partial_p
+\partial_y \partial_q ]\left(%
\begin{array}{c}
  \Phi \\
  \Psi \\
\end{array}%
\right)=0. \ee Then we conclude that the most general structure
satisfying (\ref{pito}) is given in terms of two key functions
$\Phi$ and $\Psi$ satisfying (\ref{master}). A simple calculation
shows that the metric corresponding to (\ref{duo}) is
(\ref{metropo2}) which is what we wanted to prove (Q.E.D).
\\

  It will be instructive we will show that for any structure of the proposition 12
it follows that
$$
N^i(X,Y)=0,
$$
in accordance with (\ref{imp1}). This is completely equivalent to
solve the system (\ref{genou}), that is
$$
[e_1,e_2]+[e_3,e_4]= - A_2 e_1 + A_1 e_2 - A_4 e_3 + A_3 e_4,
$$
$$
[e_1,e_3]+[e_4,e_2]= - A_3 e_1 + A_4 e_2 + A_1e_3- A_2 e_4,
$$
$$
[e_1,e_4]+[e_2,e_3]= - A_4 e_1 - A_3 e_2 + A_2 e_3 + A_1 e_4,
$$
for the basis $e_i$ corresponding to (\ref{metropo2}). From
(\ref{penu}), (\ref{ij}) and (\ref{pico}) we find that
$$
e^1=\frac{1}{\sqrt{2}}[dp +(dx -\Phi_y dp + \Phi_x dq)],
$$
$$
e^2=\frac{1}{i\sqrt{2}}[dp -  (dx - \Phi_y dp + \Phi_x dq)]
$$
\be\lb{penul} e^3=-\frac{1}{\sqrt{2}}[dq + (dy + \Psi_y dp -\Psi_x
dq)],\ee
$$
e^4=\frac{1}{i\sqrt{2}}[dq -(dy + \Psi_y dp -\Psi_x dq)].
$$
The dual basis corresponding to (\ref{penul}) is
$$
e_1=\frac{1}{\sqrt{2}}[(\partial_p +\Phi_y \partial_x - \Psi_y
\partial_y) + \partial_x],
$$
$$
e_2=-\frac{1}{i\sqrt{2}}[(\partial_p +\Phi_y
\partial_x - \Psi_y \partial_y) -  \partial_x]
$$
\be\lb{penulo} e_3=-\frac{1}{\sqrt{2}}[\partial_y +  (\partial_q -
\Phi_x\partial_x +\Psi_x\partial_y)], \ee
$$
e_4=\frac{1}{i\sqrt{2}}[\partial_y -  (\partial_q - \Phi_x
\partial_x +\Psi_x\partial_y)].
$$
 It is simple to check that the first (\ref{genou}) for
(\ref{penulo}) implies that $A_i=0$. After some tedious calculation
one obtain from the full system (\ref{genou}) that
$$ [\Phi_y
\partial_x
\partial_x + \Psi_x \partial_y \partial_y -
(\Phi_x+\Psi_y)\partial_x \partial_y+\partial_x \partial_p
+\partial_y \partial_q ]\left(%
\begin{array}{c}
  \Phi \\
  \Psi \\
\end{array}%
\right)=0,
$$
which is exactly (\ref{master}).

     We conclude that the general form of an HKT metric is
$$
g= (dx -\Phi_y dp + \Phi_x dq)\otimes dp +(dy + \Psi_y dp -\Psi_x
dq)\otimes dq
$$
up to a conformal transformation.  The form $\alpha$ in
(\ref{fingo}) can be expressed as \be\lb{cho} \alpha=\Xi_x E^1-\Xi_y
E^4, \ee where $\Xi=\Phi_x-\Psi_y$. Then it is direct to prove using
(\ref{penu}) that \be\lb{usefull} \ast E^1=-E^2\wedge E^3\wedge
E^4,\qquad \ast E^4=-E^1\wedge E^2\wedge E^3. \ee Formula
(\ref{usefull}) together with (\ref{alfo}) and (\ref{cho}) gives
explicitly the torsion, namely \be\lb{takea} T=-\Xi_x E^2\wedge
E^3\wedge E^4+\Xi_y E^1\wedge E^2\wedge E^3. \ee  All the results of
this section are stated in the propositions of the introduction.

  The geometries presented till now are weak heterotic.
Due to conformal invariance all the elements of the conformal family
$[g]$ corresponding to (\ref{metropo2}) are also weak. Therefore in
order to find an strong geometry we should not impose $dT=0$ for the
torsion (\ref{mono}) corresponding to (\ref{metropo2}) but instead,
we must solve $d\widetilde{T}=0$ where $\widetilde{T}= T + \ast_{g}
2 d \log (\omega)$ is the torsion corresponding to the
representative $\widetilde{g}=\omega^2 g$. After some calculation we
obtain the following \emph{linear} equation defining $\omega$
\be\lb{laplo} \Delta_{g} log(\omega)=\partial_{[1}T_{234]}, \ee
where $\Delta_{g}$ is the Laplace operator for (\ref{metropo2}). The
left side (\ref{laplo}) does not depend on $\omega$ and acts as an
inhomogeneous source. Then the spaces $\widetilde{g}=\omega^2 g$
will be a strong representatives of the family if $\omega$ solve
(\ref{laplo}). The spaces (\ref{metropo2}) will be strong only in
the case that constant $\omega$ is a solution of the resulting
equations, which implies that \be\lb{cono}
\partial_{[1}T_{234]}=\Xi_{xp}-(\Xi_x \Phi_y)_y+(\Xi_x \Psi_y)_x +
 \Xi_{yq} -(\Xi_y \Phi_x)_x + (\Xi_y \Psi_x)_y=0.
\ee It also follows from the result of this section that if we deal
with a minimal weak structure, then (\ref{laplo}) is reduced to
\be\lb{Calas} \Delta_{\widehat{g}} log(\omega)=0, \ee being
$\widehat{g}$ the hyperkahler representative of the family
\cite{Callan}.

\subsection{Discussion}

   In the present work we attacked the problem of constructing
weak and strong 4-dimensional hyperkahler torsion geometries without
isometries. We have shown that the most general local form for a
weak metric is defined by two potential functions. Weak condition is
conformally invariant and the corresponding metrics are determined
up to a conformal transformation.

   We argue that the present work have practical applications,
although the system of equations for the potentials is highly
non-linear and difficult to solve. We have seen that in four
dimensions hypercomplex structures and weak torsion hyperkahler
geometries are totally equivalent concepts. Concretely speaking, we
have shown that integrability of the complex structures implies and
is implied by the other weak HKT properties. Therefore the problem
to find weak heterotic geometries is reduced to find the solutions
of an Ashtekar-Jacobson-Smolin like system without the volume
preserving condition, which has the advantage to be quadratic.

   From this discussion is clear that a
weak metric will be of the Callan-Harvey-Strominger type (i.e.
conformal to an hyperkahler one) if there is a volume form preserved
by the corresponding tetrad. Nevertheless not all the hypercomplex
structures preserves a 4-form and the metrics of references
\cite{Valent}-\cite{Tod} should be counterexamples. We have seen
also that a space is of Callan-Harvey-Strominger type if the torsion
is the dual of a gradient. Therefore both conditions should be
equivalent.

   The strong condition (i.e. the closure of the torsion form)
is not invariant under general conformal transformation.
Nevertheless the problem to find the strong representatives of a
weak heterotic family $[g]$ is reduced to solve a \emph{linear} non
homogeneous laplacian equation over an arbitrary element $g$. The
homogeneous part can be taken to zero if the metric is conformal to
a hyperkahler one.

   At first sight it looks that the "Ashtekar like" formulation for HKT is
the most convenient formulation. Instead we suggest that the other
variant could be more useful in the context dualities, in which the
presence of isometries is needed \cite{Rocek}. To make this
suggestion more concrete, let us recall that self-dual spaces are
described in terms of a Kahler potential satisfying the heavenly
equation \cite{Plebanski}. Although this equation is non-linear and
the general solution is not known, the analysis of the Killing
equations and the possible isometries allowed to classify the
hyperkahler metrics with one Killing vector \cite{Boyero}. In
particular it was found that they are divided in two types, the
first described for by a monopole equation and the second by some
limit of a Toda system \cite{Gibbons}-\cite{Boyero}. We propose that
is worthy to generalize, if is possible, the mathematical machinery
to classify the Killing equations of gravitational instantons
\cite{Boyero}-\cite{Przanowski2} to the hyperkahler torsion case.
Perhaps it will possible in this way to find the classification of
heterotic geometry with one or more Killing vectors and in
particular, to identify the subcases that match for duality
applications. This goal has not been achieved yet, and our
suggestion could be an alternative point of view of the references
\cite{Tod}-\cite{Delduc}.

\newpage

\section{D-instanton sums for matter hypermultiplets}

   We calculate some non-perturbative (D-instanton) quantum
corrections to the moduli space metric of several $(n>1)$ identical
matter hypermultiplets for the type-IIA superstrings compactified on
a Calabi-Yau threefold, near conifold singularities. We find a
non-trivial deformation of the (real) $4n$-dimensional
hypermultiplet moduli space metric due to the infinite number of
D-instantons, under the assumption of $n$ tri-holomorphic commuting
isometries of the metric, in the hyperkahler limit (i.e. in the
absence of gravitational corrections).

\subsection{Ooguri-Vafa solution}

    The Ooguri-Vafa (OV) solution \cite{ov} describes the
D-instanton corrected moduli space metric of a single matter
hypermultiplet in type-IIA superstrings compactified on a Calabi-Yau
threefold of Hodge number ${\rm dim}H^{2,1}=1$, when both $N=2$
supergravity and UH are switched off, while five-brane instantons
are suppressed. The matter hypermultiplet low-energy effective
action in that limit is given by the four-dimensional $N=2$
supersymmetric non-linear sigma-model that has the four-dimensional
OV metric in its target space.

  Rigid (or global) $N=2$ supersymmetry of the non-linear
sigma-model requires a hyperkahler metric in its target space
\cite{book}. The OV metric has a (toric) $U(1)\times U(1)$ isometry
by construction \cite{ov}. There always exist a linear combination
of two commuting abelian isometries that is tri-holomorphic, i.e. it
commutes with $N=2$ rigid supersymmetry \cite{gib}.

   Given any four-dimensional hyperkahler metric with a
tri-holomorphic isometry $\partial_t$, it can always be written down
in the standard (Gibbons-Hawking) form  \be\lb{ufa} g=
\frac{1}{V}(dt + A)^2 +V(dx^2+dy^2+dz^2), \ee that is governed by
linear equations, \be\lb{ufa2} \Delta V=\vec{\nabla}^2 V\equiv
\left( \frac{\partial^2}{\partial x^2}+ \frac{\partial^2}{\partial
y^2} +\frac{\partial^2}{\partial z^2}\right)V=0~,\quad {\rm
almost~~everywhere}, \ee and \be\lb{ufa3}
\vec{\nabla}V+\vec{\nabla}\times A=0. \ee The 1-form
$A=A_1dx+A_2dy+A_3dz$ is fixed by the monopole equation (\ref{ufa3})
in terms of the real scalar potential $V(x,y,z)$. Equation
(\ref{ufa2}) means that the function $V$ is harmonic (away from
possible isolated singularities) in three Euclidean dimensions ${\bf
R}^3$. The singularities are associated with the positions of
D-instantons.

  Given extra $U(1)$ isometry, after being rewritten in the
cylindrical coordinates ($\rho=\sqrt{x^2+y^2}$, $\theta =
\arctan(y/x)$, $\eta=z$), the hyperkahler potential
$V(\rho,\theta,\eta)$ becomes independent upon $\theta$. Equation
(\ref{ufa}) was used by Ooguri and Vafa \cite{ov} in their analysis
of the matter hypermultiplet moduli space near a conifold
singularity. The conifold singularity arises in the limit of the
vanishing CY period, \be\lb{ufa4} \int_{\cal C}\O \to 0 \ee where
the CY holomorphic (nowhere vanishing) 3-form $\O$ is integrated
over a non-trivial 3-cycle ${\cal C}$ of CY. The powerful
singularity theory \cite{sin} can then be applied to study the
universal behaviour of the hypermultiplet moduli space near the
conifold limit, by resolving the singularity.

In the context of the CY compactification of type IIA superstrings,
the coordinate $\rho$ represents the `size' of the CY cycle ${\cal
C}$ or, equivalently, the action of the D-instanton originating from
the Euclidean D2-brane wrapped about the cycle ${\cal C}$. The
physical interpretation of the $\eta$ coordinate is just the
expectation value of another (RR-type) hypermultiplet scalar. The
cycle ${\cal C}$ can be replaced by a sphere $S^3$ for our purposes,
since the D2-branes only probe the overall size of ${\cal C}$.

   The OV potential $V$ is periodic in the RR-coordinate $\eta$ since
the D-brane charges are quantized \cite{bbs}. We normalize the
corresponding period to be $1$, as in reference \cite{ov}. The
Euclidean D2-branes wrapped $m$ times around the sphere $S^3$ couple
to the RR  expectation value on $S^3$ and thus should produce
additive contributions to $V$, with the factor of $\exp(2\pi i
m\eta)$ each.

In the classical hyperkahler limit, when both N=2 supergravity and
all the D-instanton contributions are suppressed, the potential
$V(\rho,\eta)$ of a single matter hypermultiplet cannot depend upon
$\eta$ since there is no perturbative superstring state with a
non-vanishing RR charge. Accordingly, the classical pre-potential
$V(\rho)$ can only be the Green function of the two-dimensional
Laplace operator, i.e. \be\lb{ufa5} V_{\rm classical} =
-\frac{1}{2\pi}\log\rho + {\rm const.} \ee whose normalization is
also in agreement with the appearing in \cite{ov}.

The calculation given in \cite{ov} to determine the exact
D-instanton contributions to the hyperkahler potential $V$ is based
on the idea \cite{bbs} that the D-instantons should resolve the
singularity of the classical hypermultiplet moduli space metric at
$\rho=0$. A similar situation arises in the standard
(Seiberg-Witten) theory of a quantized N=2 vector multiplet (see
reference\cite{book} for a review).

Equation (\ref{ufa2}) formally defines the electrostatic potential
$V$ of electric charges of unit charge in the Euclidean upper
half-plane $(\rho,\eta)$, $\rho>0$, which are distributed along the
axis $\rho=0$ in each point $\eta=n\in {\bf Z}$, while there are no
two charges at the same point \cite{ov}. A solution to (\ref{ufa2})
obeying all these conditions is unique, \be\lb{ufa6} V_{\rm
OV}(\rho,\eta)= \frac{1}{4\pi} \sum^{+\infty}_{n=-\infty}(
\frac{1}{\sqrt{\rho^2+ (\eta-n)^2}}-\frac{1}{|n|})+ {\rm const.} \ee
After Poisson summation (\ref{ufa6}) takes the desired form of
singularity resolution \cite{ov} \be\lb{ufa7} V_{\rm
OV}(\rho,\eta)=\frac{1}{4\pi} \log(\frac{\mu^2}{\rho^2})+
\sum_{m\neq 0}\frac{1}{2\pi}e^{2\pi i m\eta}\,K_0(2\pi |m|\rho), \ee
where the modified Bessel function $K_0$ of the 3rd kind has been
introduced, \be\lb{ufa8}
K_s(z)=\frac{1}{2}\int^{+\infty}_0\frac{dt}{t^{s-1}}\exp[
-\,\frac{z}{2}(t+\frac{1}{t})] \ee valid for all Re\,$z>0$ and
Re\,$s>0$, while $\mu$ is a constant (modulus).

Inserting the standard asymptotical expansion of the Bessel function
$K_0$ near $\rho=\infty$ into (\ref{ufa7}) yields \cite{ov}
$$
 V_{\rm OV}(\rho,\eta)=\frac{1}{4\pi}
\log(\frac{\mu^2}{\rho^2}) + \sum_{m=1}^{\infty} exp(-2\pi m\rho)
\cos(2\pi m \eta)
$$
\be\lb{ufa9} \times
\sum_{n=0}^{\infty}\frac{\Gamma(n+\frac{1}{2})}{\sqrt{\pi}n!
\G(-n+\frac{1}{2})}(\frac{1}{4\pi m \rho})^{n+\frac{1}{2}} \ee

The string coupling constant $g$ can be easily reintroduced into
(\ref{ufa9}) by a substitution $\rho\rightarrow\rho/g$. The factors
of $\exp{(-2\pi m\rho/g)}$ in (\ref{ufa9}) are the contributions due
to the multiple D-instantons \cite{ov}.

The OV potential (\ref{ufa6}) is given by a (regularized) T-sum over
the T-duality transformations, $\eta\rightarrow \eta+1$, being
applied to the fundamental solution $V_0\equiv\frac{1}{4\pi r}\equiv
\frac{1}{4\pi\sqrt{\rho^2+\eta^2}}$ of (\ref{ufa2}), \be\lb{ufa10}
V_{\rm OV}(\rho,\eta)=A + \sum_{\rm T}V_0(\rho,\eta)= A + \sum_{\rm
T} \frac{1}{4\pi\sqrt{\rho^2+\eta^2}} \ee where $A$ is a constant.
The fundamental solution $V_0(\rho,\eta)$ is just the Green function
of the three-dimensional Laplace operator $\Delta$ in (\ref{ufa2}).

\subsection{Pedersen-Poon Ansatz}

   The natural generalization of the Gibbons-Hawking
Ansatz (\ref{ufa}) to higher-dimensional toric hyperkahler spaces
are the {\it Pedersen-Poon} (PP) metrics (\ref{gengibbhawk}).
Namely, given $n$ commuting tri-holomorphic isometries of a (real)
$4n$-dimensional hyperkahler space, there exists a coordinate system
$(x^i_a,t_i)$, with $i,j=1,2,\ldots,n$ and $a,b,c=1,2,3$, where the
hyperkahler metric takes the form \cite{Poon} \be\lb{ufa11} g=
U_{ij}dx^i\cdot dx^j+ U^{ij}(dt_i+A_i)(dt_j +A_j) \ee Here the dot
means summation over the $a$-type indices, all metric components are
supposed to be independent upon all $t_i$ (thus reflecting the
existence of $n$ isometries), $U^{ij}=(U_{ij})^{-1}$,  while
$U_i=(U_{i1},\ldots,U_{in})$ and $A_i$ are to be solutions to the
generalized monopole (or BPS) equations
$$
F_{x_{\mu}^i
x_{\nu}^j}=\epsilon_{\mu\nu\lambda}\nabla_{x_{\lambda}^i}U_j,
$$
\be\lb{ufa12} \nabla_{x_{\lambda}^i}U_j=\nabla_{x_{\lambda}^j}U_i,
\ee
$$
U_i=(U_{i1},.., U_{in}),
$$
where the field strength $F$ of the gauge fields $A$ has been
introduced. The PP metric (11) can be completely specified by its
real PP-potential $P(x, w, \bar{w})$ that generically depends upon
$3n$ variables, \be\lb{ufa13}
 x^j=x^j_3~,\qquad w^j=\frac{x^j_1 + i x^j_2}{2},\qquad
\bar{w}^j=\frac{x^j_1-ix^j_2}{2}, \ee because the BPS equations
(\ref{ufa12}) allow a solution \cite{Anguelova}\be\lb{ufa14}
U_{ij}=P_{x^ix^j}\quad {\rm and}\quad
A_j=i\left(P_{w^kx^j}dw^k-P_{\bar{w}^kx^j}d\bar{w}^k\right), \ee
provided $P$ itself obeys (almost everywhere) a linear Laplace-like
equation \cite{Poon} \be\lb{ufa15} P_{x^ix^j} + P_{w^i\bar{w}^j}=0.
\ee The subscripts of $P$ denote partial differentiation with
respect to the given variables.

  Equation (\ref{ufa15}) is apparently the multi-dimensional
hyperkahler generalization of (\ref{ufa2}). The existence of the
PP-potential $F$ essentially amounts to integrability (or
linearization) of the BPS equations (\ref{ufa12}), because the
master equation (\ref{ufa15}) is linear, whose solutions are not
difficult to find. For instance, a general solution to (\ref{ufa15})
can be written down as the contour $(C)$ integral of an arbitrary
potential $G(\eta^j(\xi),\xi)$  \cite{Poon}, \be\lb{16} P= {\rm Re}
\oint_C \frac{dx}{2\pi i \xi}G(\eta^j(\xi),\xi)~,\quad {\rm where}
\quad \eta^j(\xi) = \bar{w}^j+x^j\xi-w^j\xi^2. \ee

\subsection{D-instanton sums}

   Our technical assumptions are essentially the same as in
reference \cite{ov}, namely,
\begin{itemize}
\item periodicity in all $x^i$ with period $1$, due to the D-brane
charge quantization, \item the classical potential $F$ near a CY
conifold singularity should have a logarithmic behaviour (elliptic
fibration), being independent upon all $x^i$ (when all
$w^i\to\infty),$ \item the classical singularity of the metric
should be removable, i.e. an exact metric should be complete (or
non-singular), \item the metric should be symmetric under the
permutation group of $n$ sets of hypermultipet coordinates
$(x^j,w^j,\bar{w}^j)$, where $j=1,2,\ldots,n$.
\end{itemize}
The last assumption means that we only consider {\it identical}
matter hypermultiplets. Together with our main assumption about $n$
tri-holomorphic isometries (see e.g., our Abstract and sect. 3), it
is going to lead us to
 an explicit solution.

   We first confine ourselves to the case of {\it two} identical
matter hypermultiplets $(n=2)$, and then quote our result for an
arbitrary $n>2$. The problem amounts to finding a solution to the
master equations (\ref{ufa15}) subject to the conditions above. The
known OV solution $V_{\rm OV}(w,x)\equiv V_{\rm OV}(x,w)$, defined
by equation(\ref{ufa6}) or (\ref{ufa7}), can be used to introduce
another function $P(x,w)$ as a solution to two equations
\be\lb{ufa17} P_{xx}=-P_{w\bar{w}}=V_{\rm OV}(x,w). \ee We do not
need an explicit solution to equation (\ref{ufa17}) in what follows.
Since equations (\ref{ufa15}) are linear, the superposition
principle applies to their solutions. It is now straightforward to
verify that there exist a `trivial' solution to equations
(\ref{ufa15}), given by a PP potential \be\lb{ufa18} P_0= c_0
\left[P(x_1,w_1) + P(x_2,w_2)\right] \ee where $c_0$ is a real
constant. A particular non-trivial solution in the $n=2$ case is
given by \be\lb{ufa19} P_{\rm mixed} = c_{+} P(x_1+x_2,w_1+w_2) +
c_{-}P(x_1-x_2,w_1-w_2) \ee where $c_{\pm}$ are two other real
constants. Though the linear partial differential equation
(\ref{ufa15}) has many other solutions, we argue at the end of this
section that the most general solution, satisfying all our
requirements, is actually given by \be\lb{ufa20} P=P_0+P_{\rm
mixed}. \ee To this end, we continue with the particular solution
(\ref{ufa20}). To write down our result for the hyperkahler moduli
space metric of two matter hypermultiplets, we merely need the
second derivatives of the PP potential, because of (\ref{ufa14}).
Using equations (\ref{ufa6}), (\ref{ufa14}), (\ref{ufa17}),
(\ref{ufa18}), (\ref{ufa19}) and
 (\ref{ufa20}), we find
$$
4 \pi U_{11}= c_ +
\sum^{+\infty}_{n=-\infty}\left(\frac{1}{\sqrt{(x_1+x_2-n)^2
+(w_1+w_2)(\bar{w}_1+\bar{w}_2)/\lambda^2}}-\frac{1}{n}\right)
$$
$$
+c_ - \sum^{+\infty}_{n=-\infty}\left(\frac{1}{\sqrt{(x_1-x_2-n)^2 +
(w_1-w_2)(\bar{w}_1-\bar{w}_2)/\lambda^2}}-\frac{1}{n}\right)
$$
\be\lb{ufa21} + c_0 \sum^{+\infty}_{n=-\infty}\left(\frac{1}{\sqrt{
(x_1-n)^2 + w_1\bar{w}_1/\lambda^2}}-\frac{1}{n}\right) \ee with a
modulus parameter $\lambda$. The component $U_{22}$ has the the form
as that of (\ref{ufa21}), but the indices $1$ and $2$ have to be
exchanged. The remaining two components of the matrix $U$ in
(\ref{ufa11}) are given by
$$
4 \pi U_{12}=4 \pi U_{21} = c_ + \sum^{+\infty}_{n=-\infty}\left(
\frac{1}{\sqrt{(x_1+x_2-n)^2
+(w_1+w_2)(\bar{w}_1+\bar{w}_2)/\lambda^2}}- \frac{1}{n}\right)
$$
\be\lb{ufa22} + c_-\sum^{+\infty}_{n=-\infty}\left(\frac{1}{\sqrt{
(x_1-x_2-n)^2
+(w_1-w_2)(\bar{w}_1-\bar{w}_2)/\lambda^2}}-\frac{1}{n}\right). \ee
The $A$-field components in the PP metric (\ref{ufa11}) are given by
the second (\ref{ufa14}).

   The physical interpretation of our solution as the infinite
D-instanton sum is evident after rewriting it as an asymptotical
expansion like that of (\ref{ufa9}), by using(\ref{ufa7}) and
(\ref{ufa8}), now being applied to the two-hypermultiplet metric
components (\ref{ufa21}) and (\ref{ufa22}). In particular, their
classical behavior is given by \be\lb{ufa23} U_{11}\sim
\frac{c_+}{4\pi}\ln\frac{\lambda^2}{|w_1+w_2|^2}+
\frac{c_-}{4\pi}\ln\frac{\lambda^2}{|w_1-w_2|^2}+
\frac{c_0}{4\pi}\ln\frac{\lambda^2}{|w_1|^2} \ee and similarly for
$U_{22}$ after exchanging the indices $1$ and $2$, and $U_{12}$
after dropping the last term in (\ref{ufa23}).

As regards the case of an arbitrary $n>2$, the easiest picture
arises in terms of the PP potential $P$ --- see sect.~3. We define
this potential as a linear
 combination of $P(x_1+x_2+\ldots +x_n, w_1+w_2+\ldots +w_n)$,
$P(-x_1+x_2+\ldots +x_n, -w_1+w_2+\ldots +w_n)$, $\ldots$,
$P(-x_1-x_2+\ldots -x_n, -w_1-w_2-\ldots -w_n)$, $P(x_1,w_1)$,
$P(x_2,w_2)$, $\ldots$ and $P(x_n,w_n)$, each one being constructed
out of the OV potential, as in (\ref{ufa17}). Next, we define the
metric components of the matrix $U_{ij}$ in (\ref{ufa11}) by the
second derivatives of the total PP potential $F$, as in
(\ref{ufa14}). The rest is fully straightforward, {\it cf.}
equations (\ref{ufa21}) and (\ref{ufa22}).

        Finally, some comments about the uniqueness of our
solution (\ref{ufa20}) are in order. First, we notice that all
functions $U^{ij}$ also obey (\ref{ufa15}) by our construction
(\ref{ufa14}). Given any other non-trivial solution, it should lead
(after Poisson resummation) to {\it the same} classical behaviour
(\ref{ufa23}), while it should also reduce to the OV solution
(sect.~2) for a single hypermultiplet. According to the Poisson
resummation formula, \be\lb{ufa24}\sum^{\infty}_{n=0}f(n) =
\sum^{\infty}_{n=0}\tilde{f}(n)~,\quad{\rm where} \quad
\tilde{f}(u)=\int^{+\infty}_{-\infty}dvf(v)e^{2\pi i u v} \ee the
logarithmic terms can only come from the $n=0$ term in the sum, e.g.
when using (\ref{ufa6}), choosing $n=v$ in the Fourier integral
(\ref{ufa24}), and then taking $u=0$ after the Fourier transform.

 Hence, a difference between ours and any other solution to the
potentials $F$ or $U^{ij}$ cannot contribute to equation
(\ref{ufa23}). Since equation (\ref{ufa23}) is $x$-independent, we
first investigated the class of $x$-independent real potentials $F$,
without any reference to the metric (\ref{ufa11}). By solving
(\ref{ufa15}) in this case we found an unique solution subject to
our symmetries, namely, the one given by (\ref{ufa23}). Hence, the
anticipated uniqueness of our solution boils down to the uniqueness
of the periodic (of period $1$) extension of a given $w$-dependent
solution to a solution depending upon both $w$ and $x$, and having a
generic form $\sum_n G(x+n,w)$ with some basic function $G$. It
seems to be quite plausible that the square root in (\ref{ufa21}) is
the {\it only} basic function to yield the logarithmic terms out of
the $n=0$ term in the D-instanton sum.

\subsection{Discussion}

In our final results (\ref{ufa21}) and (\ref{ufa22}) the quantum
corrections to the classical (logarithmic) terms (\ref{ufa23}) are
exponentially suppressed by the D-instanton factors (in the
semiclassical description). Unlike the case of a single matter
hypermultiplet, the multi-hypermultiplet moduli space metric is also
sensitive to the phases of $w_i$ (i.e. not only to their absolute
values). There are no perturbative (type IIA superstring)
corrections, while all D-instanton numbers contribute to the
solution.

   Our results can also be applied to an explicit construction of
{\it quaternionic} metrics in real $4(n-1)$ dimensions out of known
hyperkahler metrics in real $4n$ dimensions, which is of particular
importance to a description of D-instantons in N=2 supergravity
\cite{Swann}. For instance, the four-dimensional quaternionic
manifolds with toric isometry can be fully classified
\cite{Pedersen} in terms of the PP-potential $F$ obeying equation
(\ref{ufa15}).

  It would be also interesting to connect our results to
perturbative superstring calculations in the D-instanton background,
which is still a largely unsolved problem.

\newpage

\section{Hyperkahler spaces with local triholomorphic $U(1)\times U(1)$ isometry
and supergravity solutions}

    In this section some examples of toric hyperkahler metrics in
eight dimensions are constructed. The Swann construction of
hyperkahler metrics is applied to the Calderbank Pedersen spaces in
order to find hyperkahler examples with $U(1)\times U(1)$ isometry.
The connection with the Pedersen-Poon toric hyperkahler metrics is
explained by finding the momentum map coordinate system. This
hyperkahler examples are lifted to solutions of the D=11
supergravity with and without fluxes. Type IIA and IIB backgrounds
are found by use of dualities.

\subsection{Toric hyperkahler metrics of the Swann type}

    The Swann metric (\ref{abuela}) is in general quaternion
Kahler. In the special case with $b=0$ corresponds to an hyperkahler
metric with explicit form \be\lb{abuela2}
g=|u|^2\overline{g}+[(du_0-x_i\omega_{-}^i)^2 +(du_i+
u_0\omega_{-}^i + \epsilon_{ijk}u_k\omega_{-}^k)^2]. \ee By use of
(\ref{abuela2}) it is direct to extend the Calderbank-Pedersen
metrics (\ref{metric}) to an hyperkahler one. The quaternionic base
is (\ref{metric}) and the corresponding $\omega_{-}^i$ are
\cite{Pedersen} \be\lb{threefor}
\omega_{-}^1=\frac{1}{F}((\frac{1}{2}F +\rho
F_{\rho})\frac{d\eta}{\rho}-\rho
F_{\eta}\frac{d\eta}{\rho}),\;\;\;\;
\omega_{-}^2=-\frac{\alpha}{F},\;\;\;\;\;
\omega_{-}^3=\frac{\beta}{F}. \ee The resulting expression for the
hyperkahler metric is
$$
g=[(du_0-x_i\omega_{-}^i)^2 +(du_i+ u_0\omega_{-}^i +
\epsilon_{ijk}u_k\omega_{-}^k)^2]+|u|^2
[\frac{F^2-4\rho^2(F^2_{\rho}+F^2_{\eta})}{4F^2}(\frac{d\rho^2+d\eta^2}{\rho^2})
$$
\be\lb{merenge} +\frac{[(F-2\rho F_{\rho})\alpha-2\rho
F_{\eta}\beta]^2+[(F+2\rho F_{\rho})\beta - 2\rho F_{\eta}\alpha
]^2}{F^2[F^2-4\rho^2(F^2_{\rho}+F^2_{\eta})]}]. \ee The
Calderbank-Pedersen metrics (\ref{metric}) have two commuting
Killing vectors $\partial/\partial\theta$ and
$\partial/\partial\varphi$. Clearly this are also Killing vectors of
the spaces (\ref{merenge}).

   The Kahler triplet corresponding to (\ref{metric}) is
$$
\overline{J}^1=\frac{1}{F^2}(F^2-4\rho^2(F^2_{\rho} +
F^2_{\eta}))(\frac{d\rho\wedge d\eta}{\rho^2}+ \alpha\wedge\beta),
$$
\be\lb{hyperform} \overline{J}^2=\frac{1}{F^2}(\rho F_{\eta}\beta
+(\rho F_{\rho}-\frac{1}{2}F)\alpha )\wedge \frac{d\rho}{\rho}, \ee
$$
\overline{J}^3=\frac{1}{F^2}(\rho F_{\eta}\alpha-(\rho
F_{\rho}+\frac{1}{2}F)\beta )\wedge \frac{d\eta}{\rho}.
$$
Then the hyperkahler triplet $\Omega^i$ of (\ref{abuela2}) can be
expressed in compact form as \be\lb{quak} \Omega= u\overline{J}
\overline{u}+ (du + \omega_{-} u)\wedge \overline{(du + \omega_{-}
u)}, \ee where we have defined
$$
\overline{J}=\overline{J}^1 I + \overline{J}^2 J + \overline{J}^3
K,\qquad \Omega=\Omega^1 I + \Omega^2 J + \Omega^3 K
$$
$$
\omega_{-}=\omega_{-}^1 I +\omega_{-}^2 J + \omega_{-}^3 K.
$$
A direct calculation using (\ref{hyperform}) show us that
\be\lb{consi}  {\cal L}_{\frac{\partial}{\partial\varphi}}
\Omega^i=0,\;\;\;  {\cal L}_{\frac{\partial}{\partial\theta}}
\Omega^i=0. \ee This mean that $\frac{\partial}{\partial\varphi}$
and $\frac{\partial}{\partial\theta}$ are triholomorphic and
therefore (\ref{merenge}) is a toric hyperkahler metric of the type
of the Proposition 9, where we should idenfity $t_1=\theta$ and
$t_2=\varphi$.

   For applications to supergravity we will write the metric
(\ref{merenge}) in the form (\ref{gengibbhawk}). Therefore we should
find the momentum maps
$$
dx_{\theta}^k=i_{\theta}\overline{J}^k,\;\;\;\;dx_{\phi}^k=i_{\phi}\overline{J}^k
$$
for (\ref{abuela2}). The contraction of $\partial/\partial\theta$
with the hyperkahler form (\ref{quak}) gives
$$
dx_{\theta}^{1}=\frac{1}{\sqrt{\rho}F}(2q_0dq_2+2q_2dq_0-2q_1dq_3-2q_3dq_1
-(\frac{1}{2\rho}+\frac{F_{\rho}}{F})d\rho-\frac{F_{\eta}}{F}d\eta),
$$
$$
dx_{\theta}^{2}=\frac{1}{\sqrt{\rho}F}(2q_2dq_3+2q_3dq_2-2q_0dq_1-2q_1dq_0
-(\frac{1}{2\rho}+\frac{F_{\rho}}{F})d\rho-\frac{F_{\eta}}{F}d\eta),
$$
$$
dx_{\theta}^{3}=\frac{1}{\sqrt{\rho}F}(2q_0dq_0+2q_1dq_1-2q_2dq_2+2q_3dq_3
-(\frac{1}{2\rho}+\frac{F_{\rho}}{F})d\rho-\frac{F_{\eta}}{F}d\eta).
$$
The last expressions can be integrated to obtain \be\lb{momentumap}
x_{\theta}^{1}=\frac{2(q_{0}q_{2} +
q_{1}q_{3})}{\sqrt{\rho}F},\;\;\; x_{\theta}^{2}=\frac{2(q_{2}q_{3}
- q_{0}q_{1})}{\sqrt{\rho}F},\;\;\; x_{\theta}^{3}=\frac{q_{0}^2 -
q_{1}^2 - q_{2}^2 + q_{3}^2}{\sqrt{\rho}F}. \ee Similarly for
$\partial/\partial \varphi$ it is found
$$
x_{\varphi}^{1}=\eta x_{\theta}^{1}+\frac{2\sqrt{\rho}(
q_{1}q_{2}-q_{0}q_{3} )}{F},\;\;\;\;\; x_{\varphi}^{2}=\eta
x_{\theta}^{2}+\frac{\sqrt{\rho}(q_{0}^2 - q_{1}^2 + q_{2}^2 -
q_{3}^2)}{F},
$$
\be\lb{momentumap2} x_{\varphi}^{3}=\eta
x_{\theta}^{3}+\frac{2\sqrt{\rho}(q_{0}q_{1} + q_{2}q_{3})}{F}, \ee
in accordance with \cite{Pedersen}.

     The next step is to determine the matrix $U_{ij}$ for (\ref{explico}).
This is easily found by noticing that from (\ref{gengibbhawk}) it
follows that \be\lb{conditions}
U^{ij}=\overline{g}(\frac{\partial}{\partial t^i},
\frac{\partial}{\partial t^j}),
\;\;\;U^{ij}A_{j}=\overline{g}(\frac{\partial}{\partial t^i},\cdot).
\ee Then introducing the expression for (\ref{merenge}) into the
first (\ref{conditions}) gives \be\lb{solmonop}
U^{ij}=\frac{|q|^2}{F(\frac{1}{4}F^2-\rho^2
(F_{\rho}^2+F_{\eta}^2))}\left(\begin{array}{cc}
  \frac{1}{2}F-\rho F_{\rho} & -\rho F_{\eta} \\
  -\rho F_{\eta} & \frac{1}{2}F+\rho F_{\rho}
\end{array}\right),
\ee with inverse \be\lb{impi}
U_{ij}=\frac{F}{|q|^2}\left(\begin{array}{cc}
  \frac{1}{2}F+\rho F_{\rho} & \rho F_{\eta} \\
  \rho F_{\eta} & \frac{1}{2}F-\rho F_{\rho}
\end{array}\right).
\ee To find $A_i$ one should obtain from (\ref{merenge}) and
(\ref{metric}) that
$$
\overline{g}(\frac{\partial}{\partial
\theta},\cdot)=\frac{1}{F}[(q_2\sqrt{\rho} +
q_3\frac{\eta}{\sqrt{\rho}})(2q_1 \widetilde{A}^1-dq_0) +
2(q_2\frac{\eta}{\sqrt{\rho}}+ q_3\sqrt{\rho})(dq_1 + q_0
\widetilde{A}^1) -
2(q_0\sqrt{\rho}+q_1\frac{\eta}{\sqrt{\rho}})(dq_2 + q_3
\widetilde{A}^1)
$$
$$
+ 2( q_0\frac{\eta}{\sqrt{\rho}}-q_1\sqrt{\rho})(dq_3-q_2
\widetilde{A}^1)],
$$
$$
\overline{g}(\frac{\partial}{\partial\varphi},\cdot)=\frac{1}{F}[\frac{q_3}{\sqrt{\rho}}(2q_1
\widetilde{A}^1-dq_0) + \frac{2q_2}{\sqrt{\rho}}(dq_1 + q_0
\widetilde{A}^1) - \frac{2q_1}{\sqrt{\rho}}(dq_2 + q_3
\widetilde{A}^1) + \frac{2q_0}{\sqrt{\rho}}(dq_3-q_2
\widetilde{A}^1)],
$$
where $\widetilde{A}^1$ is given in (\ref{threefor}) in terms of
$F$. Then from the second (\ref{conditions}) it is obtained
\be\lb{unoformas} A_1=\frac{F}{|q|^2}[ (\frac{F}{2}+\rho F_{\rho})
\overline{g}(\frac{\partial}{\partial \theta},\cdot) +\rho F_{\eta}
\overline{g}(\frac{\partial}{\partial\varphi},\cdot) ], \ee
\be\lb{unoformas2} A_2=\frac{F}{|q|^2}[ \rho
F_{\eta}\overline{g}(\frac{\partial}{\partial \theta},\cdot) +
(\frac{F}{2}-\rho
F_{\rho})\overline{g}(\frac{\partial}{\partial\varphi},\cdot) ]. \ee
Therefore we have reduced (\ref{merenge}) with this data to the form
(\ref{gengibbhawk}). Formulas (\ref{impi}), (\ref{unoformas}) and
(\ref{unoformas2}) define a class of solutions of the Pedersen Poon
equations (\ref{genmonop}) and an hyperkahler metric
(\ref{gengibbhawk}) simply described in terms of an unknown function
$F$ satisfying (\ref{backly}). We must recall however that this
simplicity is just apparent because $U_{ij}$ depends explicitly on
the coordinates $(\rho, \eta, |q|^2)$, which depends implicitly on
the momentum maps $(x_{\varphi}^i, x_{\theta}^i)$ by
(\ref{momentumap}) and (\ref{momentumap2}). Therefore $U_{ij}$ is
given only as an implicit function of the momentum maps.

     For physical applications it is important to find solutions
which in this limit tends to \be\lb{infinito} \overline{g}=
U^{\infty}_{ij}dx^i\cdot dx^j + U^{ij}_{\infty}dt_i dt_j, \ee for a
constant invertible matrix $U^{\infty}_{ij}$ \cite{Gibbin}. Formulas
(\ref{momentumap}) and (\ref{momentumap2}) shows that the asymptotic
limit $x_{\theta}\rightarrow \infty$ or $x_{\varphi}\rightarrow
\infty$ corresponds to $q\rightarrow \infty$ or $\sqrt{\rho} F
\rightarrow 0$. In consequence from (\ref{impi}) and
(\ref{solmonop}) it follows that $U^{ij}\rightarrow \infty$ and
$U_{ij}\rightarrow 0$ asymptotically, which is not the desired
result. This problem can be evaded defining a new metric
(\ref{gengibbhawk}) with \be\lb{evade} \overline{U}_{ij}=U_{ij} +
U_{ij}^{\infty}, \ee and with the same one-forms (\ref{unoformas})
and (\ref{unoformas2}) and coordinate system (\ref{momentumap}) and
(\ref{momentumap2}). Clearly to add this constant do not affect the
solution and this data is again a solution of the Pedersen-Poon
equation (\ref{genmonop}) for which $\overline{U}_{ij}\rightarrow
U_{ij}^{\infty}$ and $\overline{U}^{ij}\rightarrow U^{ij}_{\infty}$.
Explicitly we have \be\lb{evade2}
\overline{U}_{ij}=\frac{F}{|q|^2}\left(\begin{array}{cc}
  \frac{1}{2}F+\rho F_{\rho}+\frac{U_{11}^{\infty}|q|^2}{F} & \rho F_{\eta}+\frac{U_{12}^{\infty}|q|^2}{F} \\
  \rho F_{\eta}+\frac{U_{12}^{\infty}|q|^2}{F} & \frac{1}{2}F-\rho F_{\rho}+\frac{U_{22}^{\infty}|q|^2}{F}
\end{array}\right),
\ee with inverse \be\lb{evade3}
\overline{U}^{ij}=\frac{1}{\\det(\overline{U}_{ij})}\left(\begin{array}{cc}
  \frac{1}{2}F-\rho F_{\rho}+\frac{U_{22}^{\infty}|q|^2}{F} & -\rho F_{\eta}-\frac{U_{12}^{\infty}|q|^2}{F} \\
  -\rho F_{\eta}-\frac{U_{12}^{\infty}|q|^2}{F} & \frac{1}{2}F+\rho F_{\rho}+\frac{U_{11}^{\infty}|q|^2}{F}
\end{array}\right).
\ee This modified metric is more suitable for physical purposes, but
do not correspond to a Calderbank-Pedersen base and this show that
the converse of the Swann theorem is not necessarily true.

\subsection{Supergravity solutions related to
hyperkahler manifolds}

         The hyperkahler spaces defined by (\ref{momentumap}),
(\ref{momentumap2}), (\ref{impi}), (\ref{unoformas}) and
(\ref{unoformas2}) can be extended to 11-dimensional supergravity
solutions and to IIA and IIB backgrounds by use of dualities. This
section present them following mainly \cite{Gibbin}, more details
can be found there and in references therein.

   We consider $D=11$ supergravity solutions with vanishing fermion fields and
$F_{\mu\nu\alpha\beta}$. Such solutions are of the form
\be\lb{11sugra} g=g(E^{2,1})+U_{ij}dx^i\cdot dx^j+
U^{ij}(dt_i+A_i)(dt_j+A_j), \ee that is, a direct sum of the flat
metric on $R^3$ and an hyperkahler space in eight dimensions. If all
the fields in 11 dimensional supergravity are invariant under the
action of an $U(1)$ Killing vector then the 11-dimensional
supergravity can be reduced to the type IIA one along the isometry.
In order to make such reduction we should write the 11-dimensional
metric (\ref{11sugra}) as \cite{Campbell} \be\lb{KKanzatz}
g=e^{-\frac{2}{3}\varphi(x)}g_{\mu\nu}(x)dx^{\mu}dx^{\nu}
+e^{\frac{4}{3}\varphi(x)} (dy + C_{\mu}(x)dx^{\mu})^2 , \ee
\be\lb{theform} A_{11}=A(x) + B(x)\wedge dy . \ee The field $A_{11}$
is the 3-form potential and $x^{\mu}$ are the coordinates of the
D=10 spacetime. The $NS \otimes NS$ sector of the IIA supergravity
is $(\phi, g_{\mu\nu}, B_{\mu\nu})$ and the $R \otimes R$ sector is
$(C_{\mu}, A_{\mu\nu\rho})$. After reduction of (\ref{11sugra})
along the isometry $t_1=\varphi$ we find that the the non-vanishing
fields are \be\lb{10sugra} g_{10}=(\frac{U_{11}}{\det
U})^2[g(E^{2,1})+U_{ij}dx^i\cdot dx^j] +(\frac{1}{U_{11}\det
U})^{1/2}(d\theta + A_{1})^2 , \ee \be\lb{dilaton}
\phi=\frac{3}{4}\log(U_{11})-\frac{3}{4}\log(\det U) , \ee \be\lb{C}
C=A_2-\frac{U_{12}}{U_{11}}(d\theta + A_{1}). \ee The Killing spinor
of the 11 dimensional supergravity was also independent of $t_1$ and
the reduced spinor is again a IIA Killing spinor.

   The field $\phi$ is independent of $\theta$ and $C$ satisfies
$\pounds_{k} C=0$. Then one can use T-duality rules to construct a
IIB supergravity solution \cite{Bergshoeff} \be\lb{Tdual} g=
[g_{mn}-g_{\theta\theta}^{-1}(g_{m\theta}g_{n\theta}-B_{m\theta}B_{n\theta})]dx^m
dx^n +2g_{\theta\theta}^{-1}B_{\theta n}d\theta
dx^{n}+g^{-1}_{\theta\theta}d\theta^2 \ee \be\lb{noseque}
\widetilde{B}=\frac{1}{2}dx^m \wedge dx^n [B_{mn}+
2g_{\theta\theta}^{-1}(g_{m\theta}B_{n\theta})] +
g_{\theta\theta}^{-1}g_{\theta m}d\theta \wedge dx^m \ee
\be\lb{modilat} \widetilde{\phi}=\phi-\log(g_{\theta\theta}) \ee
where the tilde indicates the transformed fields. The restrictions
$$
B=0,\;\;\; i_{k}A=0,
$$
gives the T-dual fields
$$
\textit{l}=C_{\theta}
$$
\be\lb{restrict} B'=[C_{mn}-
(g_{\theta\theta})^{-1}C_{\theta}g_{\theta m}]dx^{m} \wedge d\theta
, \ee
$$
i_{k}D=A ,
$$
where $\textit{l}$ is the IIB pseudoscalar, $B'$ is the
Ramond-Ramond 2-form potential and D is the IIB 4-form potential.
The non vanishing IIB fields resulting from the application of the
T-duality are \be\lb{obtained} g_{10}= (\det U)^{3/4}[(\det
U)^{-1}g(E^{2,1})+(\det U)^{-1} U_{ij}dx^i\cdot dx^j + d\theta^2] ,
\ee \be\lb{B} B_i=A_i\wedge d\theta , \ee \be\lb{tau}
\tau=-\frac{U_{12}}{U_{11}}+i\frac{\sqrt{\det(U)}}{U_{11}} , \ee
where
$$
\tau=\textit{l}+i e^{-\phi_B},\;\;\; B_{1}=B,\;\;\; B_{2}=B' ,
$$
and $g_{10}$ is the Einstein frame metric satisfying
$$
g_{10}=e^{-\phi_B/2}g_{IIB}.
$$
The examples that we have presented till know are obtained by use of
the isometries of the internal spaces. But more backgrounds can be
obtained by reducing (\ref{11sugra}) along one of the space
directions $E^{2,1}$ and it is obtained the IIA solution
\be\lb{flatred} g=g(E^{1,1})+U_{ij}dx^i\cdot dx^j+
U^{ij}(dt_i+A_i)(dt_j+A_j), \ee with the other fields equal to zero.
After T-dualizing in both angular directions it is obtained
$$
g=g(E^{1,1}) + U_{ij} dX^i \cdot dX^j ,
$$
\be\lb{termi} B=A_i \wedge dt_i, \ee
$$
\phi=\frac{1}{2}\log(\det U) ,
$$
where $X^i={x^i, t^i}$. This solution can be lifted to a D=11
supergravity solution \be\lb{extun} g_{11}=(\det U)^{2/3}[(\det
U)^{-1}g(E^{1,1})+(\det U)^{-1} U_{ij}dX^i\cdot dX^j + d\theta^2],
\ee \be\lb{extun2} F=F_i\wedge dt_i \wedge d\theta. \ee It is
possible to generalize (\ref{11sugra}) to include a non vanishing
4-form $F$. The result is the membrane solution \be\lb{11sugra2}
g=H^{-2/3}g(E^{2,1})+H^{1/3}[U_{ij}dx^i\cdot dx^j+
U^{ij}(dt_i+A_i)(dt_j+A_j)], \ee \be\lb{11sugraF}
F=\pm\omega(E^{2,1})\wedge dH^{-1}, \ee where $H$ is an harmonic
function on the hyperkahler manifold, i.e, satisfies
$$
U^{ij}\partial_i \cdot \partial_j H=0.
$$
We have seen in the last section that every entry of $U_{ij}$ is an
harmonic function and so such $H$ can be generated with an
hyperbolic eigenfunction F. After reduction along $\varphi$ it is
found the following IIB solution \be\lb{obtained2} g_{10}= (\det
U)^{3/4}H^{1/2}[H^{-1}(\det U)^{-1}g(E^{2,1})+(\det U)^{-1}
U_{ij}dx^i\cdot dx^j +H^{-1} d\theta^2], \ee \be\lb{B2}
B_i=A_i\wedge d\theta, \ee \be\lb{tau2}
\tau=-\frac{U_{12}}{U_{11}}+i\frac{\sqrt{\det U}}{U_{11}}, \ee
\be\lb{queseyo} i_{k}D=\pm\omega(E^{2,1})\wedge dH^{-1}. \ee If
instead (\ref{11sugra2}) is dimensionally reduced along a flat
direction it is obtained the IIA solution \be\lb{11sugra3}
g=g(E^{1,1})+U_{ij}dx^i\cdot dx^j+ U^{ij}(dt_i+A_i)(dt_j+A_j), \ee
\be\lb{sugra3} B=\omega(E^{1,1})H^{-1}, \ee \be\lb{dilaton}
\phi=-\frac{1}{2}\log(H). \ee A double dualization gives a new IIA
solution \be\lb{dualizo} g=H^{-1}g(E^{1,1})+U_{ij}dX^i\cdot dX^j,
\ee \be\lb{ah} B_i=A_i\wedge d\varphi^i+\omega(E^{1,1})H^{-1}, \ee
\be\lb{dilatame} \phi=\frac{1}{2}\log(\det U)-\frac{1}{2}\log(H).
\ee The lifting to eleven dimensions gives \be\lb{1extun} g_{11}=
H^{1/3}(\det U)^{2/3}[H^{-1}(\det U)^{-1}g(E^{1,1})+(\det U)^{-1}
U_{ij}dX^i\cdot dX^j + H^{-1}d\theta^2], \ee \be\lb{1extun2}
F=(F_i\wedge dt_i + \omega(E^{1,1})\wedge dH^{-1})\wedge d\theta.
\ee All the backgrounds presented in this section can be constructed
with a single F satisfying (\ref{backly}), but the dependence on
$(x_{\theta}, x_{\varphi})$ remains implicit.
\newpage

\section{$G_2$ toric metrics and
supergravity backgrounds}

   If the Bryant-Salamon extension is applied to the Calderbank-Pedersen
spaces the $U(1)\times U(1)$ isometry is preserved and the result is
a toric $G_2$ holonomy metric. In this subsection the $G_2$ holonomy
metrics corresponding to the examples A and B of subsection 2.4.7
will be constructed. We select such cases by simplicity, but the
procedure could be applied to any quaternion Kahler space as well.
By use of the results of the preceding section we extend this toric
$G_2$ holonomy metrics to vacuum configurations of the eleven
dimensional supergravity preserving at least one supersymmetry. Also
type IIA backgrounds will be found by reduction along one of the
isometries.

  It is straightforward to show that the explicit form
of (\ref{anzatz}) when the base space (and, in consequence, the
total one) has torus symmetry is \be\lb{quatera}
g=\frac{dr^2}{h(r)}+\frac{r^2}{2}[ U_{\phi\phi}d\phi^2+
U_{\phi\psi}d\phi d\psi+ U_{\psi\psi}d\psi^2+
Q_{\phi}d\phi+Q_{\psi}d\psi+ g_{\rho\rho}(d\eta^2+d\rho^2)+ H ], \ee
where it has been defined
$$
h(r)=1-4c/r^4
$$
$$
U_{11}=g_{\phi\phi} +h(r)\frac{u_1^2(\rho^2+\eta^2)+(u_2 \eta+u_3
\rho)^2}{2 \rho F^2},
$$
$$
U_{22}=g_{\psi\psi}+h(r)\frac{u_1^2+u_2^2}{2 \rho F^2},
$$
$$
U_{12}=U_{21}=g_{\phi\psi}+h(r)\frac{(u_1^2+u_2^2)\eta+u_2 u_3
\rho}{2 \rho F^2},
$$
$$
Q_{\phi}=h(r)\frac{1}{\sqrt{\rho}F}[u_1(\eta du_2+ \rho du_3)-(u_2
\eta+ u_3 \rho)du_1 -u_1(u_3 \eta-u_2 \rho)A^1],
$$
$$
Q_{\psi}=h(r)[\frac{u_1 du_2-u_2 du_1-u_1 u_3 A^1}{\sqrt{\rho}F}],
$$
$$
H=h(r)[|d\overrightarrow{u}|^2+(u_1^2+u_2^2)(A^1)^2-2 A^1 (u_3
du_2-u_2 du_3)].
$$

The second rank tensor $g_{ab}$ is the metric of the base manifold.
The product metric of (\ref{quatera}) with $M^4$ \be\lb{mamas}
g_{11}=ds_{M}^2+\frac{dr^2}{h(r)}+\frac{r^2}{2}[
U_{\phi\phi}d\phi^2+ U_{\phi\psi}d\phi d\psi+ U_{\psi\psi}d\psi^2+
Q_{\phi}d\phi+Q_{\psi}d\psi+ g_{\rho\rho}(d\eta^2+d\rho^2)+ H ]. \ee
is a vacuum configuration of the eleven dimensional supergravity
\cite{Duff}. With the help of the quantities
$$
\alpha_1=\frac{U_{22}Q_{\phi}-U_{12}Q_{\psi}}{\det
U},\;\;\;\alpha_2=\frac{U_{11}Q_{\psi}-U_{12}Q_{\phi}}{\det U},
$$
$$
h=H+U_{11}\alpha_{1}^2+2U_{12}\alpha_1\alpha_2+U_{22}\alpha_{2}^2,
$$
$$
\phi_1=\phi,\;\;\;\phi_2=\psi,
$$
the metric (\ref{mamas}) is expressed in more simple manner as
$$
g_{11}=ds_{M^4}^2+\frac{dr^2}{h(r)}+\frac{r^2}{2}[
U_{ij}(d\phi_i+\alpha_i)(d\phi_j+\alpha_j)+ h].
$$
The last expression takes the usual form of the Kaluza-Klein anzatz
\be\lb{colo}
g_{11}=e^{-\frac{2}{3}\varphi_{D}}G_{\mu\nu}dx^{\nu}dx^{\mu}+e^{\frac{4}{3}\varphi_{D}}(d\phi+dx^{\mu}C_{\mu}(x))^2.
\ee with the dilaton field and the RR 1-form defined by
$$
\varphi_D=\frac{3}{4}\log(\frac{r^2U_{11}}{2}),
$$
$$
C=\frac{U_{12}d\psi+Q_{\phi}}{U_{11}}.
$$
The usual reduction procedure \cite{Campbell} for(\ref{colo}) along
$\phi_1$ gives the following IIA metric in the string frame
 \be\lb{reduc}
g_{A}=(\frac{r^2U_{11}}{2})^{1/2}\{ds_{M^4}^2+\frac{dr^2}{h(r)}+\frac{r^2}{2U_{11}}
[\det U d\psi^2
+2(U_{11}Q_{\psi}-U_{12}Q_{\phi})d\psi-Q^2_{\phi}+U_{11}H ]\}. \ee
The components of (\ref{reduc}) are
$$
g_{\psi\varphi}=(\frac{r^2 U_{11}}{2})^{1/2}\frac{r^2}{4 U_{11}
F\sqrt{\rho}}h(r)[U_{11}\sin^2(\theta)
-U_{12}(\frac{\rho}{2}\sin(2\theta)\sin(\varphi)+\eta
\sin^2(\theta))]
$$
$$
g_{\psi\theta}=(\frac{r^2U_{11}}{2})^{1/2}\frac{r^2}{4
U_{11}F}h(r)U_{12}\sqrt{\rho}\cos(\varphi)
$$
$$
g_{\theta\theta}=(\frac{r^2U_{11}}{2})^{1/2}[\frac{r^2
h(r)}{4}-\frac{r^2 h^2(r)}{8 U_{11}\rho F^2}\rho^2 \cos^2(\varphi)]
$$
$$
g_{\varphi\varphi}=(\frac{r^2U_{11}}{2})^{1/2}[\frac{r^2
h(r)}{4}\sin^2(\theta)-\frac{r^2 h^2(r)}{8 U_{11}\rho F^2}
\sin^2(\theta)(\eta \sin(\theta)+\rho \cos(\theta)\sin(\varphi))^2]
$$
$$
g_{\varphi\theta}=(\frac{r^2U_{11}}{2})^{1/2} \frac{r^2 h^2(r)}{4
U_{11} F^2} \sin(\theta)\cos(\varphi)(\eta \sin(\theta)+\rho
\cos(\theta)\sin(\varphi)),
$$
where it have been introduced the spherical coordinates $\theta$,
$\varphi$ through the relations
$$
u_1=\sin(\theta)\cos(\varphi),\;\;\;u_2=\sin(\theta)\sin(\varphi),\;\;\;u_3=\cos(\theta).
$$
The range of this coordinates is $\theta \in [0,\pi]$ and $\varphi
\in [0,2\pi]$; the other components of the metric are identically
zero.

       The base metric (\ref{quaterna}) have a singularity at
$\rho\rightarrow 0$. For this case it is obtained
$$
U_{11}=\frac{\rho^2(1+3\rho)+\eta^2(9+7\rho)}{2\rho^5} +
 2h(r)[\frac{u_1^2(\rho^2+\eta^2)+(u_2 \eta+u_3 \rho)^2}{\rho^4}]\sim \frac{f(x^i)}{\rho^5},
$$
$$
U_{22}=\frac{8}{\rho^4}+2h(r)(\frac{u_1^2+u_2^2}{\rho^4})\sim
\frac{f(x^i)}{\rho^4},
$$
$$
U_{12}=U_{21}=\frac{8\eta}{\rho^4}+h(r)[\frac{(u_1^2+u_2^2)\eta+u_2
u_3 \rho}{\rho^4}] \sim \frac{f(x^i)}{\rho^4} ,
$$
$$
Q_{\phi}=h(r)\frac{2}{\rho^2}[u_1(\eta du_2+ \rho du_3)-(u_2 \eta+
u_3 \rho)du_1
$$
$$
-u_1(u_3 \eta-u_2 \rho)\frac{2 d\eta}{\rho}]\sim
\frac{f(x^i,dx^i)}{\rho^3},
$$
$$
Q_{\psi}=h(r)\frac{1}{\rho^2}(u_1 du_2-u_2 du_1-u_1 u_3 \frac{2
d\eta}{\rho})\sim \frac{f(x^i,dx^i)}{\rho^3},
$$
$$
H=h(r)[|d\overrightarrow{u}|^2+4(u_1^2+u_2^2)\frac{d\eta^2}{\rho^2}
-4\frac{d\eta}{\rho}(u_3 du_2-u_2 du_3)]\sim
\frac{f(x^i,dx^i)}{\rho^2}.
$$
where $x^i$ denotes all the coordinates except $\rho$ and the
behaviour for short distances was evaluated. The dilaton field is
given explicitly as
$$
\varphi_A=\frac{3}{4}\log\{\frac{r^2}{2}[\frac{\rho^2(1+3\rho)+\eta^2(9+7\rho)}{2\rho^5}]+
r^2h(r)[\frac{u_1^2(\rho^2+\eta^2)+(u_2 \eta+u_3
\rho)^2}{\rho^4}]\}\sim \log(\frac{f(x^i)}{\rho^5}).
$$
The expression for the R-R one form is
$$
C=\frac{2\rho\{8\eta+h(r)[(u_1^2+u_2^2)\eta+u_2 u_3 \rho
]\}d\psi}{\rho^2(1+3\rho)+\eta^2(9+7\rho) +
 4\rho h(r)[u_1^2(\rho^2+\eta^2)+(u_2 \eta+u_3 \rho)^2]}
$$
$$
+\frac{4\rho^2 h(r)[\rho u_1(\eta du_2+ \rho du_3)-\rho(u_2 \eta+
u_3 \rho)du_1 -2u_1(u_3 \eta-u_2
\rho)d\eta]}{\rho^2(1+3\rho)+\eta^2(9+7\rho) +
 4\rho h(r)[u_1^2(\rho^2+\eta^2)+(u_2 \eta+u_3 \rho)^2]}\sim f(x^i)\rho.
$$
The components of (\ref{reduc}) diverges in this case for short
$\rho$,
$$
g_{\psi\varphi}\sim \frac{f(x^i)}{\rho^5} ,\;\;\; g_{\psi\theta}\sim
\frac{f(x^i)}{\rho^4},\;\;\; g_{\theta\theta}\sim
\frac{f(x^i)}{\rho^2}
$$
$$
g_{\varphi\varphi}\sim \frac{f(x^i)}{\rho^2},\;\;\;
g_{\varphi\theta}\sim \frac{f(x^i)}{\rho}.
$$

   The quaternionic space (\ref{quaternb}) is singular too in the limit
$\rho \rightarrow 0$. Using it as a base space gives
$$
U_{11}=\frac{8\eta^4\rho^2(19+5\rho)+16\eta^6(9+7\rho)+\rho^6(1+35\rho)
+3\eta^2\rho^4(35+61\rho)}{9\rho^5(\rho^2-4\eta^2)^2(8\eta^4+6\eta^2\rho^2+3\rho^4)}
$$
$$
+8h(r)[\frac{u_1^2(\rho^2+\eta^2)+(u_2 \eta+u_3
\rho)^2}{9\rho^4(\rho^2-4\eta^2)^2}]\sim \frac{f(x^i)}{\rho^5} ,
$$
$$
U_{22}=\frac{64(4\eta^4+\rho^4)}{9\rho^4(\rho^2-4\eta^2)^2(8\eta^4+6\eta^2\rho^2+3\rho^4)}
+8h(r)[\frac{u_1^2+u_2^2}{9\rho^4(\rho^2-4\eta^2)^2}]\sim
\frac{f(x^i)}{\rho^4},
$$
$$
U_{12}=U_{21}=\frac{32(8\eta^5-4\eta^3\rho^2+3\eta\rho^4)}{9\rho^4(\rho^2-4\eta^2)^2(8\eta^4+6\eta^2\rho^2+3\rho^4)}
+8h(r)[\frac{(u_1^2+u_2^2)\eta+u_2 u_3
\rho}{9\rho^4(\rho^2-4\eta^2)^2}]\sim \frac{f(x^i)}{\rho^4},
$$
$$
Q_{\phi}=4h(r)\frac{1}{3\rho^2(4\eta^2-\rho^2)}\{u_1(\eta du_2+ \rho
du_3)-(u_2 \eta+ u_3 \rho)du_1
$$
$$
-u_1(u_3 \eta-u_2
\rho)[\frac{8\eta}{\rho^2-4\eta^2}d\rho+\frac{4(\rho^2-2\eta^2)}{\rho(\rho^2-4\eta^2)}d\eta]\}
\sim \frac{f(x^i,dx^i)}{\rho^3},
$$
$$
Q_{\psi}=h(r)\frac{4}{3\rho^2(4\eta^2-\rho^2)}\{u_1 du_2-u_2
du_1-u_1 u_3 [\frac{8\eta}{\rho^2-4\eta^2}d\rho+
\frac{4(\rho^2-2\eta^2)}{\rho(\rho^2-4\eta^2)}d\eta]\}\sim
\frac{f(x^i, dx^i)}{\rho^3},
$$
$$
H=h(r)\{|d\overrightarrow{u}|^2+(u_1^2+u_2^2)[\frac{8\eta}{\rho^2-4\eta^2}d\rho+\frac{4(\rho^2-2\eta^2)}{\rho(\rho^2-4\eta^2)}d\eta]^2
-2[\frac{8\eta}{\rho^2-4\eta^2}d\rho
$$
$$
+\frac{4(\rho^2-2\eta^2)}{\rho(\rho^2-4\eta^2)}d\eta](u_3 du_2-u_2
du_3)\}\sim \frac{f(x^i,dx^i)}{\rho}.
$$
The dilaton field is expressed through the relation
$$
e^{\frac{4}{3}\varphi_D}=\frac{r^2}{2}\{\frac{8\eta^4\rho^2(19+5\rho)+16\eta^6(9+7\rho)+\rho^6(1+35\rho)
+3\eta^2\rho^4(35+61\rho)}{9\rho^5(\rho^2-4\eta^2)^2(8\eta^4+6\eta^2\rho^2+3\rho^4)}
$$
$$
+8h(r)[\frac{u_1^2(\rho^2+\eta^2)+(u_2 \eta+u_3
\rho)^2}{9\rho^4(\rho^2-4\eta^2)^2}]\},
$$
from where follows that
$$
\varphi_D\sim \log(\frac{f(x^i)}{\rho^5}),\;\;\;\; \rho\rightarrow
0.
$$
The behavior of the RR one-form at short distances results
$$
C \sim f(x^i,dx^i)\rho.
$$
and the components of the IIA metric diverges as
$$
g_{\psi\varphi}\sim \frac{f(x^i)}{\rho^5} ,\;\;\; g_{\psi\theta}\sim
\frac{f(x^i)}{\rho^4},\;\;\; g_{\theta\theta}\sim
\frac{f(x^i)}{\rho^2}
$$
$$
g_{\varphi\varphi}\sim \frac{f(x^i)}{\rho^2},\;\;\;
g_{\varphi\theta}\sim \frac{f(x^i)}{\rho}.
$$

 A detailed analysis of the singularities of the backgrounds corresponding to
the $m$-pole solutions and their physical interpretation was given
in \cite{Angelova} and \cite{Angelito}, the interested reader may
consult those references. To finish we recall that the toric $G_2$
spaces presented here are not unique, other toric examples were also
constructed in \cite{Apostol} and match for application to M-theory
as well by an procedure analog of those presented here.
\newpage

\section{Conclusion}

    The main results presented in this dissertation are the following
\\

- We have shown that in $d=4$ weak hyperkahler torsion structures
are the same that hypercomplex structures and the same that the
Plebanski-Finley conformally invariant heavens. With the help of
this identification we have found the most general local form of an
hyperkahler torsion space in four dimensions. We also presented an
Ashtekar like formulation for them in which to finding an
hyperkahler torsion metric is reduced to solve a quadratic
differential system. We have shown that there exists more solutions
than the Callan-Harvey-Strominger ones, this are the structures for
which it does not exist a preserved volume form. Some examples with
isometries have been constructed.
\\

- It is found the most general form for the target space metric to
the moduli space metric of several $(n>1)$ identical matter
hypermultiplets for the type-IIA superstrings compactified on a
Calabi-Yau threefold, near conifold singularities, even taking into
account non-perturbative D-instanton quantum corrections. The metric
in consideration is "toric hyperkahler" if we do not take into
account the gravitational corrections. This generalize the
Ooguri-Vafa solution and has been obtained by using essentially the
same assumptions than them. Some new effects related to interference
between hypermultiplets have been presented, which do not hold in
the presence of a single hypermultiplet.
\\

- It is constructed a family of toric hyperkahler spaces in eight
dimensions by use of a matematical relation called Swann extension,
which relates quaternion Kahler and hyperkahler spaces in different
dimensions. The resulting hyperkahler space is expressed in a
coordinate system, for which it takes the generalized
Gibbons-Hawking type. The solution was lifted to a background of the
eleven dimensional supergravity preserving certain ammount of
supersymmetry, with and without the presence of fluxes. Several type
IIA and IIB backgrounds have been found by reduction along a circle
and by use of T-duality rules.
\\

- It is constructed a family of $Spin(7)$, $G_2$ and weak $G_2$
holonomy metrics. The result has been lifted to a supergravity
solution preserving one supersymmetry. The presence of a toric
symmetry allows a reduction to type IIA backgrounds by usual
reduction along one of the Killing vectors.
\\

    I acknowledge specially to my director A.P.Isaev.
I have been benefited during my studies with discussions with
A.Pashnev, S.Ketov, M.Tsulaia, C.Sochichiu, E.Ivanov, D. Mladenov,
B.Dimitrov, B.Saha and M.Paucar Acosta. Finally I would like to
acknowledge the support of CLAF and to dedicate this dissertation to
Professor L.Masperi.

\newpage

\appendix

\section{Supersymmetric sigma models}

      In this section we sketch why strong torsion hyperkahler geometry is the target
space metric of $N=4$ supersymmetric sigma models \cite{Alvarez}.
Sigma models in dimension two are described by Lagrangians of the
form
 \be\lb{sigma} I_{b}=-\frac{1}{2\pi}\int
d^2x [g_{AB}(\varphi)\partial_{\mu}\varphi^A\partial^{\mu}\varphi^B
+
 b_{AB}(\varphi)\epsilon^{\mu\nu}\partial_{\mu}\varphi^A\partial^{\nu}\varphi^B],
\ee being $g_{AB}$ an arbitrary metric tensor \footnote{In the main
text we have used latin indices (a,b,..) as flat and greek indices
($\alpha$, $\beta$,..) as curved. Instead in this appendix capital
latin indices (A,B,..) are curved, we use capital latin indices and
not greek in order to not confuse them with the greek indices used
to denote space-time derivates $\partial_{\mu}$ or gamma matrices
$\gamma^{\mu}$.}. The fields $\varphi^A$ are bosonic $(A=1,...,d)$
depending on the two coordinates $x^i$ ($i=1,2$). This mean that
$\varphi^A$ parameterize two-cycles on a d-dimensional target
manifold $M$ with a metric $g_{AB}$. The second term is a
Wess-Zumino-Witten type term.

    If we want to make the supersymmetric
extension of (\ref{sigma}) we need to include a number of fermion
degrees equal to the number of the boson ones. Thus we need to
introduce a set of spinors $\psi'^A$ and consider a supersymmetry
transformation which on general grounds should be of the form
\be\lb{sosos} \delta\varphi^A=\overline{\varepsilon}_1 J^A_B\psi'^B
\ee being $J$ some $(1,1)$ tensor. We still have the freedom to make
the redefinition $\psi'^B=(J^{-1})^B_A\psi^A$ in (\ref{sosos}) and
consider the transformation \be\lb{soso}
\delta\varphi^A=\overline{\varepsilon}_1 \psi^A \ee The $N=(1,0)$
supersymmetric extension of (\ref{sigma}) is \cite{Spindel}
\be\lb{action}
 I_{sb}=-\frac{1}{2\pi}\int d^2x
[g_{AB}(\varphi)\partial_{\mu}\varphi^A\partial^{\mu}\varphi^B +
 b_{AB}(\varphi)\epsilon^{\mu\nu}\partial_{\mu}\varphi^A\partial^{\nu}\varphi^B
 + \overline{\psi}^A
\gamma^{\mu}D_{\mu}\psi^B g_{AB}], \ee where the spinors $\psi^A$
are right-handed and $\varepsilon$ is left-handed and we have
defined the covariant derivative
$$
D_{\mu}\psi^A=\partial_{\mu}\psi^A + \Upsilon^A_{BC}\psi^B
\partial_{\mu}\varphi^C,
$$
$$
\Upsilon^A_{BC}=\Gamma^A_{BC}+\frac{1}{2}T^A_{BC}
$$
being $\Gamma^A_{BC}$ the Christofell symbols of the Levi-Civita
connection for $g_{AB}$. The torsion form $T_{ABC}$ is entirely
defined in terms of the Wess-Zumino-Witten potential $b_{AB}$ as
\be\lb{torso} 2T_{ABC}=-3b_{[AB,C]}\qquad \Longleftrightarrow \qquad
dT=0. \ee The indices $A,B$ are lowered and raised through the
metric $g_{AB}$. The action (\ref{action}) is invariant under
(\ref{soso}) and \be\lb{here} \delta
\psi^A=\gamma^{\mu}\partial_{\mu}\varphi^A
\varepsilon-\psi^B\overline{\varepsilon} S^A_{BC}\psi^C. \ee The
tensor $S_{ABC}$ is constrained by supersymmetry arguments to be
skew symmetric and covariantly constant
$$
S_{ABC;D}=S_{ABC,D}-3S_{E[AB}\Upsilon^E_{C]D}=0.
$$
The presence of a covariantly constant 3-form gives a new
restriction on the manifold. On group manifolds this condition has
solution and $S$ can be choose proportional to $T$ \cite{Spindel}.
We find the commutator \be\lb{tecnic} [\delta,
\delta']\varphi^A=\overline{\varepsilon}'\gamma^{\mu}\varepsilon
(2\partial_{\mu}\varphi^A
-\overline{\psi}^B\gamma_{\mu}S^A_{BC}\psi^C), \ee and correspond to
the usual supersymmetry algebra without the presence of central
charges.

      Let us consider now the problem to construct an $N=(2,0)$
supersymmetric extension of (\ref{sigma}). For this purpose we
should impose a new supersymmetry transformation to (\ref{action})
without spoiling (\ref{soso}) and (\ref{here}). As we saw the
general supersymmetry variation for $\varphi^A$ is of the form
\be\lb{ground} \delta^1\varphi^A=\overline{\varepsilon}_1
(J^{1})^A_B\psi^B, \ee being $J^1$ certain $(1,1)$ tensor. Due to
the invariance of (\ref{action}) with respect to the other symmetry
(\ref{here}) we have to impose \be\lb{no} I_{sb}(\varphi,
J^1\psi)=I_{sb}(\varphi,\psi). \ee Formula (\ref{no}) implies that
\be\lb{eil} g_{AB}(J^{1})_{C}^{B} + (J^{1})_{A}^{B}g_{BC} = 0, \ee
and then the metric is hermitian with respect to the tensor $J^{1}$.
Condition (\ref{no}) also implies that \be\lb{equon}
D_{A}(J^{1})_{C}^{B}=0\qquad \Longrightarrow \qquad
(J^{1})_{[AB,C]}=-2(J^{1})^{D}_{[A}T_{B,C]D}. \ee This mean that
$J^{1}$ is covariantly constant with respect to the connection with
torsion. The compatibility condition of (\ref{equon}) is
\be\lb{integrab} (J^{1})^A_{E}R^E_{BCD}=R^A_{ECD}(J^{1})^E_{B}, \ee
and more generally the tensor $J^{1}$ commute with all the
generators of the holonomy group. The new supersymmetry
transformations result
$$
\delta^{1} \varphi^A=\overline{\varepsilon}_{1}(J^{1})^A_B\psi^B.
$$
\be\lb{heru} \delta^1 \psi^A=\gamma^{\mu}\partial_{\mu}\varphi^B
(J^{1})^A_B \varepsilon_{1} -[(J^{1})^A_D \Upsilon^D_{BE}
(J^{1})^E_C + (S_{1})^A_{BC}]\psi^B\overline{\varepsilon}_{1}\psi^C.
\ee The tensor $(S_{-})^A_{BC}$ is not determined by other
quantities. The commutator of the two supersymmetries acting over
$\varphi^A$ is
$$
[\delta,
\delta^1]\varphi^A=\overline{\varepsilon}_{1}\gamma_{\mu}\varepsilon_{1}
[(J^{1})^A_B + (J^{1})^B_A]\partial_{\mu}\varphi^B
$$
\be\lb{commo} +
\overline{\varepsilon}_{1}\gamma_{\mu}\varepsilon_{1}
\overline{\psi}^B\gamma_{\mu}\psi^C (J^1)^A_{D} N^D_{BC}, \ee plus
terms depending on $(S_{1})^A_{BC}$. Then the usual supersymmetric
algebra holds if (\ref{commo}) is zero and this implies that
\be\lb{nieu} (J^{1})^A_B + (J^{1})^B_A=0,\qquad N_{AB}^C=0,\qquad
(S_{1})^A_{BC}=0. \ee The first equation (\ref{nieu}) together with
(\ref{eil}) implies that $J^{1}$ is an almost complex structure and
therefore the dimension of the target manifold should be even
(d=$2n$). The second (\ref{nieu}) show that $J^{1}$ is integrable.
The metrics $g_{AB}$ for which (\ref{equon}) and (\ref{nieu}) hold
are Kahler torsion. We conclude that $N=2$ supersymmetric sigma
models occur on \emph{Kahler torsion manifolds}.

      If there is a third supersymmetry corresponding to another
complex structure $J^2$ and to a parameter $\varepsilon_2$ then it
is obvious that the previous reasoning is true and the properties
(\ref{eil}), (\ref{equon}) and (\ref{nieu}) hold for $J^2$.  But we
obtain new restrictions by requiring that the transformation
corresponding to $\varepsilon_1$ and $\varepsilon_2$ close to a
supersymmetry. The algebra \be\lb{algebra} [\delta(\varepsilon_1),
\delta(\varepsilon_2)]\varphi^A=2
\overline{\varepsilon}^i_2\gamma^{\mu}\varepsilon^j_1 \delta_{ij}
\partial_{\mu}\varphi^A
\ee (which correspond to (\ref{tecnic}) with $S^A_{BC}=0$) is
obtained if and only if \be\lb{cuatro}
 J^i\cdot J^j + J^j\cdot J^i = \delta_{ij}I
\ee \be\lb{cruzado}
N^{ij}(X,Y)=[X,Y]+J^{i}[X,J^{j}Y]+J^{i}[J^{j}X,Y]-[J^{i}
X,J^{j}Y]=0. \ee $N^{ij}(X,Y)$ is known as the mixed Niejenhuis
tensor.

  Let us define the tensor $J^3=J^1 \cdot J^2$ for a $N=(3,0)$ supersymmetric
sigma model. It follows from (\ref{cuatro}) that
$$
(J^3)^2=-I
$$
and therefore $J^3$ is also an almost complex structure. It is
direct to check \be\lb{lie} J^i\cdot J^j=-\delta_{ij}+
\epsilon^{ijk}J^k. \ee
 The integrability of
$J^1$ and $J^2$ implies the integrability of $J^3$ and
(\ref{cruzado}). Also (\ref{equon}) for $J^1$ and $J^2$ implies that
\be\lb{shushu} D_A(J^3)^B_C=0. \ee Therefore $N=(3,0)$ supersymmetry
implies $N=(4,0)$ supersymmetry for sigma models and from
(\ref{cruzado}), (\ref{torso}), (\ref{equon}) and (\ref{lie}) it
follows that the target metric $g_{AB}$ of a $N=(4, 0)$
supersymmetric sigma model is always \emph{strong hyperkahler
torsion}. If the form $b_{AB}$ is zero, then the geometry presented
here reduce to the usual Kahler and hyperkahler ones.

   The weak cases are also of interest in the context of low energy
heterotic string actions because they contain a three form that is
not closed due to a Green-Schwarz anomaly. For more details of this
assertions see \cite{Callan}-\cite{Stromin}, and for more
applications of this geometry in the construction of sigma models
see \cite{Gobe}-\cite{Krivo2}.

\end{document}